\documentclass[a4paper,11pt,reqno]{article}
\usepackage{a4wide}

\usepackage{amsmath,amsfonts,amssymb}
\usepackage[T1]{fontenc}
\usepackage[p,osf]{cochineal}
\usepackage[varqu,varl,var0]{inconsolata}
\usepackage[scale=.95,type1]{cabin}
\usepackage[vvarbb]{newtxmath}
\usepackage[cal=boondoxo]{mathalfa}
\usepackage[mathscr]{euscript}
\usepackage{bbold} 

\usepackage{lettrine} 
\usepackage{pict2e}

\makeatletter
\DeclareRobustCommand{\loplus}{\mathbin{\mathpalette\dog@lsemi{+}}}
\DeclareRobustCommand{\lotimes}{\mathbin{\mathpalette\dog@lsemi{\times}}}
\DeclareRobustCommand{\roplus}{\mathbin{\mathpalette\dog@rsemi{+}}}
\DeclareRobustCommand{\rotimes}{\mathbin{\mathpalette\dog@rsemi{\times}}}

\newcommand{\dog@rsemi}[2]{\dog@semi{#1}{#2}{-90,90}}
\newcommand{\dog@lsemi}[2]{\dog@semi{#1}{#2}{270,90}}
\newcommand{\dog@semi}[3]{%
  \begingroup
  \sbox\z@{$\m@th#1#2$}%
  \setlength{\unitlength}{\dimexpr\ht\z@+\dp\z@\relax}%
  \makebox[\wd\z@]{\raisebox{-\dp\z@}{%
    \begin{picture}(1,1)
    \linethickness{\variable@rule{#1}}
    \roundcap
    \put(0.5,0.5){\makebox(0,0){\raisebox{\dp\z@}{$\m@th#1#2$}}}
    \put(0.5,0.5){\arc[#3]{0.5}}
    \end{picture}%
  }}%
  \endgroup
}
\newcommand{\variable@rule}[1]{%
  \fontdimen8  
  \ifx#1\displaystyle\textfont3\else
    \ifx#1\textstyle\textfont3\else
      \ifx#1\scriptstyle\scriptfont3\else
        \scriptscriptfont3\relax
  \fi\fi\fi
}
\makeatother

\usepackage{nicefrac}
\usepackage{setspace}
\usepackage{datetime}
\usepackage[nosort]{cite}
\usepackage{upgreek}

\usepackage{comment}

\usepackage[euler]{textgreek}
\usepackage[breaklinks=true]{hyperref}
\usepackage{cleveref}

\numberwithin{equation}{section}

\hypersetup{
			colorlinks=true,
			urlcolor=blue,
			citecolor=magenta,
			linkcolor=blue,
     }

\let\oldsqrt\sqrt
\def\sqrt{\mathpalette\DHLhksqrt}
\def\DHLhksqrt#1#2{%
\setbox0=\hbox{$#1\oldsqrt{#2\,}$}\dimen0=\ht0
\advance\dimen0-0.2\ht0
\setbox2=\hbox{\vrule height\ht0 depth -\dimen0}%
{\box0\lower0.4pt\box2}}

\def\crbig{\\\noalign{\vspace{1.1mm}}}
\newcommand{\RNum}[1]{\uppercase\expandafter{\romannumeral #1\relax}}

\author{
  \begin{minipage}{.97\linewidth}
    \vspace{1cm}
       \begin{center}
      \begin{small}
             \textbf{Andrea Campoleoni},$^{1}$\footnote{Research associate of the Fund  for Scientific Research -- FNRS, Belgium.} \hspace{1pt} 
             \textbf{Arnaud Delfante},$^{1}$\footnote{FRIA grantee  of the Fund  for Scientific Research -- FNRS, Belgium.}
 \hspace{1pt}  
             \textbf{Simon Pekar},$^{1\hspace{0.25pt}\text{\textdagger}, 2}$ \hspace{1pt}
      \textbf{P. Marios Petropoulos},$^2$\\ 
      \textbf{David Rivera-Betancour}$^2$ and 
      \textbf{Matthieu Vilatte}$^{2,3}$
              \end{small}
    \end{center}
    \vspace{0.5cm}
    \hspace{2.4cm}\begin{minipage}{.7\linewidth}
\begin{center}     {\it \begin{footnotesize}
\hbox{\kern-1.8cm\vbox{\vskip0cm
 \begin{itemize}
               \item[$^1$]Service de Physique de l'Univers,\\ 
               Champs et Gravitation\\
               Universit{\'e} de Mons -- UMONS\\
               20 place du Parc, 7000 Mons, Belgium\\ 
                           \vskip0.3cm
      \end{itemize}}
\kern-3.2cm\vbox{
\begin{itemize}
 \item[$^2$]Centre de Physique Th\'eorique -- CPHT\\ 
        \'Ecole polytechnique, CNRS\footnote{\emph{Centre National de la Recherche Scientifique}, Unit\'e Mixte de Recherche UMR 7644.}\\
        Institut Polytechnique de Paris\\
        91120 Palaiseau Cedex, France         
      \end{itemize}
      \vskip0.cm
}}
     \end{footnotesize}}
\end{center}
    \end{minipage}
    \vspace{0.5cm}\begin{minipage}{.7\linewidth}
\begin{center}     
{\it \begin{footnotesize}
\hbox{\kern0.6cm\vbox{\vskip2cm
}
\kern3.9cm\vbox{
\begin{itemize}
 \item[$^3$]Division of Theoretical Physics\\School of Physics\\
  Aristotle University of Thessaloniki\\ 
  54124 Thessaloniki, Greece
      \end{itemize}\vskip0.05cm
}
}
     \end{footnotesize}}
\end{center}
     \end{minipage}
  \end{minipage}
}

\title{\vspace{1.5cm}
 \boldmath \begin{LARGE}
    \textbf{\textsc{Flat from anti de Sitter}}
  \end{LARGE} \unboldmath
}

\date{}

\begin{document}

%\blinddocument
%\blindmathpaper

\begin{titlepage}
\maketitle
\thispagestyle{empty}

 \vspace{-12.cm}
  \begin{flushright}
  CPHT-RR054.082023\\
  \end{flushright}
 \vspace{11.cm}

\begin{center}
\textsc{Abstract}\\  
\vspace{1. cm}	
\begin{minipage}{1.0\linewidth}

Ricci-flat solutions to Einstein's equations in four dimensions are obtained as the flat limit of Einstein spacetimes with negative cosmological constant. In the limiting process, the anti-de Sitter energy--momentum tensor is expanded in Laurent series in powers of the cosmological constant, endowing the system with the infinite number of boundary data, characteristic of an asymptotically flat solution space. The governing flat Einstein dynamics is recovered as the limit of the original energy--momentum conservation law and from the additional requirement of the line-element finiteness, providing at each order the necessary set of flux-balance equations for the boundary data. This analysis is conducted using a covariant version of the Newman--Unti gauge designed for taking advantage of the boundary Carrollian structure emerging at vanishing cosmological constant and its Carrollian attributes such as the Cotton tensor. 

\end{minipage}
\end{center}

%\vspace{2cm} 

\end{titlepage}

\onehalfspace

%\noindent\rule{\textwidth}{1.2pt}
%\vspace{-1cm}
\begingroup
\hypersetup{linkcolor=black}
\tableofcontents
\endgroup
\noindent\rule{\textwidth}{0.6pt}

\section{Introduction}

The solution space of Einstein's equations and the corresponding asymptotic symmetries are severely altered by the presence of a cosmological constant $\Lambda$.\footnote{See \cite{Ruzziconi:2019pzd} for a recent review and further reading suggestions.} Firstly, asymptotically flat spacetimes support incoming and outgoing gravitational radiation, which are harder to accommodate in asymptotically anti-de Sitter --- unless leaky boundary conditions are assumed \cite{Fiorucci:2021pha}. Secondly, the number of free
functions on the
boundary characterizing the solution space is finite for non-zero $\Lambda$ and infinite for $\Lambda=0$. Hence, investigating the holographic description of Ricci-flat spacetimes from the limit of Einstein spacetimes with non-vanishing cosmological constant seems at best a futile task, limited to special cases like three spacetime dimensions.
 
The purpose of the present work is to reconsider this statement in four dimensions and show that expanding the anti-de Sitter energy--momentum tensor in Laurent series in $k^2=\nicefrac{-\Lambda}{3}$, one recovers the full Ricci-flat solution space in a $\nicefrac{1}{r}$-expansion together with its evolution dynamics captured in flux-balance equations.

In order to perform the above analysis, a choice of gauge is required, as usual. Baring in mind potential further developments towards flat holography, it is desirable to privilege null infinity in the asymptotically flat instance, which plays the role of a conformal boundary hosting all 
independent functions of the solution space, often referred to as degrees of freedom in the following. The null boundary is a three-dimensional Carrollian manifold and it is therefore convenient to select a bulk gauge making the corresponding boundary general and Weyl covariance manifest. This has prompted to choose a modified version of the Newman--Unti gauge \cite{CMPPS2}. This gauge is built upon an incomplete gauge fixing that is expected to lead to an enhancement of the asymptotic symmetries with respect to more customary gauges like the Bondi one, in analogy with what has been observed in three bulk dimensions in several examples of incomplete gauge fixings \cite{Troessaert:2013fma, Perez:2016vqo, Grumiller:2016pqb, Grumiller:2017sjh, oblak, Campoleoni:2018ltl, CMPR, CMPRpos, Alessio:2020ioh, Campoleoni:2022wmf}. We will not pursue this direction here, although it has attracted recent interest ---~see, e.g., \cite{Nguyen:2020hot,Geiller:2022vto} where in the latter reference the gauge under consideration was dubbed ``differential Newman--Unti''~--- and we will focus on the comparison between the space of solutions of Einstein's equations in the asymptotically anti-de Sitter and flat cases in the chosen gauge.
Reference \cite{CMPPS2} has set the stage for the gauge we will describe here, although it was originally circumscribed to the restricted class of algebraically special solutions (for AdS, see \cite{Haack:2008cp, Bhattacharyya:2008jc, Caldarelli:2012cm, Mukhopadhyay:2013gja, Gath:2015nxa, Petropoulos:2015fba, Petkou:2015fvh}).
 
In a nutshell,  the starting point is the anti-de Sitter case, where the solution space admittedly consists of the boundary metric and the boundary energy--momentum tensor, which is covariantly conserved as a consequence of bulk Einstein's equations. The analogue of the ``Bondi shear'' (sometimes referred to as ``dynamical shear'' later) is not an independent piece of data because Einstein's equations require it be proportional with a $k$-dependent factor to the geometric shear of the gauge congruence --- already part of the solution space. We move to the asymptotically flat instance by sending $k$ to zero, expanding the energy--momentum tensor in powers of $k^2$, trading the geometric shear for the dynamical shear along the lines of \cite{Compere:2019bua},
and requiring the bulk line element to remain finite. This imposes further evolution equations for the new degrees of freedom at every order in the radial $\nicefrac{1}{r}$ expansion, which supplement the energy--momentum conservation in the zero-$k$ limit. The resulting infinite set defines the flux-balance equations, which can otherwise be obtained directly by solving Einstein's equations without cosmological constant. 

Besides reaching the correct boundary Ricci-flat dynamics and tracing the AdS origin of the asymptotically flat solution space, the method presented here delivers Carroll-covariant flux-balance equations revealing the entire freedom for the choice of the boundary Carrollian geometry. The pattern involves the general and Weyl-covariant gauge mentioned earlier, which naturally incorporates 
the Cotton tensor of the anti-de Sitter boundary, or its Carrollian emanations in the asymptotically flat situation
(see \cite{MOPRS}).  
This tensor carries information on the gravitational radiation and plays a pivotal role for the description of magnetic charges \cite{Mittal:2022ywl}.

We begin our presentation by defining the covariant Newman--Unti gauge for asymptotically anti-de Sitter spaces. Along the way, we review its boundary Weyl covariance, as well as a useful decomposition of the boundary energy--momentum and Cotton tensors. We then provide an on-shell expression  of the line element up to order $\nicefrac{1}{r^2}$, where $r$ is a null radial coordinate, infinite at the conformal boundary. The flat limit, following the prescription summarized above, is carried out in the upcoming section, after a precise setting of the boundary Carrollian structure consecutive to the zero-$\Lambda$ limit. We show how the new degrees of freedom resulting from the expansion of the anti-de Sitter energy--momentum tensor are sorted out, how they enter the metric and how flux-balance equations are reached. 
Two appendices complement the main exposition, bringing about the necessary tools of Carrollian geometry in arbitrary frames (strong Carroll structures, connections, curvatures) as well as showing how such structures can be attained from pseudo-Riemannian geometries at vanishing speed of light. A presentation of the Carrollian descendants of the Cotton tensor closes this article.

\section{Einstein spacetimes in covariant Newman--Unti gauge}\label{ESMNU}

\subsubsection*{Choosing a boundary-covariant gauge}

Due to the Fefferman--Graham ambient metric construction \cite{PMP-FG1, Anderson:2004yi, PMP-FG2}, 
asymptotically locally anti-de Sitter four-dimensional spacetimes are determined by a set of independent boundary data, namely  a  three-dimensional  metric  
$\text{d}s^2=g_{\mu\nu}\text{d}x^\mu\text{d}x^\nu$ and a rank-two tensor
$\text{T}=T_{\mu \nu}\text{d}x^\mu\text{d}x^\nu$,
 symmetric ($T_{\mu\nu}=T_{\nu\mu}$), traceless ($T^\mu_{\hphantom{\mu}\mu}=0$) and conserved:\footnote{The covariant derivative $\nabla$ stands for the boundary Levi--Civita connection. Indices $\mu,\nu,\ldots \in \{0,1,2\}$ fill in the boundary natural frame, whereas $i,j, \ldots\in\{1,2\}$ are associated with spatial sections. }
\begin{equation}
\label{T-cons}
\nabla^\mu T_{\mu\nu}=0.
    \end{equation}
This construction is reached by setting a homonymous gauge, imposing fall-offs and solving Einstein's equations order by order in powers of the radial space-like coordinate.\footnote{ Residual symmetries constrain the possible terms entering each order in the radial expansion, thus simplifying the process of solving Einstein's equations. The constraints imposed by the boundary Weyl symmetry were studied in the Fefferman--Graham gauge in \cite{Schwimmer:2003eq} and will play an important role too in the covariant Newman--Unti gauge discussed in the following.} At every order in this expansion, the line element is determined by a tensor $G_{\mu\nu}^{(s)}$, fixed by Einstein's equations in terms of $g_{\mu\nu}=G_{\mu\nu}^{(-2)}$, $T_{\mu\nu}=\frac{3k}{16\pi G}G_{\mu\nu}^{(1)}$ and their derivatives (here the conformal boundary is located at $\rho \rightarrow +\infty$):\begin{equation}
\label{FGmet}
\text{d}s^2_{\text{Einstein}}
=\frac{\text{d}\rho^2}{(k\rho)^2} +\sum_{s\geq -2} \frac{1}{(k\rho)^s}G_{\mu\nu}^{(s)}(x) \text{d}x^\mu  \text{d}x^\nu.
    \end{equation}
The conservation \eqref{T-cons} is itself a consequence of Einstein's dynamics.

Fefferman--Graham's gauge is covariant with respect to the three-dimensional pseudo-Riemannian boundary $\mathscr{M}$. It can also be 
modified so as to make it
 Weyl-covariant \cite{ciamlei,Jia:2021hgy,Jia:2023gmk, Ciambelli:2023ott}. However, it does not admit a smooth vanishing-$k$ limit. Alternative gauges are Bondi or Newman--Unti  \cite{Bondi1962,Sachs1962-2, NU1962}, valid regardless of the cosmological constant, but not covariant with respect to the boundary. Let us consider for concreteness the Newman--Unti gauge with radial coordinate $r$.\footnote{Both in Bondi and Newman--Unti, $\partial_r$ is tangent to a null geodesic congruence. In Newman--Unti gauge this congruence is affinely parameterized, in contrast to Bondi. This enables to parallelly transport a canonical null tetrad and make contact with Newman--Penrose formalism \cite{NP68}.} The line element reads:
\begin{equation}
\label{bulkNU}
\text{d}s^2_{\text{bulk}}= \frac{V}{r}\text{d}u^2-2 
%\text{e}^{2\beta}
\text{d}u\text{d}r+G_{ij} \left(\text{d}x^i-U^i\text{d}u\right)\left(\text{d}x^j-U^j\text{d}u\right),
\end{equation}
where $V$, $U_i$ and $G_{ij}$ are functions of \emph{all} coordinates. They are treated as power series of $r$, possibly including logarithms,\footnote{Logarithms also appear in the Fefferman--Graham gauge when the boundary dimension is even \cite{Henningson:1998gx, deHaro:2000vlm}.}  with coefficients depending on boundary coordinates $x$: the retarded time $u$
%$u=\nicefrac{x^0}{k}$ 
and the angles $\mathbf{x}$. 

The bulk metric \eqref{bulkNU} is stable neither under boundary diffeomorphisms $x\to x'$, nor under Weyl rescalings $r\to r \mathcal{B}(x) $, and these are the features we would like to implement in a ``covariantized'' version of the gauge at hand.  To this end, we trade $-k^2\text{d}u$ for a boundary one-form $\text{u}=u_\mu \text{d}x^\mu$, which is an invariant object dual to a time-like vector field normalized at $-k^2$. As it will become manifest in Sec.~\ref{brrm}, where the timelike conformal boundary will become a null manifold equipped with a Carrollian structure in the limit $k \to 0$, our parameterization has been chosen such that $k=\sqrt{\nicefrac{-\Lambda}{3}}$ plays the role of effective boundary velocity of light. Therefore, the previous substitution amounts to choosing a time-like boundary congruence that could be interpreted as a hydrodynamic velocity field, if the boundary energy--momentum tensor were associated with a fluid. This is not necessarily so because the hydrodynamic regime requires constitutive relations, which are not obeyed everywhere in the Einstein solution space. The subspace where this happens is the realm of fluid/gravity correspondence  \cite{Haack:2008cp, Bhattacharyya:2008jc}. For convenience, we will nonetheless refer to $\text{u}$ as the ``velocity field'' and decompose the energy--momentum tensor accordingly. 

Introducing a boundary congruence provides also the appropriate tool for addressing Weyl invariance. In the spirit of the Fefferman--Graham ambient construction, 
the bulk geometry should be insensitive to a Weyl rescaling of the boundary metric  (weight $-2$) and of the boundary velocity form (weight $-1$) 
\begin{equation}
\label{conmet}
\text{d}s^2\to  \frac{\text{d}s^2}{{\cal B}^2}, \quad \text{u} \to  \frac{\text{u}}{\mathcal{B}} ,
\end{equation}
which should be reabsorbed into a redefinition of the radial coordinate:  $r\to r \mathcal{B}$. This requires to introduce a Weyl connection one-form $\text{A}=A_\mu \text{d}x^\mu$ transforming as 
\begin{equation}
\label{Atran}
\text{A}\to\text{A}-\text{d}\ln \mathcal{B}, 
\end{equation}
and suggests the following amendment to the Newman--Unti gauge 
\begin{equation}
\label{NUcov}
-\text{d}u\text{d}r\to \frac{\text{u}}{k^2}(\text{d}r+r \text{A}),
\end{equation}
which is indeed Weyl-invariant, as well as being boundary-general-covariant.

We can follow the suggested pattern and recast \eqref{bulkNU}, ignoring the logarithms,\footnote{Logarithms of the radial coordinate might or might not be required depending on the gauge chosen to investigate the space of solutions. In some cases, like, e.g., when choosing the Fefferman--Graham gauge in odd spacetime dimensions, they are necessary to reconstruct the full solution space. In other cases, like, e.g., in asymptotically flat spacetimes in the Bondi gauge, they describe an independent sector of the solution space that might be added or not to the polynomial expansion. Our case fits in the latter class and performing a thorough investigation of the larger space of solutions including logarithms and analyzing its interplay with residual symmetries (see, e.g., \cite{Fiorucci:2021pha,Compere:2019bua, Compere:2020lrt, Fiorucci:2020xto, Fuentealba:2022xsz}) is not part of our present agenda.} avoiding the demarcation of angular and time directions, and reorganizing the expansion in terms of boundary tensors according to their transversality with respect to the congruence $\text{u}$ as well as their conformal weights:
\begin{eqnarray}
\text{d}s^2_{\text{bulk}} &=&
2\frac{\text{u}}{k^2}(\text{d}r+r \text{A})
+r^2\text{d}s^2+ r \mathscr{C}_{\mu\nu}\text{d}x^\mu\text{d}x^\nu
+\frac{1}{k^4}\mathscr{F}_{\mu\nu}\text{d}x^\mu\text{d}x^\nu
\nonumber\\
&&+\sum_{s=1}^\infty
\frac{1}{r^s}\left( f_{(s)} \frac{\text{u}^2}{k^4} 
+2\frac{\text{u}}{k^2} f_{(s)\mu} \text{d}x^\mu+ f_{(s)\mu\nu} \text{d}x^\mu\text{d}x^\nu\right).
\label{newcovNU}
\end{eqnarray}
In this expression $ f_{(s)}$ are \emph{boundary scalars,} $f_{(s)\mu}$ \emph{boundary transverse vectors,} $ f_{(s)\mu} u^\mu=0$, and   
$f_{(s)\mu\nu}$ \emph{boundary symmetric and transverse tensors,} $ f_{(s)\mu\nu} u^\mu=0$. Their conformal weights are $s+2$, $s+1$ and~$s$. The $r^2$ term defines the boundary metric $\text{d}s^2$, which is a free piece of data in the spirit of \cite{Campoleoni:2018ltl, CMPR, CMPRpos, Campoleoni:2022wmf, Compere:2019bua}. As long as the bulk metric \eqref{newcovNU} is off-shell, the boundary symmetric tensors $\mathscr{F}_{\mu\nu}$ (weight $0$) and $\mathscr{C}_{\mu\nu}$ (weight $-1$) have no reason to be transverse with respect to $\text{u}$. 
The latter is the shear of the affine null geodesic congruence tangent to $\partial_r$ aka ``Bondi'' shear.\footnote{Strictly speaking, the Bondi shear is defined in the BMS gauge (in the expansion $G_{ij}=r^2q_{ij}+\mathcal{O}(r)$ the two-dimensional metric $q_{ij}$ is fixed to be the round sphere) with a prominent role in the asymptotically flat instance. Normally it is related to the one introduced here by an inhomogeneous transformation. \label{bondishear}} Imposing Einstein's equations will determine all the boundary tensors introduced so far in terms of basic independent functions that define the solution space. As we will see, this set of functions includes $\text{u}$ as well as the boundary metric $\text{d}s^2$ and a rank-two symmetric, traceless and conserved tensor
coinciding with the energy--momentum tensor of the Fefferman--Graham gauge.

Before moving on to Einstein's equations, a few comments are worth making to appreciate the covariant Newman--Unti gauge 
\eqref{newcovNU}. Introducing a normalized but otherwise arbitrary boundary congruence amounts to the on-set of two degrees of freedom, i.e. to a relaxation of the original Newman--Unti  gauge fixing. Incomplete gauge fixings might produce enhancements of asymptotic symmetries and materialize in extra charges --- not always integrable or conserved. They have been investigated mostly in three bulk dimensions \cite{Troessaert:2013fma, Perez:2016vqo, Grumiller:2016pqb, Grumiller:2017sjh, oblak, Campoleoni:2018ltl, CMPR, CMPRpos, Alessio:2020ioh, Campoleoni:2022wmf},
%(see footnote \ref{incomplete}), 
where the introduction of an arbitrary congruence\footnote{Whenever the energy--momentum tensor empowers a fluid interpretation, the conruence at hand is interpreted as the fluid lines and its arbitrariness portrays the relativistic hydrodynamic-frame invariance \cite{Landau, Eckart,  BigFluid}. This feature is however strictly  \emph{local} because the bulk diffeomorphisms associated with the boundary hydrodynamic-frame transformations are possibly charged. These properties have been thoroughly investigated in two boundary dimensions  \cite{Campoleoni:2018ltl, CMPR, CMPRpos, Campoleoni:2022wmf} and would undoubtedly deserve a generalization in higher dimensions, which is outside our scope here. It would better fit a broader study where frame orthonormality would be downsized, probing general boundary linear transformations.} combined with the freedom of choosing the boundary metric restores the boundary local Lorentz symmetry and its realization as bulk diffeomorphisms, augmenting the asymptotic symmetry group \cite{Campoleoni:2018ltl, CMPR, CMPRpos, Alessio:2020ioh, Campoleoni:2022wmf}.  
Following \cite{CMPR}, an elegant way of taming this information without redundancy is to express the boundary metric in an arbitrary 
orthonormal Cartan coframe,\footnote{We use $A,B, \ldots \in \left\{\hat 0,\hat 1, \hat 2\right\}$ as  boundary ``flat'' indices with $a,b, \ldots \in \left\{\hat 1, \hat 2\right\}$. The parameterization of the coframe in terms of $8$ arbitrary functions, suitable for the Carrollian limit, is provided in App. \ref{carman}, Eq. \eqref{metgenu}.}
\begin{equation}
\label{metgenuortho}
\text{d}s^2=\eta_{A B}\uptheta^{A}\uptheta^{B}
=-\left(\uptheta^{\hat 0}\right)^2 + \delta_{a b}\uptheta^{a}\uptheta^{b},
\end{equation}
and set 
\begin{equation}
\label{veloform}
\text{u}=-k\uptheta^{\hat 0}.
\end{equation}
 The dual frame vectors are $\left\{\text{e}_A\right\}=\left\{\text{e}_{\hat 0}, \text{e}_a\right\}$ with $\uptheta^{B}\left(\text{e}_A\right)=\delta^B_A$. A possible parameterization of the frame, which we will not use explicitly though, is displayed in Eqs. \eqref{mutvrel}, 
\eqref{upsilontvarel} and \eqref{metgenu}. 
  
We will not delve into the analysis of asymptotic symmetries in the present note.  Due to the partial relaxation of the gauge this complementary task is more intricate and deserves a separate and thorough treatment  \cite{Campoleoni:2018ltl, CMPR, CMPRpos, Alessio:2020ioh, Campoleoni:2022wmf, Nguyen:2020hot,Geiller:2022vto}. 

In order to proceed with the covariant Newman--Unti gauge \eqref{newcovNU} and impose Einstein's equations, it is desirable to list the available tensors with the correct conformal weights at each order $s$ of the radial expansion. To achieve this, we need to cope with Weyl covariance and decompose the energy--momentum tensor with respect to the chosen congruence. 

\subsubsection*{Kinematics, Weyl covariance and transverse duality}

Covariantization with respect to Weyl transformations requires to introduce a connection one-form $\text{A} = A_A \theta^A$, built on the congruence $\text{u}= u_A \theta^A$:
\begin{equation}
\label{Wconc}
\text{A}=\frac{1}{k^2}\left(\text{a} -\frac{\Theta}{2} \text{u}\right),
\end{equation} 
which transforms as anticipated in \eqref{Atran}. In this expression $\text{a}= a_A \theta^A$ and $\Theta$ are the acceleration and expansion of the congruence $ \text{u}$, defined together with the shear and the vorticity as\footnote{Our conventions for (anti-) symmetrization are:
$A_{(AB)}=\frac{1}{2}\left(A_{AB}+A_{BA}\right)$ and $ 
A_{[AB]}=\frac{1}{2}\left(A_{AB}-A_{BA}\right)$.}
\begin{eqnarray}
&a_A =u^B \nabla_B u_A , \quad
\Theta=\nabla_A  u^A ,& \label{def21}\\
&\sigma_{A B}= \nabla_{(A } u_{B )} + \frac{1}{k^2}u_{(A } a_{B )} -\frac{1}{2} \Theta h_{A B}  ,&
\label{def23}
\\ &\omega_{A B}= \nabla_{[A } u_{B ]} +  \frac{1}{k^2}u_{[A }a_{B] },&\label{def24}
\end{eqnarray}
where 
$h_{A B} $ is the projector onto the space transverse to the velocity field:
\begin{equation}
\label{relproj}
h_{AB}=\dfrac{u_{A}u_{B}}{k^2}+\eta_{AB}
\end{equation}
(remember we work in an orthonormal Cartan mobile frame --- metric displayed in \eqref{metgenuortho}).
The above vectors are transverse, whereas the tensors are transverse and traceless. 

The Weyl connection $\text{A}$ enters the Weyl covariant derivative  $\mathscr{D}_{A}$ acting on a weight-$w$ tensor as e.g. a scalar $\Phi$:
\begin{equation}
\label{WCscalar}
\mathscr{D}_A \Phi=\text{e}_A (\Phi) +w A_A \Phi,
\end{equation}
or a form $v_A$:
\begin{equation}
\label{Wv}
\mathscr{D}_B v_A=\nabla_B v_A+wA_B v_A + A_A v_B-\eta_{AB} A^C v_C. 
\end{equation}
The resulting tensors have  weight $w+1$.\footnote{Special caution is advised in comparing the present expressions with those appearing e.g. in Refs. \cite{CMPPS2, BigFluid,CMPPS1}, where a natural frame was used. 
%The developments on the anti-de Sitter side will be carried out in natural (coordinate) frame rather than in orthonormal frame, to which we %will switch for the asymptotically flat analysis. The expressions are valid irrespective of this choice. However, special caution is advised
When dealing with Weyl covariance in orthonormal frame, the metric components have weight zero. Hence for any tensor, covariant and contravariant components have the same weights. The coframe form elements, however, have weight $-1$, whereas the frame vectors have weight $+1$. If a weight-$w$ tensor has $p$ contravariant and $q$ covariant indices, its Weyl-covariant derivative reads:
\begin{eqnarray}
\mathscr{D}_C K_{B\ldots}^{\hphantom{B\ldots}A\ldots}&=&\nabla_C K_{B\ldots}^{\hphantom{B\ldots}A\ldots}+(w+p-q)A_CK_{B\ldots}^{\hphantom{B\ldots}A\ldots}
\nonumber
\\ &&+\left(
\eta_{CD} A^A-\delta^A_C A_D-\delta^A_D A_C
\right)K_{B\ldots}^{\hphantom{B\ldots}D\ldots}+\cdots
\nonumber
\\&&-\left(
\eta_{CB} A^D-\delta^D_C A_B-\delta^D_B A_C
\right)K_{D\ldots}^{\hphantom{D\ldots}A\ldots}-\cdots
\nonumber
\end{eqnarray}
and this has now weight $w+1$. \label{weylcartan}}
The form field $\text{u}$  has weight $-1$ i.e. $u_A$ are weight-zero, whereas $\omega_{AB}$ and  $ \sigma_{AB}$ have all weight $1$.    
The explicit form of $\text{A}$ \eqref{Wconc}  is obtained by demanding 
\begin{equation}
\label{conDU}
\mathscr{D}_{A}u^{A}=0\quad \text{and} \quad u^{C}\mathscr{D}_{C}u_{A}=0.
\end{equation}

The Weyl covariant derivative is metric-compatible with effective torsion: 
\begin{eqnarray}
\label{aaa}
\mathscr{D}_C \eta_{AB}&=&0,\\
\left(\mathscr{D}_A\mathscr{D}_B -\mathscr{D}_B\mathscr{D}_A\right) \Phi&=& w \Phi F_{AB},
\label{bbb}
\end{eqnarray}
where  
\begin{equation}
\label{F}
\text{F}=\frac{1}{2}F_{AB}\uptheta^A\wedge \uptheta^B=\text{d}\text{A}
\end{equation}
is Weyl-invariant ($F_{AB}$ are weight-$2$). Metric compatibility and \eqref{conDU} imply
\begin{equation}
 u^{C}\mathscr{D}_{C}h_{AB}=0, \label{trancomp}
\end{equation}
 infering that the operator $ u^{C}\mathscr{D}_{C}$ respects transversality. 

Commuting the Weyl-covariant derivatives acting on vectors, one defines the Weyl covariant Riemann tensor
\begin{equation}
\left(\mathscr{D}_A\mathscr{D}_B -\mathscr{D}_B\mathscr{D}_A\right) V^C=
\mathscr{R}^C_{\hphantom{C}DAB} V^D+ (w+1) V^C F_{AB}
\end{equation}
($V^C$ are weight-$w$ whereas $\text{V}=V^C\text{e}_{C}$ has weight $w+1$)
and the usual subsequent quantities. 
In three (boundary) spacetime dimensions, the covariant Ricci and the scalar (both weight-$2$) curvatures read: 
\begin{eqnarray}
\mathscr{R}_{AB}&=&{R}_{AB} + \nabla_B A_A + A_A A_B +
\eta_{AB}\left(\nabla_C A^C-A_C A^C\right)
-F_{AB},
\label{curlRic}
\\
\mathscr{R}&=&R +4\nabla_A A^A- 2 A_A A^A ,\label{curlRc}
\end{eqnarray}
where ${R}_{AB}$ is the Ricci tensor of the boundary Levi--Civita connection and $R$ the corresponding scalar curvature.
The Weyl-invariant Schouten tensor is 
\begin{equation}
\label{WIsch}
\mathscr{S}_{AB} = \mathscr{R}_{AB}-\frac{1}{4} \mathscr{R} \eta_{AB}
=R_{AB}-\frac{1}{4} R \eta_{AB}+\nabla_B A_A + A_A A_B -\frac{1}{2}A_C A^C \eta_{AB}-F_{AB}.
\end{equation} 

It is customary to introduce the
vorticity two-form 
\begin{equation}
\label{def25}
\upomega=\frac{1}{2}\omega_{AB }\, \text{d}x^A\wedge\text{d}x^B  
=\frac{1}{2}\left(\text{d}\text{u} +\frac{1}{k^2}
\text{u} \wedge\text{a} \right),
\end{equation}
as well as the Hodge dual of this form, which is proportional to  $\text{u}$:
\begin{equation}
\label{omdual}
k\gamma \text{u}=\star\upomega\quad\Leftrightarrow\quad 
k\gamma u_A=\frac{1}{2}\epsilon_{ABC}\omega^{BC}.
\end{equation}
In this expression  $\gamma$ is a scalar of  weight $1$.

In three spacetime dimensions and in the presence of a vector field $\text{u}$, one naturally defines a fully antisymmetric two-index tensor:\footnote{This hatted two-index tensor should not be confused with Minkowski metric.}
\begin{equation}
\label{eta2}
\hat \eta_{AB}=-\frac{u^C}{k}\epsilon_{CAB}, 
\end{equation} 
obeying
\begin{equation}
\label{eta2contr}
\hat\eta^{\vphantom{B}}_{AC}\hat\eta_{B}^{\hphantom{B}C}=h_{AB}, \quad \hat\eta^{AB}\hat\eta_{AB}=2.
\end{equation} 
With this tensor the vorticity reads:
\begin{equation}
\label{eta2vort}
\omega_{AB}=k^2\gamma \hat\eta_{AB}. 
\end{equation} 

The two-index tensor $\hat\eta_{AB}$ defines a duality map within the space of symmetric, transverse (with respect to $\text{u}$) and traceless tensors. If $V^A$ is transverse, so is
\begin{equation}
\label{eta2dualv}
\ast V^A=\hat\eta^B_{\hphantom{B}A}V_B.
\end{equation} 
Similarly with a symmetric, transverse and traceless tensor $W_{AB}$:
\begin{equation}
\label{eta2dualw}
\ast W_{AB}=\hat\eta^C_{\hphantom{C}A}W_{C B}
\end{equation} 
is symmetric, transverse and traceless.

\subsubsection*{The energy--momentum tensor and the Cotton tensor}

Given a normalized congruence $\| \text{u} \|^2=-k^2$ we can  decompose the energy--momentum tensor as in hydrodynamics:
\begin{equation}\label{T} 
T_{A B}=(\varepsilon+p) \frac{u_A  u_B}{k^2} +p  \eta_{AB}+   \tau_{A B}+ \frac{u_A  q_B}{k^2}+ \frac{u_B  q_A}{k^2} ,
\end{equation}
where 
\begin{equation}
\label{long} 
\varepsilon=\frac{1}{k^2}T_{A B} u^A u^B
\end{equation}
is the energy density and $p$ the the analogue of a perfect stress.  The symmetric viscous stress tensor $\tau_{A B}$ and the heat current $q_A$ are purely transverse:
\begin{equation}\label{trans}
u^A    \tau_{A B}=0, \quad u^A  q_A =0, \quad 
q_B= -{\varepsilon} u_B-u^A  T_{A B}.
\end{equation}

In three dimensions, a conformal energy--momentum tensor has weight-$1$ covariant components in the coordinate basis, and weight-$3$ components in the orthonormal frame. Consequently, the pressure and energy density, the heat-current $q_A$ and  the viscous stress tensor $\tau_{AB}$ have all weight $3$. Furthermore, since the splitting of the stress tensor into $p$ and $\tau_{AB}$ is arbitrary, we choose to implement the absence of trace  as
\begin{equation}\label{confluid}
\varepsilon=2p, \quad \tau_A^{\hphantom{A} A}=0.
%-\varepsilon+2p+\tau_A^{\hphantom{A} A}=0.
\end{equation}
 Due to the absence of  trace, the conservation equation \eqref{T-cons} can be traded for
\begin{equation}
\label{T-cons-con}
\mathscr{D}_C T^C_{\hphantom{\rho}B}
=0.
\end{equation}

In the gauge under consideration, the energy--momentum tensor comes along with the boundary Cotton tensor. They both enter the bulk metric, playing dual, electric versus magnetic, roles in various instances, as e.g. in the bulk Weyl tensor.
The Cotton tensor is generically a three-index tensor with mixed symmetries.\footnote{From the bulk viewpoint, the boundary energy--momentum and Cotton tensors play dual roles. Notice that the energy--momentum tensor in \eqref{T} has an extra factor of $k$ with respect to the Cotton tensor in \eqref{C}, due to their different dimensions.\label{k-fac}}
 In three dimensions, which is the case for our boundary geometry, the Cotton tensor can be dualized into a two-index, symmetric and traceless tensor:\footnote{We use a plain font for the Cotton $C_{AB}$ versus a curly font for the shear $\mathscr{C}_{AB}$.}
\begin{equation}
C_{AB}=\epsilon_{A}^{\hphantom{A}CD}
\mathscr{D}_C \left(F_{BD} +\mathscr{S}_{BD}\right)
=\epsilon_{A}^{\hphantom{A}CD}
\nabla_C \left(R_{BD}-\dfrac{R}{4}\eta_{BD} \right),
\label{cotdef}
\end{equation}
where we recall that $F_{BD}$ and $\mathscr{S}_{BD} $ are respectively the Weyl curvature and the Weyl-covariant Schouten tensor defined 
in \eqref{F} and \eqref{WIsch}.
The Cotton tensor $C_{AB}\uptheta^A\uptheta^B$ is Weyl-covariant of weight  $1$, and is \emph{identically} conserved: 
\begin{equation}
\label{C-cons}
\mathscr{D}_C C^C_{\hphantom{C}B}
=
\nabla_C C^C_{\hphantom{C}B} \equiv 0,
\end{equation}
sharing thereby all properties of the energy--momentum tensor.    

Following \eqref{T} we can decompose the Cotton tensor into longitudinal, transverse and mixed components with respect to the congruence $\text{u}$:
\begin{equation}
\label{C} 
\frac{1}{k}C_{A B}=\frac{3c}{2} \frac{u_A  u_B}{k^2} + \frac{c}{2} \eta_{AB}-  \frac{c_{A B}}{k^2}+ \frac{u_A  c_B}{k^2}   +\frac{u_B  c_A}{k^2}.
\end{equation}
Such a decomposition naturally defines the weight-$3$ \emph{Cotton scalar density}
\begin{equation}
\label{cottdens} 
c=\frac{1}{k^3}C_{AB}u^A u^B,
\end{equation}
as the longitudinal component. The symmetric and  traceless \emph{Cotton stress tensor} $c_{A B}$  and the \emph{Cotton current} $c_A$  (also weight-$3$) are purely transverse:
\begin{equation}
\label{cotrans}
  c_{A}^{\hphantom{A} A}=0, \quad
u^A   c_{A B}=0,\quad u^A  c_A=0,
\end{equation}
and obey
\begin{equation}
\label{cotransp}
c_{AB}=-k h^{C}_{\hphantom{C}A}h^{D}_{\hphantom{D}B}C_{CD}+\dfrac{ck^2}{2}h_{AB}
, \quad 
c_B= -c u_B-\frac{u^A  C_{A B}}{k}.
\end{equation}

One can use the definition \eqref{cotdef} to further express the Cotton density, current and stress tensor as ordinary or Weyl derivatives of the curvature. We find
\begin{eqnarray}
\label{cotdens}
 c&=&\frac{1}{k^2}u^B  \hat\eta^{DC} \mathscr{D}_C \left(\mathscr{S}_{BD}+F_{BD} \right),\\
\label{cotcur}
c_B&=&\hat\eta^{CD}\mathscr{D}_C \left(\mathscr{S}_{BD}+F_{BD} \right)-c u_{B} ,\\
c_{AB} &=& -h^{E}_{\hphantom{E}A}\left( k \epsilon_{B}^{\hphantom{B}CD}-u_{B}\hat\eta^{CD}\right)\mathscr{D}_C \left(\mathscr{S}_{ED}+F_{ED}\right)+\dfrac{c k^2}{2}h_{AB}.
\label{cotvis}
\end{eqnarray}

\subsubsection*{Solving Einstein's equations}

Einstein's equations are\footnote{We use $M,N, \ldots \in \left\{r, \text{boundary} \right\}$ as bulk indices.}
\begin{equation}
\label{bulkeinstein}
\mathcal{E}_{MN}\equiv R_{MN}^{\text{bulk}}-\frac12 R^{\text{bulk}}g_{MN}^{\text{bulk}}-3k^2g_{MN}^{\text{bulk}}=0,
\end{equation}
and we must probe them in the covariant Newman--Unti gauge. 
Assuming a boundary metric given in \eqref{metgenuortho}, the bulk line element \eqref{newcovNU} reads:
\begin{eqnarray}
\text{d}s^2_{\text{bulk}} &=&
2\frac{\text{u}}{k^2}(\text{d}r+r \text{A})
+r^2\text{d}s^2+ r \mathscr{C}_{AB}\uptheta^A\uptheta^B
+\frac{1}{k^4}\mathscr{F}_{AB}\uptheta^A\uptheta^B
\nonumber\\
&&+\sum_{s=1}^\infty
\frac{1}{r^s}\left( f_{(s)} \frac{\text{u}^2}{k^4} 
+2\frac{\text{u}}{k^2} f_{(s)A} \uptheta^A+ f_{(s)AB} \uptheta^A\uptheta^B\right),
\label{newcovNUortho}
\end{eqnarray}
where all boundary tensors are now defined in the orthonormal frame at hand.\footnote{In this frame $\mathscr{C}_{AB}$ has weight one,  $ f_{(s)} $, $ f_{(s)A} $ and $ f_{(s)AB}$ have all weight $s+2$ whereas $\mathscr{F}_{AB}$ is weight-$2$.}
The summation over $A$ and $B$ in the last terms of \eqref{newcovNUortho} is actually reduced to a summation over the transverse components
$a$ and $b$ thanks to the transversality of $ f_{(s)A}$ and $ f_{(s)AB}$ with respect to the velocity field \eqref{veloform}.

We will limit here the analysis to the order $\nicefrac{1}{r^2}$, which is sufficient for illustrating accurately later the asymptotically flat pattern. 
\begin{description}
\item[Order  \boldmath$r$\unboldmath] The important output here is that the Bondi shear $\mathscr{C}_{AB}$ \emph{is not} free, but settled by the shear of the congruence $\text{u}$, which is of geometric nature:
\begin{equation}
\label{Cshearmunu}
k^2\mathscr{C}_{AB}
=-2\sigma_{AB}.
\end{equation}
On shell, the Bondi shear is thus manifestly \emph{traceless and transverse} with respect to $\text{u}$. 
Anticipating the usage of the present formalism in describing general solutions of vacuum Einstein's equations, we also introduce  a \emph{news tensor} (similarly defined in arbitrary dimension). As opposed to the usual definitions, the present tensor is \emph{boundary-covariant, Weyl-invariant, symmetric, traceless and transverse:}
\begin{equation}
\label{news-rel}
\mathscr{N}_{AB}=u^{C}\mathscr{D}_{C}\mathscr{C}_{AB}.
\end{equation}
Equation \eqref{Cshearmunu} will be assumed when moving to the next orders.

\item[Order  \boldmath$1$\unboldmath] Unsurprisingly from the Feffermam--Graham experience, we learn that $\mathscr{F}_{AB}$ is related to the boundary Weyl-invariant  Schouten tensor displayed in Eq.  \eqref{WIsch}:\footnote{The tensor defined in \eqref{St} is slightly different from the analogous tensor $S_{AB}$ introduced in \cite{CMPPS2}, Eq. (2.42). It contains extra shear terms. The reason is that in Ref. \cite{CMPPS2}, when writing (2.41), the authors wanted to stress that shear terms were present, but ultimately the shear was vanishing. The present definition accounts for all shear terms.} 
\begin{equation}
\label{St}
\begin{array}{rcl}
\displaystyle{\mathscr{F}_{AB}}&=&\displaystyle{2 u^{C}\left(
\mathscr{S}_{C(A}+F_{C(A}
\right)u_{B)}+
\mathscr{D}_A u_C\, 
\mathscr{D}_B u^C} , \crbig
&=&\displaystyle{2u_{(A}\mathscr{D}_C \left(\sigma^{\hphantom{B)}C}_{B)}+\omega^{\hphantom{B)}C}_{B)}\right)-\frac{\mathscr{R}}{2}u_{A}u_{B}+
\left(\sigma^2+k^4\gamma^2\right)h_{AB}+2 \omega_{(A}^{\hphantom{(A}C} \sigma_{B)C}^{\vphantom{C}}} ,
\end{array}
\end{equation}
where
\begin{equation}
\label{gammasigma}
\gamma^2 = \frac{1}{2k^4} \omega_{AB} \omega^{AB}, \quad 
\sigma^2 = \frac{1}{2} \sigma_{AB} \sigma^{AB} 
\end{equation} 
($\gamma$ was defined alternatively in Eq. \eqref{omdual}). At this stage, the only independent and free data are those defining the boundary geometry (as stressed in \eqref{veloform}, the congruence $\text{u}$ is aligned with the observers at rest with respect to  \eqref{metgenuortho}). 

\item[Orders  \boldmath $\nicefrac{1}{r}$ and  $\nicefrac{1}{r^2}$ \unboldmath] At order $\nicefrac{1}{r}$ new information is expected to come up in the form of a boundary conformal energy--momentum tensor. In contrast with the Fefferman--Graham gauge, the energy--momentum enters through its decomposition with respect to the congruence $\text{u}$, i.e. $\varepsilon$, $q_A$ and $\tau_{AB}$, rather than $T_{AB}$. Furthermore, it comes accompanied with the transverse-dual of the Cotton current and stress, $\ast c_A$ and $\ast c_{AB}$, see Eqs. \eqref{eta2dualv}, \eqref{eta2dualw} and \eqref{cotcur},  \eqref{cotvis} --  yet another motivation to split the energy--momentum tensor as discussed earlier. This trait is new, both compared to the Fefferman--Graham gauge, where the Cotton tensor does not appear explicitly at any order, and with respect to standard Bondi or Newman--Unti gauges, where it is present in disguise.\footnote{One could not stress enough the profound versatility of the boundary Cotton tensor. Together with the boundary energy--momentum tensor, they control the asymptotic behaviour of the bulk Weyl tensor, the electric versus magnetic gravitational characteristics, the duality issues, and are natural ingredients in Newman--Penrose formalism. In the flat instance and in the current gauge, the Cotton tensor contributes to the gravitational radiation along with the Bondi shear. A recent presentation of some of these features is available in Ref. \cite{Mittal:2022ywl}.}

The functions to be determined are $f_{(1)}$, $f_{(1)A}$ and $ f_{(1)AB}$, which must have conformal weight $3$. This leaves little freedom, given the available tensors.
We find:
\begin{equation}
\label{O1}
 f_{(1)} \frac{\text{u}^2}{k^4} 
+2\frac{\text{u}}{k^2} f_{(1)A} \uptheta^A+ f_{(1)AB} \uptheta^A \uptheta^B=
 \frac{8\pi G}{k^4} \left( \varepsilon \text{u}^2
+
\frac{4}{3}\text{u}\Delta\text{q}
+\frac{2k^2}{3}\Delta\uptau
\right)
\end{equation}
with $\Delta\text{q}= \Delta q_A \uptheta^A$ and $\Delta\uptau=\Delta\tau_{AB} \uptheta^A\uptheta^B$ defined as 
\begin{equation}
\label{deltaqtau}
\Delta q_A=q_A-\frac{1}{8\pi G}\ast \! c_A
, \quad 
\Delta\tau_{AB}=\tau_{AB}+\frac{1}{8\pi Gk^2}\ast \! c_{AB}.
\end{equation}
The functions $\varepsilon$, $q_A$ and $\tau_{AB}$, which merely parameterize the line element at this stage, can be packaged in a symmetric and traceless tensor $T_{AB}$ as in \eqref{T}, \eqref{confluid} and, as we shall see shortly in \eqref{einstein_extra}, Einstein's equations demand the conservation of $T_{AB}$, thus completing its identification as the boundary energy--momentum tensor as in the Fefferman--Graham gauge.

We now ought to focus on the $\nicefrac{1}{r^2}$ contribution to the line element \eqref{newcovNUortho}, i.e., on
\begin{equation}
\label{O2}
f_{(2)} \frac{\text{u}^2}{k^4} 
+2\frac{\text{u}}{k^2} f_{(2)A} \uptheta^A+ f_{(2)AB} \uptheta^A\uptheta^B,
\end{equation}
where $f_{(2)}$, $f_{(2)A}$ and $ f_{(2)AB}$ must have conformal weight $4$.
The analogy with the Fefferman--Graham expansion suggests that no new free boundary functions  should appear without spoiling Einstein's equations. Indeed, upon imposing \eqref{Cshearmunu} and \eqref{St}, one finds
\begin{equation}
\begin{cases} 
\displaystyle{\mathcal{E}_{rr} = -\frac{3}{r^5}\eta_{AB}f_{(1)}^{AB}-6\left(\eta_{AB}f_{(2)}^{AB}+\frac{3}{2k^2}\sigma_{AB}f_{(1)}^{AB}\right)\frac{1}{r^6}+\mathcal{O}\left(\frac{1}{r^7}\right)}
\crbig
\displaystyle{k\mathcal{E}_{r\hat 0} =\left(-f_{(2)}-2k^2\eta_{AB}f_{(2)}^{AB}+\frac12h_{AB}\mathscr{D}^{A}f_{(1)}^{B}-\frac52\sigma_{AB}f_{(1)}^{AB}+c\gamma\right)\frac{1}{r^4}+\mathcal{O}\left(\frac{1}{r^5}\right)}\crbig
\displaystyle{\mathcal{E}_{ra}  = \left(2f_{(2)a}-\frac32h_{aB}\mathscr{D}_{C}f_{(1)}^{BC}+\frac{1}{k^2}\left(\sigma_{aB}+4\omega_{aB}\right)f_{(1)}^{B}\right)\frac{1}{r^4}+\mathcal{O}\left(\frac{1}{r^5}\right)}\crbig
\displaystyle{\mathcal{E}^{ab}  = \left(-f_{(2)}h^{ab}+c\gamma h^{ab}+4\omega_{C}^{\hphantom{C}(a}f_{(1)}^{b)C}+2k^2\hat{\eta}_{C}^{\hphantom{C}a}\hat{\eta}_{D}^{\hphantom{D}b}f_{(2)}^{CD}-2u^{C}\mathscr{D}_{C}f_{(1)}^{ab}\right.}\\
 \displaystyle{\qquad\left.+\, \hat{\eta}_{C}^{\hphantom{C}a}\hat{\eta}_{D}^{\hphantom{D}b}\mathscr{D}^{(C}_{\vphantom{(1)}}f_{(1)}^{D)}
   +\frac{1}{k^2}\left(c\hat{\eta}_{C}^{\hphantom{C}a}\sigma^{C b}-f_{(1)}\sigma^{ab}\right)+4\sigma_{C}^{\hphantom{C}(a}f_{(1)}^{b)C}\right)\frac{1}{r^2}+\mathcal{O}\left(\frac{1}{r^3}\right)}
\end{cases}
\label{einst-rec-2}
\end{equation}
for the often referred to as constraint Einstein's equations. These equations fix algebraically all terms at the $\nicefrac{1}{r^2}$ order in the expansion of the bulk metric, thus confirming the absence of any new free function. When rewritten in terms of the basic quantities parameterizing the space of solutions, the three coefficients in \eqref{O2} read:
\begin{eqnarray}
\label{f2sca}
f_{(2)} &=& \frac{8 \pi G}{3k^{2}} \left( \sigma_{CD}\Delta\tau^{CD}+\mathscr{D}_{C}\Delta q^{C}\right)
 +
 c \gamma,
\\
\label{f2vec}
f_{(2)A} &=&- \frac{8\pi G}{3k^{4}}\sigma_{AC} \Delta q^{C} + \frac{4\pi G}{k^{2}} \left(h_{AC}\mathscr{D}_{D}\Delta\tau^{CD}
+ \frac{8}{3}\gamma \ast\!  \Delta q_{A} 
\right),
\\
\label{f2ten}
f_{(2)AB} &=&- \frac{4\pi G}{k^{4}}\left(\frac{4}{3} u^{C} \mathscr{D}_{C}\Delta \tau_{A B}+ \frac{2}{3}h_{A C} h_{B D} \mathscr{D}^{(C}\Delta q^{D)}- \frac{1}{3}h_{AB}h^{CD}\mathscr{D}_{C}\Delta q_{D}  + 2 \sigma_{(A}^{\hphantom{A}C}\Delta\tau_{B)C}^{\vphantom{C}}  \right)
\nonumber
\\
&& - \frac{1}{2k^{4}}\left(8 \pi G \varepsilon \sigma_{AB} - c \ast\! \sigma_{AB}  \right) 
+ \frac{32\pi G}{3k^{2}}\gamma \ast\! \Delta \tau_{AB} .
\end{eqnarray}
These expressions contain all possible combinations of the shear and of the vorticity together with adequately projected Weyl covariant derivatives of the energy--momentum and Cotton tensors,\footnote{The covariant Newman--Unti gauge has often been referred to as the \emph{derivative-expansion gauge} for this reason. This was borrowed from the original fluid/gravity literature, where the derivative expansion was inspired by the fluid constitutive relations.}
carrying the right tensorial structure and conformal weight. Substituting Eqs.~\eqref{f2sca}, \eqref{f2vec}, \eqref{f2ten} into the remaining Einstein's equations \eqref{bulkeinstein} one obtains:
\begin{equation}
\label{einstein_extra}
	\frac{k}{8\pi G}\mathcal{E}_{\hat 0\hat 0}=
	\frac{1}{r^{2}} \mathscr{D}_B T^{B}_{\hphantom{B}\hat 0}+\mathcal{O}\left(\frac{1}{r^{3}}\right),
	\quad
	\frac{k}{8\pi G}\mathcal{E}_{\hat 0a}=
	\frac{1}{r^{2}} \mathscr{D}_B T^{B}_{\hphantom{B}a}+\mathcal{O}\left(\frac{1}{r^{3}}\right)
\end{equation}
(since $T_{AB}$ is traceless, $\mathscr{D}_A\equiv \nabla_A$, the Levi--Civita boundary connection for the frame metric $\eta_{AB}$).
The omitted terms contain the tensors $f_{(3)}$, $f_{(3)A}$ and $f_{(3)AB}$. This confirms that no additional constraints are imposed on the quantities parameterizing the solution space identified at the previous orders, i.e., the velocity field, the boundary metric (frame in the present formalism) and the boundary energy--momentum tensor.
 
\item[Higher orders and possible resummation]
The above pattern can be repeated ad nauseam at the cost of a substantial growth in admissible terms. The third order would be interesting as it is expected to host the Newman--Penrose charges in the flat limit. This is beyond our motivations, but raises the issue of resummability under conditions of the series \eqref{newcovNU}. This question is usually immaterial  in Bondi or Newman--Unti gauges, where due to the absence of boundary vorticity\footnote{An explicit realization of the boundary frame of Newman--Unti gauge is displayed in \eqref{mutvrel}, 
\eqref{upsilontvarel} with $\Delta_i=0$, as mentioned earlier in this section. The boundary vorticity always vanishes then as it is proven by comparing \eqref{reldiffrNU} with \eqref{reldiffr}.} simple solutions such as Kerr's are embodied in the form of infinite series. In the covariant Newman--Unti gauge, the explicit appearance of the boundary Cotton tensor allows to tune the bulk Weyl tensor and select algebraically special Einstein spacetimes, for which the series is resummable.  This is achieved by imposing 
\begin{equation}
\label{rescon}
\sigma_{AB}=0, \quad \Delta q_{A}=0, \quad  \Delta \tau_{A B} =0,
\end{equation}
which implies that 
\begin{equation}
\label{resconseq}
 f_{(s)A}=0,\quad f_{(s)AB}=0 
\end{equation}
and 
\begin{equation}
\label{resconseqsc}
 f_{(2s+1)}= (-)^{s} 8\pi G\varepsilon \gamma^{2s},
 \quad f_{(2s+2)}= (-)^{s} c \gamma^{2s+1}.
\end{equation}
The boundary metric is still a free variable, but only the energy density $\varepsilon(x)$ remains from the energy--momentum tensor, whose heat current and stress are fixed by those of the Cotton:
\begin{equation}
\label{deltaqtaures}
q_A=\frac{1}{8\pi G}\ast \! c_A
, \quad 
\tau_{AB}=-\frac{1}{8\pi Gk^2}\ast \! c_{AB}.
\end{equation}
As a consequence, assuming that \eqref{T-cons} is satisfied, one finds
\begin{equation}
\text{d}s^2_{\text{res. Einstein}} =
2\frac{\text{u}}{k^2}(\text{d}r+r \text{A})+r^2\text{d}s^2+\frac{\mathscr{F}}{k^4}
+ \frac{\text{u}^2}{k^4\rho^2} \left(8\pi G \varepsilon r+
c \gamma\right)
\label{papaefgenrescrec}
\end{equation}
with
\begin{equation}\label{rho2c}
 \rho^2= r^2 +\gamma^2
\end{equation} 
and $\mathscr{F}=\mathscr{F}_{AB}  \uptheta^A\uptheta^B$ given in \eqref{St} imposing zero shear. The Petrov analysis of \eqref{papaefgenrescrec} has been discussed in Refs. \cite{CMPPS2,Gath:2015nxa}.

\end{description}

\section{The flat avatars}
\label{brrm}

\subsubsection*{First things first}

Handling the flat limit is a triptych. At the first place stands the boundary geometry, which becomes Carrollian as the time-like conformal boundary of asymptotically anti-de Sitter spacetimes is traded for the null infinity of their
asymptotically flat relatives. Secondly, the energy--momentum tensor should be expanded in Laurent series with respect to $k^2$ and embrace all extra degrees of freedom of the flat solution space. Finally comes the bulk line element that should remain finite in the zero-$k$ limit, imposing to this end constraints and evolution equations on the functions defining the solution space, besides the Carrollian limit of the already available Eqs. \eqref{T-cons}.

Given the  relativistic boundary metric and the velocity field, \eqref{metgenuortho} and \eqref{veloform}, the starting point of our analysis is the bulk line element \eqref{newcovNUortho}, which we reproduce here bearing in mind the transversality properties:
\begin{eqnarray}
\text{d}s^2_{\text{bulk}} &=&
2\frac{\text{u}}{k^2}(\text{d}r+r \text{A})
+r^2\text{d}s^2+ r \mathscr{C}_{ab}\uptheta^a\uptheta^b+\frac{1}{k^4}\mathscr{F}_{AB}\uptheta^A\uptheta^B
\nonumber\\
&&+\sum_{s=1}^\infty
\frac{1}{r^s}\left( f_{(s)} \frac{\text{u}^2}{k^4} 
+2\frac{\text{u}}{k^2} f_{(s)a} \uptheta^a+ f_{(s)ab} \uptheta^a\uptheta^b\right).
\label{newcovNUortho-tran}
\end{eqnarray}

The Carrollian limit of the boundary geometry is reached as follows:\footnote{Carrollian quantities will often be distinguished with hats.
However, in order to avoid cluttering of indices and symbols, we do make the distinction amongst relativistic and Carrollian attributes, only when it is necessary. This will not be the case e.g. for the Bondi shear and news.} 
\begin{equation}
\label{limudc}
\upmu
%=-\hat{\uptheta}^{\hat u}
=\lim_{k\to 0}\frac{\text{u}}{k^2}=-\lim_{k\to 0}\frac{{\uptheta}^{\hat 0}}{k}
 ,
\quad
\hat{\uptheta}^a
=\lim_{k\to 0}{\uptheta}^a,
\end{equation}
so that the Carrollian degenerate metric spells
\begin{equation}
\label{carmetlim}
\text{d}\ell^2=\lim_{k\to 0}\text{d}s^2=\delta_{ab}\hat{\uptheta}^a\hat{\uptheta}^b.
\end{equation}
For the frame vectors, the prescription is 
\begin{equation}
\label{limuvecc}
\upupsilon
=\lim_{k\to 0}\text{u}=\lim_{k\to 0}k\text{e}_{\hat 0},
\quad
\hat{\text{e}}_a
=\lim_{k\to 0}\text{e}_a.
\end{equation}
It should be stressed that the limit may not be necessary, because the parameterization of the diads $\uptheta^a$ in terms of the natural-coframe components $\text{d}x^\mu$ can be chosen so as not to depend on $k$, and that $\uptheta^{\hat 0}$ could simply be $\text{d}x^0=k \text{d}u$ in which case  $\upmu= -\text{d}u$
(some further examples are displayed in App. \ref{carman}, Eqs. \eqref{mutv}, 
\eqref{upsilontva}, 
\eqref{carmetexpl}
and
Eqs. \eqref{mutvrel}, 
\eqref{upsilontvarel}, 
\eqref{metgenu}). This will be definitely our viewpoint here.

The kernel of the degenerate metric \eqref{carmetlim} is the \emph{field of observers} $\upupsilon$, and $\upmu$ is its dual  \emph{clock form} embracing also the Ehresmann connection, as explained in App.  \ref{carman}. All these obey 
\begin{equation}
\label{carfrcofrlim}
\upmu(\upupsilon)=-1,\quad
\hat{\uptheta}^a(\hat{\text{e}}_b)=\delta^a_b,\quad
\hat{\uptheta}^a(\upupsilon)=0,\quad
\upmu(\hat{\text{e}}_a)=0.
\end{equation}
The Carrollian geometric data are part of the solution space of Ricci-flat spacetimes in the flat covariant Newman--Unti gauge. Compared to the standard flat Newman--Unti gauge, the extra piece of data is the clock form $\upmu$, which echoes the velocity congruence of the AdS relative. More accurately, the additional piece of information carried by the covariant Newman--Unti gauge is the boundary vorticity $\ast \varpi $, as discussed in App.  \ref{carman}, footnote \ref{NUvssNU}. 

The vanishing-$k$ limit the AdS-boundary Weyl connection $\text{A}$ is readily reached due to its $k$-independence. 
As described explicitly in  App. \ref{carman}, one effortlessly expresses $\text{A}$ in Carrollian terms, Eq. \eqref{Weyl-con-gen-car}:
\begin{equation}
\label{Weyl-con-gen-car-text}
\text{A}=\varphi_a \hat{\uptheta}^a-\frac{\theta}{2}\upmu
\end{equation}
with $\varphi_a$ and $\theta$ given in \eqref{cardiffr} or  \eqref{reldiffr}  and \eqref{excurvexpshear}. Therefore, the first two terms in \eqref{newcovNUortho-tran} have a well-defined limit without the need of imposing Einstein's equations. 

The next term in \eqref{newcovNUortho-tran}  plays an essential role in gravitational physics. Indeed,  
Einstein's equation \eqref{Cshearmunu}, reproduced here for the spatial components --- the only non-zero due transversality combined with our choice of congruence $\text{u}$,
\begin{equation}
\label{Cshearmunurep}
k^2\mathscr{C}_{ab}
=-2\sigma_{ab},
\end{equation}
 implies that $\sigma_{ab}=0$ at vanishing $k$. As explained in Eq. \eqref{carlim3}, the latter 
translates in Carrollian terms into 
\begin{equation}
\label{no-shear}
\xi_{ab}
=0,
\end{equation}
where $\xi_{ab}$ is defined in \eqref{excurvexpshear} as the traceless component of 
 the extrinsic curvature.
On the one hand, the geometrical shear $\xi_{ab}$ of the boundary Carrollian geometry \emph{must vanish} ---  an extrinsic-curvature condition for the conformal null boundary. On the other hand, the dynamical shear $\mathscr{C}_{ab}$ \emph{is free} and carries information on the gravitational radiation. No equation will constrain it or make it evolve, but it will source the evolution of other degrees of freedom.

In summary, till the order $r$, the Ricci-flat bulk metric reads:
\begin{equation}
\label{RFr}
\left.\text{d}s^2_{\text{Ricci-flat}}\right\vert_r =
 \upmu\left[2\text{d}r+r\left(
2\varphi_a \hat{\uptheta}^a-\theta\upmu
\right)\right]
+r^2\text{d}\ell^2+ r \mathscr{C}_{ab}\hat{\uptheta}^a\hat{\uptheta}^b,
\end{equation}
where $ \mathscr{C}_{ab}(u,\mathbf{x})$ is an arbitrary traceless Carrollian tensor, referred to as the Bondi shear (cf. the footnote~\ref{bondishear}). The Bondi news is another traceless Carrollian tensor obtained as the Carrollian limit of Eq.~\eqref{news-rel}:
\begin{equation}
\label{news-car}
\hat{\mathscr{N}}_{ab}=\hat{\mathscr{D}}_\upupsilon\mathscr{C}_{ab}.
\end{equation}

In order to pursue the study of the next orders, we must be careful with the  zero-$k$ limit. Both in the line element and in the conservation equations \eqref{T-cons}, the geometric shear $\sigma_{ab}= \xi_{ab}$ must be substituted by $-\frac{k^2}{2} \mathscr{C}_{ab}$ on account of Eq. \eqref{Cshearmunurep} before the limit is taken.
Often this won't have any effect and the term in consideration will drop. Sometimes, however, due to the presence of negative powers of $k$, finite terms will survive or divergences will impose further requirements.

The first and simplest application of the rule just stated concerns the order-$1$ term $\frac{\mathscr{F}}{k^4}=\frac{1}{k^4}\mathscr{F}_{AB}\uptheta^A\uptheta^B
$. Expressing \eqref{St}  in Carrollian terms we find:
\begin{eqnarray}
\frac{\mathscr{F}}{k^4}&=&\frac{\xi^2}{k^4}\text{d}\ell^2+\frac{1}{k^2}\left(3\xi^2\upmu^2+2\hat{\mathscr{D}}_b\xi^{b}_{\hphantom{b}a}\upmu\hat{\uptheta}^a-2\ast \! \varpi \ast\!\xi_{ab}\hat{\uptheta}^a\hat{\uptheta}^b\right)
\nonumber
\\&&+\ast\!\varpi^2\text{d}\ell^2-2\ast\!\hat{\mathscr{D}}_{a}\ast\!\varpi\upmu\hat{\uptheta}^a-\hat{\mathscr{K}}\upmu^2-5k^2\ast\varpi^2\upmu^2\nonumber
\\
&=&\left(\frac{\mathscr{C}^2}{4}+\ast\varpi^2\right)\text{d}\ell^2-\hat{\mathscr{K}}\upmu^2-\hat{\mathscr{D}}_b\mathscr{C}^{b}_{\hphantom{b}a}\upmu\hat{\uptheta}^a
-2\ast\!\hat{\mathscr{D}}_{a}\ast\!\varpi\upmu\hat{\uptheta}^a
\nonumber
\\
&&+\ast \! \varpi \ast\!\mathscr{C}_{ab}\hat{\uptheta}^a\hat{\uptheta}^b+k^2\left(\frac{3}{4}\mathscr{C}^2-5\ast\!\varpi^2\right)\upmu^2
\end{eqnarray}
with $\mathscr{C}^2=\frac{1}{2}\mathscr{C}^{ab}\mathscr{C}_{ab}$ and quantities like $\xi^2$, $\ast\varpi$, $\hat{\mathscr{K}}$ defined in App. \ref{carcot3}. The asterisk  stands for the relativistic congruence-transverse or Carrollian-basis duality introduced in   Eqs. \eqref{eta2},  \eqref{eta2contr},  \eqref{eta2dualv},  \eqref{eta2dualw}
or \eqref{etaortho},  \eqref{eta2dualvwcar}.
Some terms drop in the zero-$k$ limit but no divergence occurs and \emph{we are left with a piece in the line element, which now contains explicitly the Bondi shear:}
\begin{equation}
\label{S-car}
\lim_{k\to0} \frac{\mathscr{F}}{k^4}=\left(\frac{\mathscr{C}^2}{4}+\ast\varpi^2\right)\text{d}\ell^2-\hat{\mathscr{K}}\upmu^2-\hat{\mathscr{D}}_b\mathscr{C}^{b}_{\hphantom{b}a}\upmu\hat{\uptheta}^a
-2\ast\!\hat{\mathscr{D}}_{a}\ast\!\varpi\upmu\hat{\uptheta}^a+\ast \varpi \ast\!\mathscr{C}_{ab}\hat{\uptheta}^a\hat{\uptheta}^b
.
\end{equation}

Before moving on to the next order, which uncovers the method of expanding the anti-de Sitter energy--momentum tensor as a mean of reconstructing Ricci-flat spacetimes, it is fair to give credit to the authors of Ref. \cite{Compere:2019bua}, where the pioneering idea of substituting the Bondi for the geometric shear with the accompanying power of the cosmological constant was initiated.

\boldmath
\subsubsection*{Order $\nicefrac{1}{r}$ and the advent of the energy--momentum tensor}
\unboldmath

Let us assume that in the course of the bulk flat limit, the boundary energy--momentum tensor is analytic in $k^2$. It can thus be represented as a Laurent series about $k=0$:
 \begin{eqnarray}
\label{varepsilonlim}
\varepsilon&=& 
\sum_{n\in \mathbb{Z}}k^{2n}  
\varepsilon_{(n)}
,
\\
\label{qlim}
q^a &=&\sum_{n \geq 2} \frac{\zeta^{a}_{(n)}}{ k^{2n}} + \frac{\zeta^{a} }{k^2}+Q^{a}+k^2 \pi^{a}+\sum_{n\geq2}  k^{2n}  \pi^{a}_{(n)} , 
\\
\label{taulim}
\tau^{ab}&=&
-\sum_{n\geq3}  \frac{\zeta^{ab}_{(n)} }{k^{2n} }
-\frac{\zeta^{ab} }{k^4}-\frac{\Sigma^{ab} }{k^2}- \Xi^{ab}-k^2 E^{ab}
-\sum_{n\geq2}  k^{2n} E^{ab}_{(n)} 
.
\end{eqnarray}
Each function in these series (some have been singled out for reasons that will be clarified later) is a Carrollian tensor (scalar, vector, or symmetric and traceless two-tensor) that is possibly one of the boundary degrees of freedom, which we call \emph{Chthonian} to recall  they encode the asymptotically flat Einstein dynamics probing the bulk metric \emph{in depth} from the boundary. These tensors are expected to obey flux-balance equations, which are Carrollian avatars of vacuum Einstein's equations, and that we will attain using anti-de Sitter dynamics and imposing a regular behaviour at zero~$k$.

As an introductory statement, it is important to stress that we have no proof for the proclaimed  analyticity. The latter is a working framework, resulting in a consistent description of asymptotically flat spacetimes, and this end justifies the means. The rules are simple: insert \eqref{varepsilonlim}, \eqref{qlim}, \eqref{taulim} in the line element \eqref{newcovNUortho-tran} at each order, and impose regularity at $k=0$ after trading $\xi_{ab}$ for $-\frac{k^2}{2} \mathscr{C}_{ab}$ . This process starts with $\nicefrac{1}{r}$, since this is the first term sensitive to the energy--momentum tensor, but the substitution of $\mathscr{C}_{ab}$ will be performed systematically, in the line element, in Einstein's equations, or in the further definition of the complex mass aspect, without raising any order ambiguity.

At order $\nicefrac{1}{r}$ we should probe \eqref{O1}, which spells 
\begin{equation}
\label{O1car}
 f_{(1)} \frac{\text{u}^2}{k^4} 
+2\frac{\text{u}}{k^2} f_{(1)a} \uptheta^a+ f_{(1)ab} \uptheta^a \uptheta^b=
 8\pi G \left( \varepsilon \upmu^2
+
\frac{4}{3}\upmu\frac{\Delta q_a}{k^2} \hat{\uptheta}^a
+\frac{2}{3}\frac{\Delta\tau_{ab}}{k^2} \hat{\uptheta}^a \hat{\uptheta}^b
\right)
\end{equation}
with $\varepsilon$ given in \eqref{varepsilonlim} and
 \begin{eqnarray}
\label{Delqlimcar}
\frac{\Delta q^a}{k^2} &=&\sum_{n \geq 2} \frac{\zeta^a_{(n)}}{ k^{2n+2}} + \frac{1 }{k^4}\left(\zeta^a-\frac{\ast z^a}{8\pi G}\right)
+\frac{1}{k^2}\left(Q^a-\frac{\ast \chi^a}{8\pi G}\right)+\left(\pi^a-\frac{\ast \psi^a}{8\pi G}\right)
\nonumber
\\
&&+\sum_{n\geq2}  k^{2n-2}  \pi^a_{(n)} , 
\\
\label{Deltaulimcar}
\frac{\Delta \tau^{ab}}{k^2}&=&
-\sum_{n\geq3}  \frac{\zeta^{ab}_{(n)} }{k^{2n+2} }
-\frac{1}{k^6}
\left(\zeta^{ab}-\frac{\ast Z^{ab}}{8\pi G}\right)
-\frac{1}{k^4}
\left(\Sigma^{ab} -\frac{\ast X^{ab}}{8\pi G}\right)- \frac{1}{k^2}
\left(\Xi^{ab}-\frac{\ast \Psi^{ab}}{8\pi G}\right)
\nonumber
\\
&&- E^{ab}
-\sum_{n\geq2}  k^{2n-2} E^{ab}_{(n)} 
,
\end{eqnarray}
where we have used \eqref{qlim}, \eqref{taulim}, the definitions  \eqref{deltaqtau} of $\Delta q^a$ and $\Delta \tau^{ab}$, as well as the Carrollian Cotton tensors $z^a$, $\chi^a$, $\psi^a$, $Z^{ab}$, $X^{ab}$, $\Psi^{ab}$ displayed in  \eqref{cotveca},  \eqref{cottenab}.
Finiteness in the flat limit sets up two sorts of requirements on the Carrollian descendants of the energy--momentum tensor.
\begin{itemize}
\item Infinite subsets of Laurent coefficients are required to vanish:
\begin{equation}
\label{vancoef}
\begin{cases}
\varepsilon_{(n)}=0 \quad \forall n<0\\
\zeta^a_{(n)}=0 \quad \forall n\geq 2 \\
\zeta^{ab}_{(n)}=0 \quad \forall n\geq 3;
\end{cases}
\end{equation}

\item Five Laurent coefficients are locked in terms of the Carroll Cotton tensors defined in \eqref{c6fr}, \eqref{c5fr}, \eqref{c10fr}, \eqref{c9fr}, \eqref{c8fr}:
\begin{equation}
\label{cotfixcoef}
\zeta^a=\frac{\ast z^a}{8\pi G},
\quad
Q^a=\frac{\ast \chi^a}{8\pi G},
\quad
\zeta^{ab}=\frac{\ast Z^{ab}}{8\pi G}
,\quad
\Sigma^{ab} =\frac{\ast X^{ab}}{8\pi G}
,\quad
\Xi^{ab}=\frac{\ast \Psi^{ab}}{8\pi G}
.
\end{equation}
Hence a finite subset of energy--momentum Carrollian descendants \emph{are not independent but are instead of geometric nature, determined by the boundary Carroll structure via its Cotton tensor.} 

\end{itemize}
No more constraints show on the Chthonian degrees of freedom at $\nicefrac{1}{r}$ order. 

Defining  
\begin{equation}
\label{angasp}
N^a=\ast\psi^a-8\pi G \pi^a,
\end{equation}
we recast the order-$\nicefrac{1}{r}$ term \eqref{O1car} in the flat limit as:
\begin{eqnarray}
\lim_{k\to 0} \left(f_{(1)} \frac{\text{u}^2}{k^4} 
+2\frac{\text{u}}{k^2} f_{(1)a} \uptheta^a+ f_{(1)ab} \uptheta^a \uptheta^b\right)
&=&
 8\pi G  \varepsilon_{(0)} \upmu^2
-
\frac{4}{3}\upmu N_a\hat{\uptheta}^a
-\frac{16 \pi G}{3}E_{ab}\hat{\uptheta}^a \hat{\uptheta}^b
\nonumber
\\
&\equiv&\hat f_{(1)} \upmu^2
+2 \upmu \hat f_{(1)a}\hat{\uptheta}^a+\hat f_{(1)ab} \hat{\uptheta}^a \hat{\uptheta}^b.
\label{O1carlim}
\end{eqnarray}
The latter expression calls for two remarks. Firstly, the Carrollian tensors $\varepsilon_{(n\geq 1)}$, $\pi^a_{(n\geq 2)}$ and $E^{ab}_{(n\geq 2)}$ are absent. We should refrain from interpreting this as a sign that those aren't genuine degrees of freedom. Some of them ought to appear in the line element in the next orders and therefore participate in the dynamics. Only when one is guaranteed that a Laurent coefficient is absent from the line element at any order, can we declare it is irrelevant and set it consistently to zero. The order-$\nicefrac{1}{r^2}$ analysis will significantly underpin this statement.

Secondly comes an important question: \emph{what is the dynamics of the boundary degrees of freedom $ \varepsilon_{(0)}$, $N_a$ and $E_{ab}$ that remain in the $\nicefrac{1}{r}$ term of the bulk line element?} Ensuing our philosophy, this dynamics is encoded (\romannumeral1) in the zero-$k$ limit of anti-de Sitter Einstein's equations and (\romannumeral2) in the finiteness requirement  of the line element. The latter has already been exploited at the order under consideration, while the former is the energy--momentum conservation \eqref{T-cons} on which we will elaborate now. Our treatment consists in the four steps summarized below.
\begin{enumerate}
\item In the frame at use, we consider the relativistic energy--momentum tensor conservation equations \eqref{T-cons} recast in Carrollian terms as in App. \ref{carman}, Eqs.  \eqref{newlongrelconf} and  \eqref{newtrrelaconf}, 
which we redisplay here for convenience:
\begin{eqnarray}
\label{newlongrelconftext}
&\mathcal{L}=\hat{\mathscr{D}}_{\upupsilon} \varepsilon+\hat{\mathscr{D}}_a q^a
+
\xi_{ab}\tau^{ab}=0,&
\\
&\mathcal{T}^{a}= \frac{1}{d} \hat{\mathscr{D}}^a \varepsilon+ \hat{\mathscr{D}}_b
\tau^{ab}
 +  2q_b\varpi^{ba}
+\frac{1}{k^2}\left(
 \hat{\mathscr{D}}_{\upupsilon}q^a
+\xi^{ab}q_b
\right)=0
.&
\label{newtrrelaconftext}
\end{eqnarray}

\item We insert in these equations the variables $\varepsilon$, $q^a$ and $\tau^{ab}$ in their expanded forms \eqref{varepsilonlim}, \eqref{qlim}  and \eqref{taulim}, taking into account the finiteness requirements \eqref{vancoef} and \eqref{cotfixcoef}. 

\item The requirements \eqref{vancoef} and \eqref{cotfixcoef} bring the Cotton tensor inside the boundary energy--momentum conservation equations $\mathcal{L}=0$ and $\mathcal{T}^a=0$. At this stage we must exploit the Cotton identities 
$\left\{\mathcal{D}_{\text{Cot}}=0, \mathcal{I}_{\text{Cot}}^a =0\right\}$, 
$\left\{\mathcal{E}_{\text{Cot}}=0, \mathcal{G}_{\text{Cot}}^a =0\right\}$, 
$\left\{\mathcal{F}_{\text{Cot}}=0, \mathcal{H}_{\text{Cot}}^a =0\right\}$
and $\left\{\mathcal{W}_{\text{Cot}}=0, \mathcal{X}_{\text{Cot}}^a =0\right\}$ set in App. \ref{carcot3} --- see Eqs. \eqref{carDcot}, 
\eqref{carEcot}, \eqref{carFcot}, \eqref{carWcot}, \eqref{carIcot}, \eqref{carGcot}, \eqref{carHcot}, \eqref{carXcot} 
--- and recast for our needs as:
\begin{eqnarray}
   \hat{\mathscr{D}}_b  \ast\!\Psi^{ab}
+2\ast\! \varpi  \chi^{a}
&=&\hat{\mathscr{D}}_\upupsilon \ast\!\psi^a  + \frac{1}{2} \ast\!\hat{\mathscr{D}}^a c_{(0)}- \ast\psi_b\xi^{ab} 
,
  \label{carGcotgen}\\
 \hat{\mathscr{D}}_b  \ast\! X^{ab}
 +2\ast\! \varpi z^{a} -\hat{\mathscr{D}}_\upupsilon  \ast\! \chi^a&=& \frac{1}{2}\ast\!\hat{\mathscr{D}}^a  c_{(1)}- \ast \chi_b\xi^{ab} 
,
 \label{carHcotgen} \\
  \hat{\mathscr{D}}_b \ast\! Z^{ab}-\hat{\mathscr{D}}_\upupsilon \ast\! z^a
&=& \frac{1}{2}\ast\!\hat{\mathscr{D}}^a c_{(2)}-\ast z_b\xi^{ab} 
.
  \label{carXcogent}
  \end{eqnarray}
With this we reach the following:
\begin{eqnarray}
\mathcal{L}&=&k^2\hat{\mathscr{D}}_a\pi^{a}+\sum_{n\geq2}  k^{2n} \hat{\mathscr{D}}_a \pi^{a}_{(n)}
-\xi_{ab}\left(k^2E^{ab}
+\sum_{n\geq2}  k^{2n} E^{ab}_{(n)} 
\right)
\nonumber \\
 &&+\hat{\mathscr{D}}_\upsilon \varepsilon_{(0)} 
 + \sum_{n\geq 1} k^{2n}   \hat{\mathscr{D}}_\upsilon
\varepsilon_{(n)}
-\frac{1}{8\pi G}\left(\ast  \Psi^{ab}\xi_{ab}-\hat{\mathscr{D}}_a  \ast \! \chi^{a} \right)
 \nonumber \\
&&-\frac{1}{8\pi Gk^2} \left(\ast  X^{ab}\xi_{ab}-
\hat{\mathscr{D}}_a  \ast\!  z^{a}
\right)-\frac{1}{8\pi Gk^4} \ast\!  Z^{ab}\xi_{ab},
\label{eqgenC0}
\end{eqnarray}
and
\begin{eqnarray}
\mathcal{T}^a&=&
- \hat{\mathscr{D}}_b\left(k^2E^{ab}
+\sum_{n\geq2}  k^{2n} E^{ab}_{(n)} 
\right)
+2\ast \!\varpi \left(k^2 \ast\! \pi^{a}
+ \sum_{n\geq2}  k^{2n}\ast\! \pi^{a}_{(n)}\right)+\frac{1}{2}\hat{\mathscr{D}}^a \varepsilon_{(0)}
\nonumber \\
 &&
 +\frac{1}{2} \sum_{n\geq 1} k^{2n}   \hat{\mathscr{D}}^a
\varepsilon_{(n)}+ \hat{\mathscr{D}}_\upupsilon \left( \pi^{a}
+ \sum_{n\geq2}  k^{2n-2} \pi^{a}_{(n)}\right)
+\xi^{a}_{\hphantom{a}b}  \left( \pi^{b}
+ \sum_{n\geq2}  k^{2n-2} \pi^{b}_{(n)}\right)
 \nonumber
 \\
&&
 -\frac{1}{8\pi G}\left(\hat{\mathscr{D}}_\upupsilon \ast\!\psi^a  + \frac{1}{2}\ast\! \hat{\mathscr{D}}^a c_{(0)}- \ast\psi_b\xi^{ab} \right)
  \nonumber \\
&&-\frac{1}{8\pi Gk^2} \left( \frac{1}{2}\ast\!\hat{\mathscr{D}}^a  c_{(1)}- \ast \chi_b\xi^{ab} \right)
-\frac{1}{8\pi Gk^4}\left( \frac{1}{2}\ast\!\hat{\mathscr{D}}^a c_{(2)}-\ast z_b\xi^{ab} \right) .
\label{eqgenCa}
\end{eqnarray}

\item Lastly we express the geometric Carrollian shear as  $\xi_{ab}=-\frac{k^2}{2} \mathscr{C}_{ab}$ inside Eqs. \eqref{eqgenC0} and \eqref{eqgenCa}. This is a juggernaut due to the heavy presence of $\xi_{ab}$ in the Carrollian Cotton tensors $c_{(1)}$, $c_{(2)}$, $\chi^a$, $z^a$, $\Psi^{ab}$, $X^{ab}$ and $Z^{ab}$ --- see their definitions \eqref{cfr},  \eqref{c6fr},  \eqref{c5fr},  \eqref{c10fr},  \eqref{c9fr} and  \eqref{c8fr}. This operation regularizes the otherwise singular behaviour of the last lines in  \eqref{eqgenC0} and \eqref{eqgenCa} at vanishing $k$, which instead produce  a wealth of finite terms, \emph{all rooted in the Carrollian Cotton tensor. } 

The flat limit of the boundary energy--momentum conservation can now be safely taken and yields: 
\begin{equation}
\label{eqgenC0car}
\lim_{k\to 0}\mathcal{L}=\hat{\mathscr{D}}_\upupsilon \varepsilon_{(0)}+\frac{1}{8\pi G}\hat{\mathscr{D}}_a  \ast \! \chi^{a}
 -\frac{1}{16\pi G} \left(
\hat{\mathscr{D}}_a  \hat{\mathscr{D}}_b \hat{\mathscr{N}}^{ab}+ \mathscr{C}^{ab}
\hat{\mathscr{D}}_a \hat{\mathscr{R}}_b 
+\frac{1}{2} \mathscr{C}_{ab}\hat{\mathscr{D}}_\upupsilon  \hat{\mathscr{N}}^{ab}
\right),
\end{equation}
and 
\begin{eqnarray}
\lim_{k\to 0}\mathcal{T}^a&=&\frac{1}{2}\hat{\mathscr{D}}_b \left(\delta^{ab}\varepsilon_{(0)}
+\frac{1}{8\pi G} \eta^{ab} c_{(0)}
\right)
+ \hat{\mathscr{D}}_\upsilon \left( \pi^a 
 -\frac{1}{8\pi G}\ast\psi^a \right)
  \nonumber \\
&&+\frac{1}{16\pi G}\bigg[\mathscr{C}^{ab}\hat{\mathscr{D}}_b\hat{\mathscr{K}}
+\ast\mathscr{C}^{ab}\hat{\mathscr{D}}_b\hat{\mathscr{A}}
-4\ast\!\varpi \ast\!\mathscr{C}^{ab}\hat{\mathscr{R}}_b\
  \nonumber \\
&&-\frac{1}{2}\hat{\mathscr{D}}^b\left(\hat{\mathscr{D}}_b\hat{\mathscr{D}}_c \mathscr{C}^{ac}
-\hat{\mathscr{D}}^a\hat{\mathscr{D}}^c \mathscr{C}_{bc}
\right)
 \nonumber \\
&&+\mathscr{C}^{ab}\hat{\mathscr{D}}^c\hat{\mathscr{N}}_{bc}+\frac{1}{2}\hat{\mathscr{D}}^b\left( \mathscr{C}^{ac}\hat{\mathscr{N}}_{bc}\right)-\frac{1}{4}\hat{\mathscr{D}}^a\left( \mathscr{C}^{bc}\hat{\mathscr{N}}_{bc}\right)\bigg]
.
\label{eqgenCicar}
\end{eqnarray}

\end{enumerate}

Equations \eqref{eqgenC0car} and \eqref{eqgenCicar} are one of our main achievements and deserve further discussion. We would like to insist that there is neither magic nor ambiguity in reaching them. We have followed a plain zero-$k$ limit informed about the regularity conditions \eqref{cotfixcoef}, which involve the Carrollian Cotton tensor and its identities, and instructed with Einstein's equation $\xi_{ab}=-\frac{k^2}{2} \mathscr{C}_{ab}$. Although long and technical, the method reveals the central role of the Cotton tensor: \emph{all terms responsible for the gravitational radiation, involving among others the shear and the news tensors, originate from the Carrollian Cotton tensors.} Because of the vanishing $\xi_{ab}$, only six of those remain --- see  App. \ref{carcot3}:  $c_{(-1)}$, $c_{(0)}$, $\psi^a$, Eqs. \eqref{cfr}, \eqref{c7fr}, and  $\chi^a$, $\Psi^{ab}$, $X^{ab}$
given in \eqref{c6frnt}, \eqref{c10frnt}, \eqref{c9frnt} for vanishing Carrollian shear. They
obey Eqs. \eqref{carDcotns}, \eqref{carEcotns}, \eqref{carIcotns}, 
\eqref{carGcotns}  and
\eqref{carHcotns}.\footnote{The Carrollian Cotton identities described in App. \ref{carcot3} as $\left\{\mathcal{D}_{\text{Cot}}=0, \mathcal{I}_{\text{Cot}}^a =0\right\}$, 
$\left\{\mathcal{E}_{\text{Cot}}=0, \mathcal{G}_{\text{Cot}}^a =0\right\}$, 
$\left\{\mathcal{F}_{\text{Cot}}=0, \mathcal{H}_{\text{Cot}}^a =0\right\}$
and $\left\{\mathcal{W}_{\text{Cot}}=0, \mathcal{X}_{\text{Cot}}^a =0\right\}$
are obtained from the relativistic conservation \eqref{C-cons} expressed as \eqref{Lcot} and \eqref{Tcota}. We \emph{must not} insert in these equations $\xi_{ab}=-\frac{k^2}{2}\mathscr{C}_{ab}$ before taking the flat limit as they are agnostic about bulk Einstein's equations. The Cotton identities at hand are of boundary-geometric nature, and when the Carrollian shear  vanishes, they just become simpler by setting $\xi_{ab}=0$.}
 This means in particular that once the bulk flat limit is reached i.e. the boundary Carroll structure has no geometric shear, $z^a$ and $Z^{ab}$ vanish. The Carrollian energy--momentum tensors $\zeta^a$ and $\zeta^{ab}$ do also vanish by virtue of \eqref{cotfixcoef}. Only $Q^a$, $\Sigma^{ab}$ and $\Xi^{ab}$ survive, and  \eqref{eqgenC0car}, \eqref{eqgenCicar} lead  to an alternative writing of the conservation equations:
\begin{equation}   
\label{eqgenC0caralt}
\hat{\mathscr{D}}_\upupsilon \varepsilon_{(0)}+\hat{\mathscr{D}}_a  Q^a=
 \frac{1}{16\pi G} \left(
\hat{\mathscr{D}}_a  \hat{\mathscr{D}}_b \hat{\mathscr{N}}^{ab}+ \mathscr{C}^{ab}
\hat{\mathscr{D}}_a \hat{\mathscr{R}}_b 
+\frac{1}{2} \mathscr{C}_{ab}\hat{\mathscr{D}}_\upupsilon  \hat{\mathscr{N}}^{ab}
\right)
\end{equation}
and
\begin{eqnarray}
\frac{1}{2}\hat{\mathscr{D}}^a\varepsilon_{(0)}
-\hat{\mathscr{D}}_b \Xi^{ab}+2 \ast\! \varpi\ast\! Q^a+ \hat{\mathscr{D}}_\upsilon  \pi^a 
&=&
  -\frac{1}{16\pi G}\bigg[\mathscr{C}^{ab}\hat{\mathscr{D}}_b\hat{\mathscr{K}}
+\ast\mathscr{C}^{ab}\hat{\mathscr{D}}_b\hat{\mathscr{A}}
-4\ast\!\varpi \ast\!\mathscr{C}^{ab}\hat{\mathscr{R}}_b
  \nonumber \\
&&-\frac{1}{2}\hat{\mathscr{D}}^b\left(\hat{\mathscr{D}}_b\hat{\mathscr{D}}_c \mathscr{C}^{ac}
-\hat{\mathscr{D}}^a\hat{\mathscr{D}}^c \mathscr{C}_{bc}
\right)+\mathscr{C}^{ab}\hat{\mathscr{D}}^c\hat{\mathscr{N}}_{bc}
 \nonumber \\
&&+\frac{1}{2}\hat{\mathscr{D}}^b\left( \mathscr{C}^{ac}\hat{\mathscr{N}}_{bc}\right)-\frac{1}{4}\hat{\mathscr{D}}^a\left( \mathscr{C}^{bc}\hat{\mathscr{N}}_{bc}\right)\bigg]
.
\label{eqgenCicaralt}
\end{eqnarray}
This latter form discloses a Carrollian conservation of the type \eqref{carEbiscon}, \eqref{carGcon} \emph{with a right-hand side though}. This is thus a \emph{flux-balance equation}, where the 
 source is maintained by the bulk gravitational radiation encoded in the shear and the news. Notice that the above momentum $\pi^a$
 coincides with  $P^a$ in \eqref{carGcon} and is dynamical, whereas the  traceless Carrollian stress $\Xi^{ab}$ is $-\Upsilon^{ab}$ in \eqref{carEbiscon}, \eqref{carGcon},
 and  is dictated by the Cotton due to \eqref{cotfixcoef}; similarly $Q^a $ here is the energy flux $\Pi^a $ of \eqref{carEbiscon}, \eqref{carGcon}, also locked by the Cotton in \eqref{cotfixcoef}.
  
Even in absence of Bondi shear $\mathscr{C}_{ab}$ and news $\hat{\mathscr{N}}_{ab}$, 
the presence of a non-vanishing energy flux betrays the breaking of local Carroll boost invariance (see the end of App. \ref{carcot3}) in the boundary Carrollian dynamics associated with Ricci-flat spacetimes. This breaking accounts for bulk gravitational radiation, which in the covariant Newman--Unti gauge does not originate solely in the news \eqref{news-car} but is also carried by the Carrollian energy flux $\Pi^a=Q^a=\frac{1}{8\pi G} \ast \! \chi^{a}$. In Robinson--Trautman spacetimes and in the present gauge, the gravitational radiation is exclusively rooted in this Carrollian Cotton descendant --- see Ref. \cite{CMPPS2}.

Observe in passing the Carrollian Cotton identities \eqref{carEcotns} and \eqref{carGcotns}, which we replicate here for convenience:\begin{eqnarray}
\hat{\mathscr{D}}_\upupsilon c_{(0)}+\hat{\mathscr{D}}_a \chi^{a}&=&0,
 \label{carEcotnsrep} 
   \\
 \frac{1}{2}\hat{\mathscr{D}}^a c_{(0)}- \hat{\mathscr{D}}_b \Psi^{ab}
+2 \ast\! \varpi   \ast\!\chi^{a}
+ \hat{\mathscr{D}}_\upupsilon \psi^a &=&0
.
  \label{carGcotnsrep}
\end{eqnarray}
They play dual roles with respect to Eqs. \eqref{eqgenC0caralt} and \eqref{eqgenCicaralt}, because the energy density $\varepsilon_{(0)}$
carries information on the mass of the source, while $c_{(0)}$ endorses its nut charge (monopole-like magnetic mass) (see e.g. \cite{Mittal:2022ywl} for a recent discussion on these electric--magnetic dual observables). The two sets of equations are dissymmetric though:
Eq. \eqref{carEcotnsrep} for instance is 
driven exclusively by the Cotton vector $\chi^a$ --- as opposed to its Carroll-dual $\ast\chi^a$ entering the electric-mass equation \eqref{eqgenC0caralt} through $Q^a=\frac{1}{8\pi G} \ast \! \chi^{a}$. Even though 
loss phenomena concern both the electric  and the magnetic masses,  as captured e.g.  in Eqs. (76) and (80) of \cite{Freidel:2021qpz} --- see also App. D of \cite{Barnich:2019vzx},
the time evolution of the nut is not affected by $\mathscr{C}_{ab}$ and $\hat{\mathscr{N}}_{ab}$, whereas that of the mass is, in line with an important distinction between these aspects raised in \cite{Ashtekar82}.

A useful exercise, which we will not undertake here, would be to set up a precise dictionary between the gauge at hand and the more conventional Newman--Unti or Bondi gauges, regarding the radiation observables. We can nonetheless take a few  steps towards this end using the Carrollian tensor $N^a$ introduced in \eqref{angasp},  reminiscent of the Bondi \emph{angular-momentum aspect},\footnote{As for the shear and the news, the physics conveyed by $N^a$ in the covariant Newman--Unti gauge, is slightly different compared to the standard angular-momentum aspect. For the Kerr geometry, as an example, in the gauge at hand $N^a=0$ and the angular momentum is carried by the Carrollian vorticity, as opposed to plain Newman--Unti gauge, where the boundary vorticity is  absent (see Eq. \eqref{cardiffrNU}).  This hints towards the recent progress in defining a supertranslation-invariant angular momentum and comparing the multiple routes to it (see e.g.\ \cite{Fuentealba:2022xsz, Chen:2021szm, Javadinezhad:2022hhl, Javadinezhad:2022ldc, Riva:2023xxm}).} and a Bondi \emph{mass aspect}  
\begin{equation}
\label{map}
M =4\pi G \varepsilon_{(0)}-\frac{1}{8}\mathscr{C}^{ab}\hat{\mathscr{N}}_{ab}.
\end{equation}
This definition is reached from Eq. (2.39) of \cite{Compere:2019bua} valid in anti-de Sitter, at $k=0$.\footnote{It coincides with (42) of 
\cite{Freidel:2021qpz} upon identifying $\mathcal{M}$ of this reference with our $4\pi G \varepsilon_{(0)}$. \label{24}}
What distinguishes the energy density $4\pi G \varepsilon_{(0)}$ and the mass $M$ is a radiative contribution. 

We can attempt to define a magnetic-mass aspect starting from anti-de Sitter, where the behaviour of the bulk Weyl tensor in the gauge used here exhibits the complex-mass combination $\tau=-c+8\pi \text{i} G \varepsilon$ (see \cite{Petkou:2015fvh}) with $\varepsilon$ the AdS-boundary energy density and $c$ the Cotton scalar (longitudinal component with respect to the congruence $\text{u}$). We thus define the \emph{complex mass aspect} of Ricci-flat spacetimes in covariant Newman--Unti gauge as
\begin{equation}
\label{cmap}
\hat \tau =\lim_{k\to 0} \tau = -2\nu + 8\pi \text{i}G \varepsilon_{(0)},
\end{equation}
where 
\begin{equation}
\label{mmap}
\nu  =\frac{1}{2}\lim_{k\to 0} c =\frac12 c_{(0)}-\frac14 \hat{\mathscr{D}}_a\hat{\mathscr{D}}_b \ast\! \mathscr{C}^{ab}-\frac18\mathscr{C}_{ab} \ast\!\hat{\mathscr{N}}^{ab}
\end{equation}
is the \emph{magnetic-mass aspect} reached using \eqref{cotsc} and \eqref{cfr} upon substitution of  $\xi_{ab}=-\frac{k^2}{2}\mathscr{C}_{ab}$. Subtracting the radiative contribution as in \eqref{map}, we define the \emph{nut aspect} 
\begin{equation}
\label{nmap}
N  = \nu + \frac18\mathscr{C}_{ab} \ast\!\hat{\mathscr{N}}^{ab}
=\frac12 c_{(0)}-\frac14 \hat{\mathscr{D}}_a\hat{\mathscr{D}}_b \ast\! \mathscr{C}^{ab},
\end{equation}
where $ c_{(0)}= \left(\hat{\mathscr{D}}_a\hat{\mathscr{D}}^a+2\hat{\mathscr{K}}\right)\ast\! \varpi$ is one of the four Carroll Cotton scalars displayed in  \eqref{cfr}.\footnote{Our definitions for $\nu$ and $N$ match with $-\tilde{\mathcal{M}}$ and $-\tilde M$ of \cite{Freidel:2021qpz} , Eqs. (53) and (55), for $c_{(0)}=0$ (no magnetic monopole mass). This condition pertains to the use of the Bondi gauge in the quoted reference, where no Ehresmann connection exists and thus $\ast \varpi $ vanishes (as $\varphi_a$).\label{25}} Following the case of asymptotically AdS spacetimes quoted earlier, 
the behaviour of the bulk Weyl tensor in the Ricci-flat instance does also depend on the complex mass aspect $\hat \tau$, and we find indeed
\begin{equation}
\label{Psi2}
\Uppsi_2=\frac{\text{i}\hat{\tau}}{2r^3} +\mathcal{O}\left(\nicefrac{1}{r^4}\right).
\end{equation}
The higher-order missing terms in \eqref{Psi2} are absent in the resummable, algebraically special solutions discussed in Refs. \cite{Gath:2015nxa, Petkou:2015fvh, Mittal:2022ywl}. Unsurprisingly, this expression coincides with Eq. (68c) of \cite{Freidel:2021qpz}.

With the above definitions, Eqs.  \eqref{carEcotnsrep}, \eqref{eqgenC0caralt} and \eqref{eqgenCicaralt} become:\footnote{All these computations call for abundant use of the Weyl-covariant-derivative commutators presented in the appendix, Eqs. \eqref{carphi},  
\eqref{CWcurvten},  
\eqref{CWcurvtenT2},  
\eqref{CWrsc},  
\eqref{CWrvecfull} and 
\eqref{CWrvecfullT2} 
 .}
\begin{eqnarray}
\label{nuttime}
\hat{\mathscr{D}}_\upupsilon N&=&-\frac{1}{2}\hat{\mathscr{D}}_a \chi^{a}-\frac{1}{4} \left(
\hat{\mathscr{D}}_a  \hat{\mathscr{D}}_b \ast\! \hat{\mathscr{N}}^{ab}-\ast  \mathscr{C}^{ab}
\hat{\mathscr{D}}_a \hat{\mathscr{R}}_b \right),
\\
\hat{\mathscr{D}}_\upupsilon M&=&-\frac{1}{2}\hat{\mathscr{D}}_a  \ast\! \chi^{a}+
\frac{1}{4} \left(
\hat{\mathscr{D}}_a  \hat{\mathscr{D}}_b \hat{\mathscr{N}}^{ab}+ \mathscr{C}^{ab}
\hat{\mathscr{D}}_a \hat{\mathscr{R}}_b 
-\frac{1}{2}  \hat{\mathscr{N}}_{ab} \hat{\mathscr{N}}^{ab}
\right),
\label{eqgenC0carlbms}
\\
\hat{\mathscr{D}}_\upupsilon N^a-\hat{\mathscr{D}}^a M+\ast \hat{\mathscr{D}}^a  N
&=&
\frac{1}{2}\bigg[\mathscr{C}^{ab}\hat{\mathscr{D}}_b\hat{\mathscr{K}}
+\ast\mathscr{C}^{ab}\hat{\mathscr{D}}_b\hat{\mathscr{A}}
-4\ast\!\varpi \ast\!\mathscr{C}^{ab}\hat{\mathscr{R}}_b-\frac{1}{2}\ast\! \hat{\mathscr{D}}^a  \hat{\mathscr{D}}_b  \hat{\mathscr{D}}_c \ast\! \mathscr{C}^{bc}
  \nonumber \\
&&-\frac{1}{2}\hat{\mathscr{D}}^b\left(\hat{\mathscr{D}}_b\hat{\mathscr{D}}_c \mathscr{C}^{ac}
-\hat{\mathscr{D}}^a\hat{\mathscr{D}}^c \mathscr{C}_{bc}
\right)+\mathscr{C}^{ab}\hat{\mathscr{D}}^c\hat{\mathscr{N}}_{bc}+\frac{1}{2}\hat{\mathscr{D}}^b\left( \mathscr{C}^{ac}\hat{\mathscr{N}}_{bc}\right)\bigg]
.\qquad
\label{eqgenCicarlbms}
\end{eqnarray}
The first equation phrases the loss process of the nut aspect sustained by the Carroll-dual news $\ast \hat{\mathscr{N}}_{ab}$ and the Carroll Cotton current $\chi^a$. It is actually a \emph{geometric identity }associated with the Carroll structure --- as is \eqref{carGcotnsrep}, which could have been reexpressed as well in terms of the nut aspect.
The last two flux-balance equations \eqref{eqgenC0carlbms} and \eqref{eqgenCicarlbms} for the electric-mass and angular-momentum aspects
are \emph{genuinely dynamical} and coincide with Eqs. (2.53) and (2.50) of Ref. \cite{Compere:2019bua}, where the
 approach to asymptotic flatness 
  via a limit of vanishing cosmological constant 
 was proposed, or else with (4.50) and (4.49) of \cite{Barnich:2010eb},  obtained in a plain Ricci-flat context.\footnote{In the quoted section 2.5 of  \cite{Compere:2019bua} $\upmu=-\text{d}u$ so that $\varphi_a=\varpi_{ab}=0$ (Bondi gauge with $\exp 2 \beta_0=1$). Furthermore our definition of $N^a$ is slightly different: $ N_{\text{here}}^a= N_{\text{there}}^a+\frac{1}{4}\left(\mathscr{C}^{ab}\hat \nabla^c\mathscr{C}_{bc}+\frac{3}{8}\hat\nabla^a \left(\mathscr{C}^{bc}\mathscr{C}_{bc}\right)\right) $
with $\hat \nabla^c$ being actually the ordinary two-dimensional Levi--Civita connection due to the absence of Ehresmann connection in 
 \cite{Compere:2019bua} (see Eq. \eqref{CWs-tens} where the Carroll-Weyl covariant derivative reduces to the ordinary one when $\varphi_{a}$ vanishes). This definition is in line with that of \cite{Flanagan:2015pxa}.
 Likewise   $\hat{\mathscr{N}}^{ab}_{\text{here}}=N_{\text{TF  there}}^{ab}-\frac{l_{\text{there}}}{2}\mathscr{C}^{ab}$ with $l_{\text{there}}=\theta_{\text{here}}$ and for further use we also quote that $E^{ab}_{\text{here}}=-\frac{3}{16\pi G}\left(\mathcal{E}^{ab}_{\text{there}}-\frac{1}{16}\mathscr{C}^{ab}\mathscr{C}^{cd}\mathscr{C}_{cd}\right)$. The comparison with Ref. \cite{Barnich:2010eb} is reviewed in \cite{Compere:2019bua}.\label{27}}
 
\boldmath
\subsubsection*{Order $\nicefrac{1}{r^2}$ and  the next flux-balance equation}
\unboldmath

The Carrollian symmetric and traceless two-tensor $E_{ab}$, descendant of the AdS-boundary stress, enters the line element at order 
$\nicefrac{1}{r}$. However,  the fundamental Carrollian  energy--momentum conservation equations \eqref{eqgenC0car} and \eqref{eqgenCicar} fail to capture  its dynamics. In a direct search of Ricci-flat spacetimes, Einstein's equations bring their share at each order and this is how the flux-balance equations emerge for the Chthonian degrees of freedom as $E_{ab}$. In the present method, Einstein's equations have already been imposed at the considered order. The bulk metric including the term \eqref{O2} with the $f_{(2)}$s as in \eqref{f2sca} \eqref{f2vec} and \eqref{f2ten} is thus on-shell --- assuming  \eqref{T-cons} is satisfied. However, this term is due to exhibit divergences at vanishing $k$. Removing them will impose conditions involving the Chthonian degrees of freedom as well as their longitudinal derivatives appearing explicitly in \eqref{f2ten}. This is how flat \emph{flux-balance equations are recovered in the transition from anti-de Sitter to asymptotically flat spacetimes}, and this is another laudable achievement of this note.

The protocol is by now well established: we ought to follow the four steps enumerated earlier, starting with any tensor $f_{(2)}$ --- and later on with other $f_{(s)}$. Let us open the study with the scalar contribution $f_{(2)}$, Eq.  \eqref{f2sca}. With little effort we find:
\begin{equation}
\label{f2scflat}
\lim_{k\to 0}f_{(2)} = 2\ast\! \varpi \nu -\frac{1}{3}\hat{\mathscr{D}}_{a} N^{a} \equiv \hat f_{(2)} .
\end{equation}
Next we consider the transverse vector $f_{(2)a}\uptheta^a$ in \eqref{f2vec}:
\begin{equation}
\label{f2vecflat}
\lim_{k\to 0}f_{(2)a} = -\frac{1}{6} N^b \mathscr{C}_{ba} -\frac{4}{3}\ast  \! \varpi\ast \! N_a-4\pi G \hat{\mathscr{D}}_{b} E^b_{\hphantom{b}a} \equiv \hat f_{(2)a}  .
\end{equation}

Neither the limit \eqref{f2scflat} nor  \eqref{f2vecflat} introduce any new Chthonian degree of freedom or impose any further condition on their  evolution.
As we will  now see, the situation is different for the transverse tensor \eqref{f2ten} $f_{(2)ab}\uptheta^a\uptheta^b$.  Using the numerous tools developed in this work, we find:\footnote{We define the symmetric and traceless part of a Carrollian two-tensor $s_{ab}$ as $s_{\langle ab\rangle}=s_{(ab)}-\frac{1}{d}s_c^{\hphantom{c}c}\delta_{ab}$ (here $d=2$).}
\begin{eqnarray}
f_{(2)ab}&=& \frac{1}{k^2}\left(
\frac{16\pi G}{3}  \hat{\mathscr{D}}_\upupsilon E_{ab}+\frac{1}{3} \hat{\mathscr{D}}_{\langle a} N_{b\rangle}+ 2\pi G \varepsilon_{(0)}\mathscr{C}_{ab} -  \frac{\nu }{2}  \ast \! \mathscr{C}_{ab}
\right)
\nonumber
\\
&&+2\pi G\left(
\frac{8}{3}  \hat{\mathscr{D}}_\upupsilon E_{(2)ab}-\frac{4}{3} \hat{\mathscr{D}}_{\langle a} \pi_{(2)b\rangle}+\varepsilon_{(1)} \mathscr{C}_{ab}
-2  \mathscr{C}_{(a}^{\hphantom{(a}c}E_{b)c}^{\vphantom{c}} 
\right)
-2\ast \!  \varpi^3 \ast \! \mathscr{C}_{ab}
\nonumber
\\
&&+\mathcal{O}\left(k^2\right).
\label{f2tenexp}
\end{eqnarray}
This result meets our expectations and allows us to draw significant conclusions. 
\begin{itemize}
\item The flat limit is singular unless the order-$\nicefrac{1}{k^2}$ contribution to $f_{(2)ab}$  is absent i.e.
\begin{equation}
\label{f2fluxeq}
\boxed{
 \hat{\mathscr{D}}_\upupsilon E_{ab}=\frac{3}{16\pi G} \left( -\frac{1}{3} \hat{\mathscr{D}}_{\langle a} N_{b\rangle} -2\pi G \varepsilon_{(0)}\mathscr{C}_{ab}+  \frac{ \nu }{2}\ast \! \mathscr{C}_{ab}\right)
,}
\end{equation}
which is the sought-after Carrollian  \emph{flux-balance equation for $E_{ab}$}, later referred to as $\text{FBE}_{(1)}=0$. This equation matches with Eq. (4e) of \cite{Freidel:2021qpz}.\footnote{For this we use the dictionary for Ref. \cite{Freidel:2021qpz} set up in footnotes \ref{24} and \ref{25}, together with the relations $-16\pi G E^{ab}_{\text{here}}=\mathcal{T}^{ab}_{\text{there}}$ and $N^{a}_{\text{here}}=\mathcal{P}^{a}_{\text{there}}$. Observe that Eqs. (58) and (63) of \cite{Freidel:2021qpz} are also compatible with further quantities introduced in \cite{Compere:2019bua} and mentioned in footnote \ref{27}.} 
\item Assuming Eq. \eqref{f2fluxeq} is fulfilled, the limit can be taken
\begin{eqnarray}
\lim_{k\to 0}f_{(2)ab} &=&
\frac{16\pi G}{3}\left(
 \hat{\mathscr{D}}_\upupsilon E_{(2)ab}-\frac{1}{2} \hat{\mathscr{D}}_{\langle a} \pi_{(2)b\rangle}
+\frac{3}{8} \varepsilon_{(1)} \mathscr{C}_{ab}
-\frac{3}{4} \mathscr{C}_{(a}^{\hphantom{(a}c}E_{b)c}^{\vphantom{c}} 
\right)
-2\ast \!  \varpi^3 \ast \! \mathscr{C}_{ab}
\nonumber
\\
&\equiv&
\hat f_{(2)ab} 
,
\label{f2tenflat}
\end{eqnarray}
and provides the last piece of the order-$\nicefrac{1}{r^2}$ term in the Ricci-flat line element.
\item New Chthonian degrees of freedom enter the bulk metric at this order: $E_{(2)ab}$,  $\pi_{(2)a}$ and $\varepsilon_{(1)}$ in the form of a symmetric and traceless Carrollian tensor  
\begin{equation}
\label{Fab}
F_{ab}=  \hat{\mathscr{D}}_\upupsilon E_{(2)ab}-\frac{1}{2}\hat{\mathscr{D}}_{\langle a} \pi_{(2)b\rangle}+ \frac{3}{8}\varepsilon_{(1)} \mathscr{C}_{ab} -\frac{3}{8\pi G}\ast \!  \varpi^3 \ast \! \mathscr{C}_{ab}.
\end{equation}
Their dynamics is unknown at this stage but will be unravelled in the course of the analysis at order $\nicefrac{1}{r^3}$.

\end{itemize}

We will close this paragraph exhibiting the explicit Ricci flat metric at the considered order. To this end we use the results \eqref{newcovNUortho-tran}, \eqref{RFr}, \eqref{S-car}, \eqref{O1carlim}, \eqref{f2scflat},  \eqref{f2vecflat},  \eqref{f2tenflat} and  \eqref{Fab}:
\begin{eqnarray}
\label{RFrm2}
\text{d}s^2_{\text{Ricci-flat}}&=&
 \upmu\left[2\text{d}r+\left(
2r\varphi_a-2\ast\!\hat{\mathscr{D}}_{a}\ast\!\varpi -\hat{\mathscr{D}}_b\mathscr{C}^{b}_{\hphantom{b}a}
\right)\hat{\uptheta}^a-\left(r\theta+\hat{\mathscr{K}}\right)\upmu
\right]
\nonumber
\\
&&+\left(r^2+\ast\varpi^2+\frac{\mathscr{C}^2}{4} \right)\text{d}\ell^2+\left( r \mathscr{C}_{ab}+\ast \varpi \ast\!\mathscr{C}_{ab}\right)\hat{\uptheta}^a\hat{\uptheta}^b
\nonumber
\\
&&+\frac{1}{r}\left(
 8\pi G  \varepsilon_{(0)} \upmu^2
-
\frac{4}{3}\upmu N_a\hat{\uptheta}^a
-\frac{16 \pi G}{3}E_{ab}\hat{\uptheta}^a \hat{\uptheta}^b
\right)
\nonumber
\\
&&+\frac{1}{r^2}\left(2\ast\! \varpi \nu -\frac{1}{3}\hat{\mathscr{D}}_{a} N^a
\right)\upmu^2
-\frac{1}{r^2}\upmu\left(
 \frac{1}{3} N^b \mathscr{C}_{ba} +\frac{8}{3}\ast  \! \varpi\ast \! N_a+8\pi G \hat{\mathscr{D}}_{b} E^b_{\hphantom{b}a}
\right)\hat\uptheta^a
\nonumber
\\
&&+\frac{1}{r^2}
\left(
\frac{16\pi G}{3}  F_{ab}
-4\pi G  \mathscr{C}_{(a}^{\hphantom{(a}c}E_{b)c}^{\vphantom{c}} 
\right)
\hat\uptheta^a\hat\uptheta^b
+  \mathcal{O}\left(\nicefrac{1}{r^3}\right).
\end{eqnarray}
This solution to vacuum Einstein's equations is built upon the following boundary Carrollian data: (\romannumeral1) a generic Carrollian structure with  geometric shear $\xi_{ab}=0$ (but arbitrary Ehresmann connection providing $\varphi_a$ and $\ast \varpi)$; (\romannumeral2)  a dynamical shear $\mathscr{C}_{ab}$, utterly free; (\romannumeral3) an energy density $\varepsilon_{(0)}$ i.e. a Bondi mass $M$, a heat current $N_a$ aka the Bondi angular momentum aspect and a stress $E_{ab}$, all satisfying the flux-balance equations \eqref{eqgenC0carlbms}, \eqref{eqgenCicarlbms} and \eqref{f2fluxeq};\footnote{As pointed out earlier the nut aspect $N$ --- equivalently the magnetic mass $\nu$ --- is in essence  part of the Carrollian structure
and its evolution equation
\eqref{nuttime} is a geometric identity in disguise.}
(\romannumeral4) three more degrees of freedom $E_{(2)ab}$, $\pi_{(2)a}$ and $\varepsilon_{(1)}$ encoded in $F_{ab}$ \eqref{Fab} with evolution equations yet to be uncovered.

\subsubsection*{Recursion and the fate of Chthonian degrees of freedom}

 All this has been achieved as the limit of vanishing cosmological constant within general asymptotically anti-de Sitter Einstein spacetimes, where infinitely many flat degrees of freedom originate from the Laurent expansion of the anti-de Sitter boundary energy--momentum tensor about $k^2=\nicefrac{-\Lambda}{3}=0$ and constrained through evolution equations. Compared to the anti-de Sitter solution space, the extra --- Chthonian --- functions are $\left\{\upchi_{(n\geq 2)}\right\}\equiv\left\{\varepsilon_{(n-1\geq 1)}, \pi^a_{(n\geq 2)}, E^{ab}_{(n\geq 2)}\right\}$. It is natural to wonder whether these are truly independent functions. Answering this question demands a higher-order analysis but some simple considerations allow to infer that \emph{$\varepsilon_{(n -1)}$, $\pi^a_{(n)}$ and $E_{(n)}^{ab}$ could be repackaged in a single symmetric traceless tensor $F^{ab}_{(n)}$, having the expected conformal weight.}

Indeed, one should recall that the $\upchi_{(n)}$s are all weight-$3$ and contribute to the $f_{(s)}$s (of weight $s+2$) through an appropriate number of longitudinal or transverse Weyl-covariant  derivatives $u^C\mathscr{D}_C $ or $\mathscr{D}_a$, powers of vorticity $\omega_{ab}$ or shear $\sigma_{ab}$, all raising the weight by one unit (in the Carrollian limit, the latter two bring a factor $k^2$ with $\ast \varpi\hat\eta_{ab}$ or $\mathscr{C}_{ab}$).
The analysis of Einstein's equations 
	$\mathcal{E}_{rr}$, 
	$\mathcal{E}_{r\hat 0}$,
	$\mathcal{E}_{ra}$ and
	$\mathcal{E}_{ab}$
in the radial expansion exhibits a remarkable recursion structure for the $f_{(s)}$,  $f_{(s)}^{a}$ and   $f_{(s)}^{ab}$ --- for $s=2$ these equations are sorted in \eqref{einst-rec-2}. The latter are given in terms of quantities of order $s-1$ along with one transverse Weyl derivative, one power of vorticity, or one power of shear.  
Furthermore, the scalar and the vector do not involve any net power of $k^2$, whereas the tensor does: $f_{(s)}^{ab}=\frac{1}{k^2} \left[u^C\mathscr{D}_C f_{(s-1)}^{ab}+\cdots\right]$. This shows, on the one hand, that the scalar and vector contributions to the line element remain  finite in the Carrollian limit, and do not impose any supplementary constraint. On the other hand, flux-balance equations originate exclusively from the two-index term. 

Owing to the fact that  $\upchi=\sum_{m \geq  2} k^{2m}\upchi_{(m)}$, the Chthonian degrees of freedom $\upchi_{(m)}$ persist in the Carrollian limit of the $f$s if a power of $k^2$ equal to or more negative than $-m$ is inherited from the AdS solution.
Combined with the above recursive pattern, where in particular negative powers appear solely in the tensor $f_{(s)}^{ab}$,  this suggests that once a combination of $\upchi_{(m)}$ has emerged, such as $F^{ab}\equiv F^{ab}_{(2)} $ in Eq.~\eqref{Fab} 
 for $\upchi_{(2)}$  inside
\eqref{f2tenexp}, only this precise expression will appear in the subsequent orders, along with more derivatives, powers of shear and vorticity, and increasing negative powers of $k^2$. For instance, this occurs for $F_{(1)}^{ab}\equiv E_{(1)}^{ab}\equiv E^{ab}$ in $f_{(1)}^{ab}$ as in Eq.~\eqref{O1carlim}, $f_{(2)}^{ab}$ as in Eq.~\eqref{f2tenexp}, and likewise in higher orders.

This scheme has two consequences. The first is that at order $(s+1)$, \emph{one} new flux-balance equation $\text{FBE}_{(s)}=0$  emerges, for the previously determined combination $F^{ab}_{(s)} $ of the Chthonian functions  $\upchi_{(s)}$, as it should for global evolution  consistency. Schematically this property is captured in the following: 
\begin{equation}
 f_{(s+1)}^{ab}= \sum_{n=1}^{s-1}\frac{c_{(s,n)}}{k^{2(s-n+1)}}
 \hat{\mathscr{D}}_{\upupsilon}^{s-n} \text{FBE}_{(n)}+\frac{c_{(s,s)}}{k^2}  \text{FBE}_{(s)} +  \hat f_{(s+1)}^{ab} +\mathcal{O}\left(k^2\right) ,
\end{equation}
with $c_{(s,n)}$ some immaterial coefficients, the new equation being $\text{FBE}_{(s)} =0$. The second consequence is that \emph{the triplet $\upchi_{(s)}$ counts as a single Chthonian degree of freedom materialized in $F^{ab}_{(s)} $,} the one appearing in the line element and obeying a flux-balance equation revealed at the next order in $\nicefrac{1}{r}$. The following generic structure of the solutions underpins the above reasoning:\footnote{We remind that $F_{(1)}^{ab}\equiv E_{(1)}^{ab}\equiv E^{ab}$,  $\pi_{(1)}^{a}\equiv \pi^{a}$,  and $F_{(2)}^{ab}\equiv F^{ab}$.}
\begin{itemize}
\item $\hat f_{(2s+1)}$ contains $(-)^ s 8 \pi G \varepsilon_{(0)} \ast\!\varpi^{2s}$ and $\hat{\mathscr{D}}_a \hat f_{(2s)}^a$;
\item $\hat f_{(2s+2)}$ contains $(-)^s  2 \nu \ast\!\varpi^{2s+1}$ and $\hat{\mathscr{D}}_a \hat f_{(2s+1)}^a$;
\item $\hat f_{(s+1)}^a$ contains $\hat{\mathscr{D}}_b  \hat f_{(s)}^{ab}$, $\ast\varpi \ast \hat f_{(s)}^a$, $\mathscr{C}^a_{\hphantom{a}b}\hat f_{(s)}^b$;
\item $ \hat f_{(s+1)}^{ab}= c_{(s+1,s+1)}  F_{(s+1)}^{ab} + \text{ tensors based on objects of order } s$;
\item $F_{(s+1)}^{ab}$ contains $ \hat{\mathscr{D}}_{\upupsilon}^{s} E_{(s+1)}^{ab} $,   $ \hat{\mathscr{D}}_{\upupsilon}^{s-1} \hat{\mathscr{D}}_{\vphantom{(s+1)}}^{\langle a} \pi_{(s+1)}^{b\rangle} $, $\varepsilon_{(s)} \ast\!\varpi^{s-1}\mathscr{C}^{ab}$, \dots;
\item  $\text{FBE}_{(s)}=0$ is of the form $ \hat{\mathscr{D}}_{\upupsilon}F_{(s)}^{ab}=\left\{\hat{\mathscr{D}}_{\vphantom{(s)}}^{\langle a} \hat f_{(s)}^{b\rangle} , \mathscr{C}^{ab} \hat f_{(s)}, \ldots \right\} $,
\end{itemize}
where the dots stand for other possible admissible terms. As anticipated, the actual Chthonian degrees of freedom capturing the flat dynamics are the emerging $F_{(s)}^{ab}$, which should be substituted for the  $(s-1)$th derivatives of  $E_{(s)}^{ab}$, $\pi_{(s)}^{a}$ and $\varepsilon_{(s-1)}$ delivered by the anti-de Sitter energy--momentum tensor. 

A legitimate question one may finally ask in view of our analysis pertains to the existence of other, possibly infinite, sets of Chthonian data originating from a Laurent expansion of the AdS boundary metric (see e.g. \cite{BigFluid, Laurhydro}). Direct exploration of Ricci-flat solution spaces does not seem to support such an expectation, but a definite answer requires a thorough investigation, which would bring us far from our main goal.

\subsubsection*{The flat resummation}

The anti-de Sitter resummable instance presented in Eq. \eqref{papaefgenrescrec} can be realized in the flat limit, as it was shown in \cite{CMPPS2}. In this case all Chthonian functions should vanish, together with $N^a$, $E^{ab}$ and the shear $\mathscr{C}^{ab}$, leading ultimately to 
\begin{eqnarray}
\label{RFrm2res}
\text{d}s^2_{\text{res. Ricci-flat}}&=&
 \upmu\left[2\text{d}r+\left(
2r\varphi_a-2\ast\!\hat{\mathscr{D}}_{a}\ast\!\varpi
\right)\hat{\uptheta}^a-\left(r\theta+\hat{\mathscr{K}}\right)\upmu
\right]
\nonumber
\\
&&+\left(r^2+\ast\varpi^2 \right)\text{d}\ell^2
+\frac{1}{r^2+\ast\varpi^2}\left( 8\pi G  \varepsilon_{(0)} r+\ast\! \varpi c_{(0)} 
\right)\upmu^2
.
\end{eqnarray}
This captures all algebraically special Ricci-flat spacetimes provided $\varepsilon_{(0)}$ obeys \eqref{eqgenC0caralt} and \eqref{eqgenCicaralt} which now read:
\begin{eqnarray}
\hat{\mathscr{D}}_\upupsilon \varepsilon_{(0)}
+\frac{1}{8\pi G} \hat{\mathscr{D}}_a  \ast \! \chi^{a}
&=& 0 ,
  \label{carEbiscon-res} 
  \\
\hat{\mathscr{D}}_a \varepsilon_{(0)}-\frac{1}{8\pi G} \ast\!\hat{\mathscr{D}}_a c_{(0)} 
&=&0 .
  \label{carGconres} 
  \end{eqnarray}
Equations  \eqref{carEbiscon-res} and \eqref{carGconres} coincide with Eqs. (29.16) and (29.15)
of \cite{Stephani:624239}.\footnote{For that purpose, the following identifications are necessary, in Papapetrou--Randers frame and complex coordinates $\mathbf{x}=\left\{\zeta, \bar \zeta\right\}$ with $\text{d}\ell^2=\frac{2}{P^2(u, \zeta, \bar\zeta)}\text{d}\zeta\text{d}\bar\zeta$, $\upupsilon=\frac{1}{\Omega}\partial_u$, $\upmu =-\Omega \text{d}u +b_a \text{d}x^a$ and $\hat{\text{e}}_a=\hat{\partial}_a=\partial_a+\frac{b_a}{\Omega}\partial_u$, $\hat\uptheta^a=\text{d}x^a$: $\Omega = 1$, $b_\zeta=-L$, $\ast  \varpi=-\Sigma$, $\hat \tau =2(M +\text{i}m)$,  whereas their radial coordinate is $\tilde r = r-r_0$ with $r_0(u, \zeta, \bar \zeta)$ the origin in the affine parameter of the geodesic congruence tangent to $\partial_r$.
\label{origin}} The latter are rather
complicated and it is remarkable they are tamed into simple conservation equations such as  \eqref{carEbiscon-res} and \eqref{carGconres}.  Reaching this conclusion would have been inconceivable without the null boundary perspective and the  Carrollian tools, which are the appropriate language for asymptotically flat spacetimes.

The algebraically-special nature of the metric \eqref{RFrm2res} is proven using the Goldberg--Sachs theorem with the null, geodesic, and shear-free in the resummed instance, bulk congruence tangent to $\partial_r$. The latter is part of the canonical null tetrad  parallelly transported  along  $\partial_r$ (thanks to the affine nature of $r$) introduced in \cite{CMPPS2}, which coincides with that of \cite{Stephani:624239}, Eq. (29.13a), as well as with the original Ref. \cite{NP68}. In complex celestial-sphere coordinates $\zeta$ and $\bar \zeta$ (see footnote \ref{origin}) the null tetrad reads:
\begin{equation}
\label{nultetrad}
\begin{cases}
\mathbf{k}=\partial_r \\
\mathbf{l}=\frac{1}{2}\left(\frac{8\pi G\varepsilon_{(0)} r+\ast \varpi c_{(0)}}{r^2+\ast\varpi^2}-r\theta-\hat{\mathscr{K}}\right)\partial_r+\upupsilon\\
\mathbf{m}=\frac{P}{r-\text{i}\ast\! \varpi}
\left(\hat\partial_{\bar \zeta}
+\left(\ast\hat{\mathscr{D}}_{\bar \zeta}\ast\!\varpi-r\varphi_{\bar \zeta}\right)\partial_r\right)
\end{cases}  
\end{equation}  
with $\mathbf{k}\cdot\mathbf{l}=-1 $, $\mathbf{m}\cdot\bar{\mathbf{m}}=1$ and $\text{d}s^2_{\text{res. Ricci-flat}}=-2 \mathbf{k} \mathbf{l}+2 \mathbf{m}\bar{\mathbf{m}}$. Generically, $\mathbf{k}$ is a multiplicity-two principal null direction of the Weyl tensor, and using the tetrad at hand we find the following Weyl complex scalars: ${\Uppsi}_0={\Uppsi}_1=0$ and
\begin{equation}
\label{psis}
{\Uppsi}_2=\frac{\text{i}\hat{\tau}}{2(r-\text{i}\ast\! \varpi)^3},\quad
{\Uppsi}_3=\frac{\text{i}P\chi_\zeta}{(r-\text{i}\ast\! \varpi)^2}+ \text{O}\left(\nicefrac{1}{(r-\text{i}\ast\! \varpi)^3}\right),\quad
{\Uppsi}_4=\frac{\text{i} X_\zeta^{\hphantom{\zeta}\bar \zeta}}{r-\text{i}\ast\! \varpi}+ \text{O}\left(\nicefrac{1}{(r-\text{i}\ast\! \varpi)^2}\right).
\end{equation}  
Observe that neither $\Uppsi_3$ nor $\Uppsi_4$ vanish in the instance of Petrov type D solutions, because $\mathbf{l}$ \emph{is not}  a principal null direction.  Another tetrad is reached  with a Lorentz transformation suitably adjusted for $\mathbf{l}' $ be a principal direction of multiplicity two whereas $\mathbf{k}'\propto \mathbf{k}$, and $\Uppsi_3'=\Uppsi_4'=0$. 
Unsurprisingly, all $\Uppsi$s are spelled using the Carrollian  descendants of the boundary Cotton tensor --- as well as their derivatives in the higher-order terms.

\section{Outlook}

Asymptotically anti-de Sitter and flat spacetimes subject to Einstein equations are distinguished mainly by two features. The first is gravitational radiation escaping at or arriving from null infinity in the flat instance, which is absent under the usual boundary conditions for anti-de Sitter. 
The second concerns the data required for a faithful depiction of these geometries and of their dynamics imposed by Einstein's equations: a finite versus an infinite number for asymptotically AdS or flat. 

In spite of the sharp distinctness of the solution spaces with non-vanishing and zero cosmological constant, the latter can be smoothly reached from the former in a procedure that is the core of this work. It can be outlined in three steps, performed along with the process of sending $\Lambda$ to zero, which simultaneously transmutes the pseudo-Riemannian conformal boundary of anti-de Sitter into a Carrollian descendant, carrying akin information.
\begin{itemize}

\item  Bondi's shear $\mathscr{C}_{AB}$ is substituted on-shell for the geometric shear $\sigma_{AB}$. 

\item The anti-de Sitter boundary energy--momentum tensor $T_{AB}$ is Laurent-expanded in powers of $k^2=\nicefrac{-\Lambda}{3}$ about $k^2=0$. This supplies an infinite number of replicas, which account for the awaited  
flat, Chthonian, degrees of freedom. 

\item The evolution (flux-balance) equations of the --- now Carrollian --- degrees of freedom are reached using both the limit of the conservation of the energy and momentum, as well as the requirement of finiteness for the line element in the flat limit. The latter (\romannumeral1) selects the Chthonian 
variables $\varepsilon_{(n\geq 1)}$, $\pi^a_{(n\geq 2)}$ and $E^{ab}_{(n\geq 2)}$ besides the standard energy density $\varepsilon_{(0)}$, momentum $\pi^a$ and stress $E^{ab}$, and we have argued that genuine degrees of freedom are only the $F^{ab}_{(n)}$s; (\romannumeral2)~freezes a few other components of the expanded energy--momentum tensor in terms of the boundary Carrollian Cotton tensors $\left(\zeta^a, Q^a, \zeta^{ab}, \Sigma^{ab}, \Xi^{ab}\right)$; (\romannumeral3) delivers the Chthonian dynamics, which is not captured by the energy--momentum conservation but echoes flat Einstein equations.  

\end{itemize}

The technical \emph{tour de force} of our exploration shouldn't shadow the conceptual aftermath of our findings. These bring back the boundary energy and momentum at the center of the asymptotically flat bulk reconstruction, besides the Bondi shear, under the form of a Carrollian energy density, momentum and stress, together with an infinite tower of replicas of the latter. Speculating over a flat extension of AdS gauge/gravity duality, and owing to the key role played by the energy--momentum tensor in the latter, one is led to several unescapable questions. What would  the fundamental observables be in the dual Carrollian field theory? What role would the replicas of the energy--momentum sector play? What is the interplay between the Chthonian and the shear/news sector, which has been investigated in celestial holography? Could this correspondence still be qualified as holographic --- given the seemingly infinite number of necessary data? Our approach does not yet provide any cue for answering these questions, though it hands some confidence in the zero-$\Lambda$ limit, that could be inquired within the AdS/CFT correspondence. This last point is probably the deepest our analysis conveys.  

This is the big picture. Other questions merit equal attention, starting with the ones related to symmetries and charges. What are the asymptotic symmetries in a partially unfixed gauge like the covariant Newman--Unti introduced here? What sort of charges does this extension carry? 
What is the precise combination of vorticity and angular-momentum aspect that would define the physical angular momentum? How would logarithmic terms in the radial expansion alter the analysis? In answering these questions, one could follow recent works such as, e.g., \cite{CMPR, Campoleoni:2022wmf, Geiller:2022vto} as well as \cite{Fuentealba:2022xsz, Chen:2021szm, Javadinezhad:2022hhl, Javadinezhad:2022ldc, Capone:2021ouo}. In particular, one should adress the Weyl invariance in conjunction with the boundary local Lorentz (or Carroll) gauge invariance inherited from the onset of a velocity congruence or a clock form in the boundary pseudo-Riemannian (or Carrollian) structure. In a similar fashion as the one presented in this work, a careful analysis would allow to embrace both the anti-de Sitter and flat cases.  

Regarding the charges, a thorough comparison of our method with Newman--Penrose's would be a valuable practice, reasonably accessible thanks to the affinely parameterised radial congruence $\partial_r$ present in the (covariant) Newman--Unti gauge. In the first place, this would allow to extract the famous ten non-vanishing Newman--Penrose conserved charges --- we know that these are carried by the Chthonian stress tensor descendant $F_{ab}$.\footnote{Contact with the Newman--Penrose formalism beyond the algebraically special resummable metrics mentioned at the end of Sec. \ref{brrm} starts with $\Uppsi_0^0 \propto\text{i}   E_{\zeta}^{\hphantom{\zeta}\bar\zeta}$,  $\Uppsi_0^1 \propto\text{i}   F_{\zeta}^{\hphantom{\zeta}\bar\zeta}$,
$\Uppsi_1^0 \propto\text{i}    N_\zeta$, $\Uppsi_2^0 \propto\text{i}\hat \tau  $ (see \eqref{Psi2}), $\Uppsi_3^0 \propto\text{i}   P \chi_\zeta$ and $\Uppsi_4^0 \propto\text{i} X_\zeta^{\hphantom{\zeta}\bar \zeta} $, where the adopted Carrollian frame is that of footnote \ref{origin}.
% metric is $\text{d}\ell^2=\frac{2}{P^2(u, \zeta, \bar\zeta)}\text{d}\zeta\text{d}\bar\zeta$. 
%Carrollian Cotton tensors appear in the last three. 
The higher-order terms will involve derivatives of the Cotton tensors, of the energy density, the momentum and the stress, as well as the infinite tower of Chthonian replicas $F^{ab}_{(s)}$. }
 Secondly, one could recast these charges following their general Carrollian definition, as described in Refs. \cite{Mittal:2022ywl, CM1}, giving credit to this full-fledged boundary method for the charge computation. Lastly, one may deepen concepts such as subleading charges or electric versus magnetic charges and possible dualities involving the Carrollian Cotton tensors, 
as recently undertaken in \cite{Mittal:2022ywl} from the Carrollian standpoint in the limited framework of resummable, algebraically special Ricci-flat spacetimes, and more generally discussed in Refs.  \cite{Conde:2016rom, Compere:2017wrj, Godazgar:2018vmm, Godazgar:2018qpq, Godazgar:2018dvh, Kol:2019nkc, Godazgar:2020gqd, Godazgar:2020kqd, Oliveri:2020xls, Kol:2020vet, Grant:2021hga, Seraj:2021rxd, Seraj:2022qyt, Godazgar:2022jxm, Awad:2022jgn, Godazgar:2022pbx}.

\section*{Acknowledgements}

The present investigation started in 2020 with discussions involving L. Ciambelli, C.~Marteau,  A. Petkou, R. Ruzziconi, K. Siampos and later A. Fiorucci. We would like to thank these colleagues as well as B.~Oblak and A. Seraj. We also thank each other's institutions for hospitality and financial support for the numerous collaboration visits. The work of A. Campoleoni, A. Delfante and S. Pekar 
was partially supported by the Fonds de la Recherche Scientifique -- FNRS 
under grants F.4503.20,  T.0022.19, FC.41161 and FC.36447. Starting October 1st the work of S. Pekar is funded by the \emph{Fonds Friedmann }run by the \emph{Fondation de l'\'Ecole polytechnique}.
The work of D. Rivera-Betancour was funded by Becas Chile (ANID) Scholarship No.~72200301. The work of M. Vilatte was supported by the Hellenic Foundation for Research and Innovation (H.F.R.I.) under the \textsl{First Call for H.F.R.I. Research Projects to support Faculty members and Researchers and the procurement of high-cost research equipment grant} (MIS 1524, Project Number: 96048). The graduate students D. Rivera-Betancour and M.~Vilatte thank the programme \emph{Erasmus+} of the Institut Polytechnique de Paris as well as the Kapodistrian University of Athens, the Aristotle University of Thessaloniki and the University of Mons for hosting them with these fellowships.

\appendix

\section{Carrollian geometry in Cartan frame and arbitrary dimension} \label{carman}

\subsubsection*{Frame and covariance}

Carroll structures on $\mathscr{M}= \mathbb{R} \times \mathscr{S}$ with a $d$-dimensional base $ \mathscr{S}$ were alluded to in Sec. \ref{brrm}.  They are equipped with a degenerate metric, 
\begin{equation}
\label{carmet}
\text{d}\ell^2=\delta_{ab}\hat{\uptheta}^a\hat{\uptheta}^b,
\end{equation}
as well as  a frame and a coframe, $\left\{\hat{\text{e}}_{\hat u}=\upupsilon, \hat{\text{e}}_a\right\}$ and $\left\{\hat{\uptheta}^{\hat u}=-\upmu, \hat{\uptheta}^a\right\}$
obeying
\begin{equation}
\label{carfrcofr}
\upmu(\upupsilon)=-1,\quad
\hat{\uptheta}^a\left(\hat{\text{e}}_b\right)=\delta^a_b,\quad
\hat{\uptheta}^a\left(\upupsilon\right)=0,\quad
\upmu(\hat{\text{e}}_a)=0.
\end{equation}
Here $\upupsilon$ is the field of observers, kernel of the degenerate metric, and $\upmu$ the clock form (see e.g. \cite{Duval:2014uoa}). 

A convenient parameterization in terms of $d+\frac{(d+1)(d+2)}{2}$ functions (i.e. $8$ for $d=2$) is\footnote{Here $\gamma$ is an arbitrary function and must not to be confused with \eqref{omdual} which is related to the vorticity of the timelike congruence $\text{u}$.}
\begin{eqnarray}
\label{mutv}
&\upupsilon=\gamma\left(\partial_u+ v^i\partial_i\right)
 \quad \Leftrightarrow
\quad\upmu=-\frac{\text{d}u}{\gamma}
+ \Delta_i
\left(\text{d}x^i -v^i\text{d}u\right) ,&
\\
\label{upsilontva}
&\hat{\text{e}}_a
%=e_a^{\hphantom{a}i}\upupsilon_i
=e_a^{\hphantom{a}i}\left(\partial_i
+\gamma\Delta_i\left(\partial_u+ v^j\partial_j\right)
\right) \quad \Leftrightarrow\quad\hat{\uptheta}^a
%=e^a_{\hphantom{a}i}\hat{\uptheta}^i
= e^a_{\hphantom{a}i}
\left(\text{d}x^i -v^i\text{d}u\right)&
\end{eqnarray}
with 
\begin{equation}
\Gamma^2_{ij}=\delta_{ab}e^a_{\hphantom{a}i}e^b_{\hphantom{b}j} \quad \Leftrightarrow \quad
\delta_{ab}=
e_a^{\hphantom{a}i}e_b^{\hphantom{b}j}\Gamma^2_{ij}
\end{equation}
and
\begin{equation}
e^a_{\hphantom{a}i}e_a^{\hphantom{a}j}
=\delta_i^j, \quad
e^b_{\hphantom{b}j} e_a^{\hphantom{a}j}
=\delta_a^b, \quad
\delta^{ab}e_a^{\hphantom{a}i} \Gamma^2_{ij} = e^b_{\hphantom{b}j} , \quad
\delta_{ab}e^a_{\hphantom{a}i} \Gamma^{2ij}=e_b^{\hphantom{b}j},
\end{equation}
where $(\Gamma^{2)ik}\Gamma^2_{kj} =\delta_j^i $. 
Consequently, the degenerate metric assumes the form\footnote{The degenerate metric is often spelled $\text{d}\ell^2=q_{\mu\nu}\text{d}x^\mu\text{d}x^\nu $ in the Carrollian literature, and $\text{n}=n^\mu\partial_\mu$ stands for the field of observers. }
\begin{equation}
\label{carmetexpl}
\text{d}\ell^2
=
\Gamma^2_{ij}
\left(\text{d}x^i -v^i\text{d}u\right)\left(\text{d}x^j -v^j\text{d}u\right) .
\end{equation}
In this specific parameterization,  which generalizes that of \cite{CMPR} in arbitrary dimension, the bulk Newman--Unti gauge is recovered by setting $\Delta_i=0$ in the boundary frame.\footnote{The presence of $\gamma \equiv \exp( -2 \beta_0)$, which persists in the bulk line element as $-2  \exp (2 \beta_0) \text{d}u \text{d} r$, assesses a slight redefinition of the radial coordinate before reaching stricto sensu Newman--Unti gauge. We are cavalier with this detail because the counting from the point of view of the solution space matches: the contribution of the boundary geometry is $\nicefrac{(d+1)(d+2)}{2}$. The same holds for the anti-de Sitter ascendant. \label{NUvssNU}}

Carrollian tensors have commonly spacetime indices. In the Cartan frame \eqref{carmet}, \eqref{carfrcofr}, their tensorial behaviour refers to the local Carroll group, as much as relativistic tensors in an orthonormal Cartan frame are tamed according to the local Lorentz group. Here, the metric being degenerate the spacetime indices cannot be lowered or raised. One way to manage this inconvenience is by introducing a pseudo-inverse \cite{Henneaux:1979vn}. Our strategy has been slightly different, and is hinged on \emph{separating time and space,} since this is natural in Carrollian manifolds due to the fibre structure. In the frame at hand,  the method boils down to considering tensors with solely spatial indices, organized in representations of the $d$-dimensional orthogonal local group, subgroup of the local Carroll group, and raised or lowered with $\delta^{ab}$ or $\delta_{ab}$. The fibre null-time direction supports scalars without indices.\footnote{When working in natural frames, as in Refs. \cite{CMPPS2, Campoleoni:2018ltl,
MOPRS, 
Mittal:2022ywl, BigFluid, CMPPS1, Rivera-Betancour:2022lkc}, the tensor structure is based instead on diffeomorphisms. 
The  time/space splitting sought for is realized in Papapetrou--Randers frame, i.e. setting $v^i = 0 $ in the formulas
\eqref{mutv}, 
\eqref{upsilontva}, 
\eqref{carmetexpl}, because this frame is stable under the Carrollian subset of diffeomorphisms, consisting of transformations $u \to u'(u,\mathbf{x})$ and $\mathbf{x} \to \mathbf{x}'(\mathbf{x})$. Carrollian tensors have again spatial indices and transform with the Jacobian matrices of  Carrollian diffeomorphisms.} This approach is in line with the boundary reconstruction of Ricci-flat spacetimes, where the longitudinal/transverse decomposition of the fundamental tensors coincides with the time/space reduction of the Carrollian tensors.

A strong Carroll structure comes with a metric-compatible and field-of-observers-compatible connection, which is not unique due to the metric degeneracy. The connection we use defines a parallel transport that respects the time/space splitting mentioned above, embracing distinct time and 
space Carrollian covariant derivatives $\hat{\nabla}_\upupsilon$ acting as a scalar 
and
$\hat \nabla_a$ acting as a form. We set for this purpose
\begin{equation}
\label{defcovder}
\hat{\nabla}_{\upupsilon}\upupsilon=0,
\quad
\hat{\nabla}_{\upupsilon}\hat{\text{e}}_a=\hat{\gamma}_{[ab]}\delta^{bc}\hat{\text{e}}_c,
\quad
\hat{\nabla}_{\hat{\text{e}}_a} \upupsilon=0,
\quad
\hat{\nabla}_{\hat{\text{e}}_a}\hat{\text{e}}_b=\hat\gamma^c_{ab}\hat{\text{e}}_c,
\end{equation}
from which we infer the resulting Carrollian affine connection one-form:\footnote{Remember that $\upomega^A_{\hphantom{A}B}= \Gamma^{A}_{C B}\uptheta^C$ with $\nabla_{\text{e}_A}\text{e}_B=\Gamma^C_{AB}\text{e}_C$.
The torsion and curvature two-forms are $\mathcal{T}^C=\text{d} \uptheta^C+\upomega^C_{\hphantom{C}A}\wedge  \uptheta^A=\frac{1}{2}  T^C_{\hphantom{C}AB}\uptheta^A\wedge\uptheta^B$ and $\mathcal{R}^A_{\hphantom{A}B}=\text{d}\upomega^A_{\hphantom{A}B}+\upomega^A_{\hphantom{A}C}\wedge \upomega^C_{\hphantom{C}B}=\frac{1}{2}  R^A_{\hphantom{A}BCD}\uptheta^C\wedge\uptheta^D$.
Torsion and curvature tensors can alternatively be determined using the commutator of covariant derivatives: 
$\left[\nabla_A,\nabla_B \right]W^C=R^C_{\hphantom{C}DAB}W^D-T^D_{\hphantom{D}AB}\nabla_DW^C$.
\label{cartan}
}
\begin{equation}
\label{spinco-Car}
\hat \upomega^{\hat{u}}_{\hphantom{\hat{u}}\hat{u}}=\hat \upomega^{\hat{u}}_{\hphantom{\hat{u}}b}=\hat \upomega^{a}_{\hphantom{\hat{u}}\hat{u}}=0,
\quad
\hat \upomega^{a}_{\hphantom{a}b}=
\delta^{ac}\hat{\gamma}_{[cb]} \upmu
+\hat\gamma_{cb}^a\hat{\uptheta}^c
\end{equation}
At this stage $\hat{\gamma}_{[ab]} $ and $\hat\gamma_{cb}^a$ are arbitrary, although anticipating the next step (metric compatibility), we have imposed antisymmetry for the former. 

The covariant time and space derivatives act on Carrollian scalars as time and space directional derivatives. For Carrollian vectors $\upzeta=\zeta^a\hat{\text{e}}_a$ and forms $\upzeta=\zeta_a\hat{\uptheta}^a$ we obtain:
\begin{eqnarray}
\hat{\nabla}_a \zeta^b=\hat{\text{e}}_a\left(\zeta^b\right)
+\hat{\gamma}^{b}_{a c}\zeta^c &\Leftrightarrow&
\hat{\nabla}_a \zeta_b=\hat{\text{e}}_a\left(\zeta_b\right)
-\hat{\gamma}^{c}_{ab}\zeta_c, \\
\hat{\nabla}_\upupsilon \zeta^a=\upupsilon\left(\zeta^a\right) -\hat{\gamma}^{[ab]}\zeta_b
&\Leftrightarrow&
\hat{\nabla}_\upupsilon \zeta_a=\upupsilon\left(\zeta_a\right) -\hat{\gamma}_{[ab]}\zeta^b.
\end{eqnarray}
Under a frame transformation, $\hat{\gamma}_{[ab]} $ and $\hat\gamma_{cb}^a$ transform as connection coefficients , i.e. with inhomogeneous terms.

Field-of-observers-compatibility is built in \eqref{defcovder}. Metric-compatibility translates in $\hat\upomega_{(ab)}=0$. This imposes
\begin{equation}
\label{metric-com}
\hat\gamma_{(a\vert c \vert b)}=0,
\end{equation}
where the symmetrization acts on the two extreme indices. The latter can be utterly determined by further imposing the absence of torsion in the spatial section, $T^c_{\hphantom{c}ab}=0$. In order to implement this we can use the following parameterization of the $\text{d}\hat{\uptheta}^A$s:
\begin{equation}
\label{cardiffr}
\text{d}{\upmu}-\varphi_{a}
\hat{\uptheta}^{a}
\wedge{\upmu}
-
\varpi_{a b}\hat{\uptheta}^{a}
\wedge\hat{\uptheta}^{b}=0,
\quad \text{d}\hat{\uptheta}^{c}+\hat{\gamma}^{c}_{\hphantom{c}a}
{\upmu}
\wedge\hat{\uptheta}^{a}
+
\frac{1}{2}\hat c^c_{\hphantom{c}ab}\hat{\uptheta}^{a}
\wedge\hat{\uptheta}^{b}=0,
\end{equation}
or equivalently 
\begin{equation}
\label{carcom}
\left[\upupsilon,\hat{\text{e}}_{a}
\right]=
\varphi_{a}\upupsilon- 
\hat{\gamma}^{c}_{\hphantom{c} a}\hat{\text{e}}_{c},
\quad
\left[\hat{\text{e}}_{a},\hat{\text{e}}_{b}
\right]=2\varpi_{a b}\upupsilon+ 
\hat c^{c}_{\hphantom{0}a b}\hat{\text{e}}_{c}.
\end{equation}
We have again foreseen the following action by introducing $\hat{\gamma}_{ab}$ whose antisymmetric part already appears in the affine connection one-form. Hence, the extra condition of the absence of torsion in the spatial section combined with \eqref{metric-com} delivers 
\begin{equation}
\label{concarframe}
\hat{\gamma}^a_{bc}=\frac{1}{2}\left(
\hat c^{a}_{\hphantom{a}b c}
+
\hat c^{\hphantom{b}a}_{b\hphantom{a}c}
+
\hat c^{\hphantom{b}a}_{c\hphantom{a}b}
\right).
\end{equation}
Let us also point out  the useful  integrability conditions $\text{d}^2\upmu=\text{d}^2\hat\uptheta^a=0$ associated with \eqref{cardiffr}:
\begin{equation}
\label{intconmu}
\begin{cases}
\hat\nabla_{[c} \varpi_{ab]}=\varphi_{[c} \varpi_{ab]}\\
\hat\nabla_\upupsilon\varpi_{ab} +\varpi_{a}^{\hphantom{a}c}\hat\gamma_{(cb)}-\varpi_{b}^{\hphantom{b}c}\hat\gamma_{(ca)}=\hat\nabla_{[a} \varphi_{b]}
\end{cases}
\end{equation}
and
\begin{equation}
\label{intconth}
\begin{cases}
\upupsilon\left(
\hat c^{a}_{\hphantom{a}b c} \right)
-\hat\gamma^a_{\hphantom{a}d}\hat c^{d}_{\hphantom{d}b c} 
-2\hat c^{a}_{\hphantom{a}d[b}  \hat\gamma^d_{\hphantom{d}c]}
+2\hat{\text{e}}_{[b}^{\vphantom{a}}\left(
\hat\gamma^a_{\hphantom{a}c]}
\right)
-2  \hat\gamma^a_{\hphantom{a}[b} \varphi_{c]}^{\vphantom{a}}=0
\\
\hat{\text{e}}_{[d}^{\vphantom{a}}\left(
\hat c^a_{\hphantom{a}bc]}
\right)
-\hat c^a_{\hphantom{a}e[b}
\hat c^e_{\hphantom{e}cd]}
+2 \gamma^a_{\hphantom{a}[b} \varpi_{cd]}^{\vphantom{a}}\
=0.
\end{cases}
\end{equation}

In summary, our strong Carroll connection is totally determined thanks to the information stored inside the second of Eqs. \eqref{cardiffr}, by requiring the time-and-space splitting and the absence of spatial torsion. 
The total torsion is non-zero though and we find:
\begin{equation}
\label{tor-tf-car}
\hat{\mathcal{T}}^{\hat u}=\varphi_a
\upmu
\wedge
\hat{\uptheta}^a-\varpi_{ab}
\hat{\uptheta}^a
\wedge
\hat{\uptheta}^b
,\quad\hat{\mathcal{T}}^a= \delta^{ab}\hat{\gamma}_{(bc)} 
\hat{\uptheta}^c\wedge\upmu.
\end{equation}
The torsion is thus encoded in three Carrollian tensors (i.e. transforming homogeneously), featuring three properties of the null-time fibre materialized in $\upupsilon$: the \emph{acceleration} $\varphi_a$, the \emph{vorticity} $\varpi_{ab}$ and the \emph{extrinsic curvature} $\hat{\gamma}_{(ab)} $, which can be further decomposed into the \emph{geometric shear} $\xi_{ab}$ (traceless) and the \emph{expansion} $\theta$:
\begin{equation}
\label{excurvexpshear}
\hat{\gamma}_{(ab)}=\xi_{ab}+\frac{\theta}{d}\delta_{ab}.
\end{equation}

We could consistently set the Carrollian torsion to zero. From the bulk perspective, this would significantly impoverish the range of options the covariant Newman--Unti gauge offers for Ricci-flat spacetimes, as discussed in Sec.  \ref{brrm}. It is opportune to recall that in the frame-parameterization \eqref{mutv}, \eqref{upsilontva}, ordinary Newman--Unti gauge corresponds  to $\Delta_i=0$. In more intrinsic terms, this amounts to setting 
\begin{equation}
\label{cardiffrNU}
\text{d}{\upmu}=\varphi_{a}
\hat{\uptheta}^{a}
\wedge{\upmu}
\Leftrightarrow
\left[\hat{\text{e}}_{a},\hat{\text{e}}_{b}
\right]=
\hat c^{c}_{\hphantom{0}a b}\hat{\text{e}}_{c}
\end{equation}
i.e.\ to  discarding the vorticity.

We can finally determine the curvature of the Carrollian connection under consideration using Cartan's formula, cf. footnote \ref{cartan}:
\begin{equation}
\label{cur-tf-car}
\hat{\mathcal{R}}^{\hat u}_{\hphantom{\hat u}b}=0 ,\quad\hat{\mathcal{R}}^a_{\hphantom{a}b}= \hat R^a_{\hphantom{a}cb}\upmu\wedge \hat{\uptheta}^c+\frac{1}{2}  \hat R^a_{\hphantom{a}bcd} \hat{\uptheta}^c\wedge  \hat{\uptheta}^d
\end{equation}
with 
\begin{eqnarray}
\label{carriemannaltfr}
\hat R^a_{\hphantom{a}bcd} &=& 
\hat{\text{e}}_c\left(\hat\gamma^a_{db}\right)
-\hat{\text{e}}_d\left(\hat\gamma^a_{cb}\right)
+\hat\gamma^e_{db}\hat\gamma^a_{ce}
-\hat\gamma^e_{cb}\hat\gamma^a_{de}
-\hat c^e_{\hphantom{e}cd}\hat\gamma^a_{eb}
+
2\varpi_{cd}
\hat\gamma_{[eb]}
\delta^{ae},\\
 \hat R^a_{\hphantom{a}cb}&=&
 \left(\hat\nabla^a+\varphi^a\right)\hat\gamma_{(bc)}
 -\left(\hat\nabla_b+\varphi_b\right)\hat\gamma_{(cd)}\delta^{ad} 
 .
\label{carriemannaltfr0}
 \end{eqnarray}
One can trace the above and yield the Carroll-Ricci tensor and the Carroll scalar curvature:
\begin{equation}
\label{carricci-scalar}
\hat{R}_{cd}=\hat{R}^a_{\hphantom{a}cad},\quad \hat{R}=\delta^{cd}\hat{R}_{cd}.
\end{equation}

Let us stress anew that the freedom in designing a Carrollian connection is rather wide --- see \cite{Duval:2014uoa, Bekaert:2014bwa, Bekaert:2015xua, Ciambelli:2019lap} or \cite{Herfray:2021qmp,Bergshoeff:2022eog,Adrien-solo-23} for a review --- even when conditions like Levi--Civita are imposed, which we haven't. Our guideline has been to ensure that all information ultimately stored in the Carrollian frame, connection, torsion and curvature coincides with that of the relativistic, pseudo-Riemannian ascendant, as we will shortly see: $\varphi_a$, $\varpi_{a b}$, $\hat{\gamma}_{ab} $ and $\hat c^{c}_{\hphantom{0}a b}$.

As a final comment, we would like to mention that Carrollian geometries may have isometries and in particular conformal isometries. The latter play a central role when considering the null conformal boundary, as they mirror bulk asymptotic symmetries.  A vector field $\upxi=\xi^{\hat u}\upupsilon+ \xi^a \hat{\text{e}}_a$ is a Carrollian Killing if the Lie derivative of the degenerate metric and of the field of observers vanishes. This requirement generates three conditions:
\begin{equation}
\label{carkil}
\begin{cases}
\hat\nabla_{(a} \xi_{b)}+ \xi^{\hat u} \hat \gamma_{(ab)}=0\\
\upupsilon\left(\xi^{\hat u}\right)  +\xi^a\varphi_a=0\\
\hat\nabla_\upupsilon \xi_a-\hat \gamma_{(ab)} \xi^b=0.
\end{cases}
\end{equation}
In Papapetrou--Randers frame where $\upupsilon =\frac{1}{\Omega}\partial_u$ and the degenerate metric has no time legs, the last condition selects the Carrollian diffeomorphisms, $\partial_u \xi^i=0$. In the Cartan frame at hand all diffeomorphisms are permitted; the Killing fields are nonetheless further constrained. As usual, strong Killing fields must also leave the clock form invariant, which implies
\begin{equation}
\label{carkilstrong}
\hat{\text{e}}_a \left(\xi^{\hat u}\right)  -\varphi_a \xi^{\hat u}+2 \varpi_{ab}\xi^b=0
.
\end{equation}
Bulk Killing fields of Ricci-flat spacetimes are mapped onto  strong Killings of their null boundary 
\cite{Mittal:2022ywl}.

\subsubsection*{Weyl covariance}

Following the pattern adopted for the affine connection, we introduce here a Weyl connection that respects the time and space splitting, associated with two Weyl-covariant derivatives. These act on weight-$w$ Carrollian tensors and deliver Carrollian tensors of weight $w+1$.\footnote{As already mentioned in footnote \ref{weylcartan}, when working in a Cartan frame the Weyl properties are slightly modified and there is no contradiction with the results displayed in Refs. \cite{CMPPS2, Campoleoni:2018ltl,
MOPRS, 
Mittal:2022ywl, BigFluid, CMPPS1, Rivera-Betancour:2022lkc}, where a Papapetrou--Randers frame was in use.} The Weyl connection is encoded in $\theta$ and $\varphi_a$, see \eqref{cardiffr} and \eqref{excurvexpshear}, and the  Weyl-covariant derivatives are defined as follows:
\begin{itemize}
\item on scalars
\begin{equation}
\hat{\mathscr{D}}_\upupsilon \Phi=\upupsilon (\Phi)+\frac{w}{d}\theta\Phi
, \quad 
\hat{\mathscr{D}}_a \Phi=\hat{\text{e}}_a(\Phi)+w\varphi_a \Phi
;
\end{equation}

\item on vectors $\text{v}=v^a\hat{\text{e}}_a$
\begin{equation}
\label{CWs-vecfr}
\hat{\mathscr{D}}_\upupsilon v^a=\hat{\nabla}_\upupsilon v^a+\frac{w}{d}\theta v^a, \quad \hat{\mathscr{D}}_av^b=\hat\nabla_a v^b +w \varphi_a v^b +\varphi^b v_a -\delta^b_a v^c\varphi_c;
\end{equation}
\item on rank-$2$ tensors $\text{t}=t_{ab}\hat{\uptheta}^a\otimes \hat{\uptheta}^b$:
\begin{eqnarray}
\label{CWs-tens}
\hat{\mathscr{D}}_ct_{ab}&=&\hat\nabla_c t_{ab} +w \varphi_ct_{ab}+\varphi_at_{cb}+\varphi_bt_{ac} -\delta_{ac} t_{db} \varphi^d-\delta_{cb} t_{ad} \varphi^d,
\\
\hat{\mathscr{D}}_\upupsilon t_{ab}&=&\hat{\nabla}_\upupsilon t_{ab}+\frac{w}{d}\theta t_{ab}.
\end{eqnarray}

\end{itemize}
Using Leibniz' rule one obtains the generalization for any conformal tensor.

The Riemann--Carroll--Weyl curvature is a weight-$2$ tensor defined through the commutator of the Carrollian Weyl derivatives acting on Carrollian scalars $\Phi$, vectors  $v^c$ or $2$-tensors $t^{cd}$ of weight $w$:\footnote{The use of $\hat{\mathscr{S}}$ 
is unconventional for a curvature, but is intended to avoid confusion with a slightly different definition given as  $\hat{\mathscr{R}}$  in \cite{CMPPS2, Mittal:2022ywl, BigFluid, CMPPS1}.}
\begin{eqnarray}
\label{carphi}
\left[\hat{\mathscr{D}}_a,\hat{\mathscr{D}}_b\right]\Phi&=&
2
\varpi_{ab}\hat{\mathscr{D}}_\upupsilon \Phi
+w \Omega_{ab}
\Phi,
\\
\label{CWcurvten}
\left[\hat{\mathscr{D}}_a,\hat{\mathscr{D}}_b\right]v^c&=&
\hat{\mathscr{S}}^c_{\hphantom{c}dab} 
v^d+2
\varpi_{ab}\hat{\mathscr{D}}_\upupsilon v^c
+
w \Omega_{ab} v^c
,
\\
\left[\hat{\mathscr{D}}_a,\hat{\mathscr{D}}_b\right]t^{cd}
&=&
\hat{\mathscr{S}}^c_{\hphantom{c}eab}t^{ed}+\hat{\mathscr{S}}^d_{\hphantom{d}eab}t^{ce}+2
\varpi_{ab}\hat{\mathscr{D}}_\upupsilon t^{cd}
+w \Omega_{ab}t^{cd},
\label{CWcurvtenT2}
\end{eqnarray}
where 
\begin{equation}
\label{CWOme}
 \Omega_{ab} =
 %\varphi_{ab} 
  \hat{\text{e}}_a \left(\varphi_b\right) - \hat{\text{e}}_b\left(\varphi_a\right)
 -\hat c^c_{\hphantom{c}ab}\varphi_c
 -\frac{2}{d}\varpi_{ab} \theta
\end{equation}
is yet another weight-$2$ Carrollian tensor. 
From the Riemann--Weyl--Carroll tensor, we define
\begin{equation}
\label{CWricci-scalar}
\hat{\mathscr{S}}_{cd}=\hat{\mathscr{S}}^a_{\hphantom{a}cad},\quad \hat{\mathscr{R}}=\delta^{cd}\hat{\mathscr{S}}_{cd},
\end{equation}
all weight-$2$.

We can further consider time and space derivatives:
\begin{eqnarray}
\left[\hat{\mathscr{D}}_{\upupsilon},\hat{\mathscr{D}}_a\right]\Phi&=& -
\xi^{b}_{\hphantom{b}a}\hat{\mathscr{D}}_b \Phi+w \hat{\mathscr{R}}_{a}\Phi,
\label{CWrsc}
\\
\left[\hat{\mathscr{D}}_{\upupsilon},\hat{\mathscr{D}}_a\right]v^b&=&
- \hat{\mathscr{S}}_{\hphantom{b}ac}^{b} v^c 
-\xi^{c}_{\hphantom{b}a}\hat{\mathscr{D}}_cv^b
+w\hat{\mathscr{R}}_{a} v^b
,
\label{CWrvecfull}
\\
\left[\hat{\mathscr{D}}_{\upupsilon},\hat{\mathscr{D}}_a\right]t^{bc}&=&
-\hat{\mathscr{S}}_{\hphantom{b}ad}^{b} t^{dc}-\hat{\mathscr{S}}_{\hphantom{c}ad}^{c} t^{bd} 
-\xi^{d}_{\hphantom{d}a}\hat{\mathscr{D}}_dt^{bc}+w\hat{\mathscr{R}}_{a} t^{b c},
\label{CWrvecfullT2}
\end{eqnarray}
revealing a clear pattern for any Carrollian conformal tensor. In these expressions 
\begin{equation}
\label{CWriem3in}
\hat{\mathscr{S}}^{c}_{\hphantom{c}ab}=-\hat{\mathscr{S}}_{ba}^{\hphantom{ba}c}=\hat{\mathscr{D}}^c \xi_{ab}
- \hat{\mathscr{D}}_b \xi^{c}_{\hphantom{c}a}+\delta^c_a \hat{\mathscr{R}}_b-\delta_{ab}  \hat{\mathscr{R}}^c
\end{equation}
and $\hat{\mathscr{R}}_a$ are weight-two tensors. Note that in Cartan frame, both the shear $\xi_{ab}$ and the vorticity $\varpi_{ab}$ have weight one, regardless of the position of the indices. In natural frame $\xi_{ij}$ and $\varpi_{ij}$ have weight $-1$, but raising an index augments the weight by two units.

\subsubsection*{Relation with a relativistic ascendant}

A Carrollian manifold as described earlier can be reached from a pseudo-Riemannian geometry at zero velocity of light $k$. Following
the pattern proposed in Eqs. \eqref{limudc}, 
\eqref{carmetlim} and 
\eqref{limuvecc}, 
we can express the metric  \eqref{metgenuortho} of the pseudo-Riemannian ascendant as 
\begin{equation}
\label{metgenuorthocar}
\text{d}s^2=\eta_{A B}\uptheta^{A}\uptheta^{B}
=-\left(\uptheta^{\hat 0}\right)^2 + \delta_{a b}\uptheta^{a}\uptheta^{b}
=-k^2\left(\hat\uptheta^{\hat u}\right)^2 + \delta_{a b}\hat\uptheta^{a}\hat\uptheta^{b},
\end{equation}
where we have assumed that all $k$-dependence is explicit i.e. $\uptheta^{a}=\hat \uptheta^{a}$ while $\uptheta^{\hat 0}=k\hat\uptheta^{\hat u}$. The relationship among the relativistic congruence \eqref{veloform} and the Carrollian fibre attributes, field of observers and clock form, is $\upupsilon = \text{u}=\hat{\text{e}}_{\hat u}$ for the former and $ \upmu=\frac{\text{u}}{k^2}=-\hat\uptheta^{\hat u}$ for the latter. 

If the Carrollian frame, coframe and degenerate metric are parameterized as in Eqs. \eqref{mutv}, 
\eqref{upsilontva} and
\eqref{carmetexpl}, then 
\begin{eqnarray}
\label{mutvrel}
&\text{e}_{\hat 0}=\frac{\gamma}{k}\left(\partial_u+ v^i\partial_i\right)
 \quad \Leftrightarrow
\quad\uptheta^{\hat 0}=k\left(\frac{\text{d}u}{\gamma}
- \Delta_i
\left(\text{d}x^i -v^i\text{d}u\right)\right) ,&
\\
\label{upsilontvarel}
&\text{e}_a
=e_a^{\hphantom{a}i}\left(\partial_i
+\gamma\Delta_i\left(\partial_u+ v^j\partial_j\right)
\right) \quad \Leftrightarrow\quad\uptheta^a
= e^a_{\hphantom{a}i}
\left(\text{d}x^i -v^i\text{d}u\right)&
\end{eqnarray}
and the relativistic metric reads:
\begin{equation}
\label{metgenu}
\begin{array}{rcl}
\displaystyle{\text{d}s^2}&=&\displaystyle{-k^2 \left(\frac{\text{d}u}{\gamma}
-\Delta_i
\left(\text{d}x^i -v^i\text{d}u\right) 
\right)^2
+
\Gamma^2_{ij}\left(\text{d}x^i -v^i\text{d}u\right) \left(\text{d}x^j-v^j\text{d}u\right)} , \crbig
\displaystyle{}&=&\displaystyle{ -\frac{k^2}{\gamma^2}\left(\text{d}u^2-2\gamma \Delta_i \text{d}u\left(
\text{d}x^i-v^i \text{d}u
\right)
\right)
+
\left(\Gamma^2_{ij}-k^2 \Delta_i \Delta_j \right)\left(\text{d}x^i -v^i\text{d}u\right) \left(\text{d}x^j-v^j\text{d}u\right)} ,
\end{array}
\end{equation}
where the normalized vector congruence is
\begin{equation}
\label{utv}
\text{u}=\gamma\left(\partial_u+ v^i\partial_i\right).
\end{equation}
We will not explicitly operate with this frame, which coincides at $v^i=0$ with the Papapetrou--Randers form employed in Refs. \cite{CMPPS2, Campoleoni:2018ltl,
MOPRS, 
Mittal:2022ywl, BigFluid, CMPPS1}, where $\Omega = \nicefrac{1}{\gamma}$, $b_i=\Delta_i$ and $a_{ij}=\Gamma^2_{ij} $. 

At $\Delta_i=0$, one recovers the boundary frame of bulk Newman--Unti anti-de Sitter gauge (modulo a remark stated in footnote \ref{NUvssNU} and valid here), and 
\begin{equation}
\label{reldiffrNU}
\text{d}{\uptheta}^{\hat 0}=\varphi_{a} {\uptheta}^{a} \wedge{\uptheta}^{\hat 0},
\end{equation}
which resonates with the Carrollian relative \eqref{cardiffrNU}.
Hence the boundary vorticity vanishes following Eq. \eqref{reldiffr} below. 

The pseudo-Riemannian manifold is equipped with a Levi--Civita connection. We would like to express the latter in terms of the Carrollian tensors appearing in Eqs. \eqref{spinco-Car} and \eqref{cardiffr} or \eqref{carcom}. The purpose of this exercise is to provide the suitable tools for reaching the $k\to 0$ limit in relativistic dynamical equations such as \eqref{T-cons}.  We reckon that in the parameterization of $\left\{\text{d}\uptheta^A\right\}=\left\{\text{d}\uptheta^{\hat 0}, \text{d}\uptheta^a\right\}$, Eqs. \eqref{cardiffr} and 
\eqref{carcom}, hold:
\begin{equation}
\label{reldiffr}
\text{d}{\uptheta}^{\hat 0}-\varphi_{a}
{\uptheta}^{a}
\wedge{\uptheta}^{\hat 0}
+k
\varpi_{a b}{\uptheta}^{a}
\wedge{\uptheta}^{b}=0,
\quad \text{d}{\uptheta}^{c}+\frac{1}{k}\hat{\gamma}^{c}_{\hphantom{c}a}
{\uptheta}^{a}
\wedge
{\uptheta}^{\hat 0}
+
\frac{1}{2}\hat c^c_{\hphantom{c}ab}{\uptheta}^{a}
\wedge{\uptheta}^{b}=0.
\end{equation}
Thus the Levi--Civita affine connection one-form reads:
\begin{equation}
\label{spinco-rel-ab}
\begin{array}{rcl}
\displaystyle{\upomega_{ab}}&=&\displaystyle{-\left(k \varpi_{ab}+\frac{1}{k}\hat\gamma_{[ab]}\right) \uptheta^{\hat 0}+\delta_{ad}\hat\gamma_{cb}^d\hat\uptheta^c} \crbig
&=&\displaystyle{\left(k^2 \varpi_{ab}+\hat\gamma_{[ab]}\right)\upmu+\delta_{ad}\hat\gamma_{cb}^d\hat\uptheta^c}  
 \crbig
&=&\displaystyle{k^2 \varpi_{ab}\upmu}+\hat{\upomega}_{ab},
\end{array}
\end{equation}
and
\begin{equation}
\label{pinco-rel-0a}
\upomega^{\hat{0}}_{\hphantom{\hat{0}}a}=\displaystyle{\varphi_a \uptheta^{\hat 0}-k\varpi_{ab}\uptheta^b+\frac{1}{k}\hat\gamma_{(ab)}\uptheta^b}  
=-k\left(\varphi_a \upmu+\varpi_{ab}\hat\uptheta^b\right)+\frac{1}{k}\hat\gamma_{(ab)}\hat\uptheta^b
\end{equation}
with $\hat\gamma_{ab}^c$ as in \eqref{concarframe}. It has zero torsion and the curvature reads:
\begin{eqnarray}
\label{cur-tf-riem0a}
\mathcal{R}^{\hat 0}_{\hphantom{\hat 0}a}&=&
\left[\frac{1}{k}\left(
\hat{\nabla}_{\upupsilon}\hat{\gamma}_{(ab)}+\hat{\gamma}_{(ac)}\hat{\gamma}_{(bd)}\delta^{cd}
\right)-k\left(
\varpi_a^{\hphantom{a}c}\hat\gamma_{(cb)}
+
\varpi_b^{\hphantom{b}c}\hat\gamma_{(ca)}
+\hat\nabla_{(a} \varphi_{b)}
+\varphi_a \varphi_b
\right)\right.
\nonumber
\\
\nonumber
&&+k^3\varpi_a^{\hphantom{a}c}\varpi_{bc}^{\vphantom{c}}\bigg] 
\hat{\uptheta}^b\wedge\upmu +\frac{1}{2}\left[\frac{1}{k} \left(\hat R_{bac}-\varphi_b \hat\gamma_{(ac)}+\varphi_c \hat\gamma_{(ab)}\right)\right.
\nonumber
\\
&&
-k\left( \hat\nabla_a \varpi_{bc} +\varphi_a \varpi_{bc} 
+\varphi_b \varpi_{ac} 
-\varphi_c \varpi_{ab} 
\right)
\bigg]\hat{\uptheta}^b\wedge \hat{\uptheta}^c,
\\
 \mathcal{R}^a_{\hphantom{a}b}&=&
  \hat{\mathcal{R}}^a_{\hphantom{a}b}
 +\delta^{ad}\left[\varphi_d \hat\gamma_{(cb)}-\varphi_b \hat\gamma_{(cd)} + k^2\left(
\hat\nabla_c \varpi_{db} +\varphi_c \varpi_{db} 
+\varphi_d \varpi_{cb} 
-\varphi_b \varpi_{cd} 
\right)\right]
\hat{\uptheta}^c\wedge\upmu \nonumber
\\
&&+\frac{1}{2}\delta^{ae}\left[\frac{1}{k^2}
\right.
\left(\gamma_{(ec)}\gamma_{(bd)}-\gamma_{(ed)}\gamma_{(bc)}\right)-\gamma_{(ec)}\varpi_{bd}+\gamma_{(ed)} \varpi_{bc}
\nonumber
\\
&&-\gamma_{(bd)}\varpi_{ec}+
\gamma_{(bc)}\varpi_{ed}
 +k^2\left(
 2 \varpi_{eb}
 \varpi_{cd}
-\varpi_{ed}
 \varpi_{bc}+\varpi_{ec}
 \varpi_{bd}
 \right)
\bigg]\hat{\uptheta}^c\wedge \hat{\uptheta}^d,
\label{cur-tf-riemab}
\end{eqnarray}
where we have used the Carrollian expressions available in \eqref{cur-tf-car}, \eqref{carriemannaltfr} and \eqref{carriemannaltfr0}.

We would like now to make the contact with the Carrollian descendants. The relativistic congruence is  $\text{u}=-k\uptheta^{\hat 0}$ see \eqref{veloform}. Given the connection , we can determine its kinematical properties: the expansion $\Theta$, the acceleration $a_A$, the shear $\sigma_{AB}$ and the vorticity $\omega_{AB}$ as defined in Eqs. \eqref{def21},  \eqref{def23}, \eqref{def24}. The latter tensors are all transverse (and traceless for the last two) and have thus non-vanishing components  in spatial directions only (indices $a,b, \ldots$). We find
\begin{equation}
\label{carlim2}
\Theta = \theta=\hat{\gamma}^{c}_{\hphantom{c}c},\quad 
a_a=k^2 \varphi_a,
\end{equation}
and 
\begin{equation}
\label{carlim3}
\quad 
\sigma_{ab}=\xi_{ab}=\hat{\gamma}_{(ab)}-\frac{\theta}{d}\delta_{ab},
\quad
\omega_{ab}=k^2\varpi_{ab}.
\end{equation}
We can furthermore determine the Weyl connection \eqref{Wconc} (where we must trade the $2$ for $d$)
\begin{equation}
\label{Weyl-con-gen-car}
\text{A}=\varphi_a \hat{\uptheta}^a-\frac{\theta}{d}\upmu,
\end{equation}
and its curvature \eqref{F}:
\begin{equation}
\label{Weyl-curv-gen-car}
\text{F}=\text{d}\text{A}=\frac{1}{2} \Omega_{ab}\hat{\uptheta}^a\wedge \hat{\uptheta}^b+
 \hat{\mathscr{R}}_a \hat{\uptheta}^a\wedge\upmu,
\end{equation}
where $ \Omega_{ab}$ and $\hat{\mathscr{R}}_a$ are defined in Eqs. \eqref{CWOme} and \eqref{CWrsc} --- explicitly
\begin{equation}
\hat{\mathscr{R}}_{a}=
\hat\nabla_{\upupsilon}\varphi_a+\xi_{ab}\varphi^b-\frac{1}{d} \hat{\text{e}}_a (\theta).
\label{CWRvec}
\end{equation}
All the above quantities are relativistic, but expressed in terms of the Carrollian descendants describing the properties of the manifold reached at vanishing-$k$.

We can finally convey the relativistic conservation equations \eqref{T-cons} for an arbitrary energy--momentum tensor $T^{AB}$ as in \eqref{T}, stated in Carrollian language. Given the choice of congruence, the transverse heat current and stress tensor have only spatial components: $q^{a}$ and $\tau^{a b}$. We then define as usual the longitudinal and transverse components of the conservation equations,
\begin{equation}
\mathcal{L}=-u^{B}\nabla_{C}^{\vphantom{B}}  T^{C}_{\hphantom{C}B}=-k\nabla_{C}^{\vphantom{\hat 0}} T^{C}_{\hphantom{C}\hat 0}=-\nabla_{C}^{\vphantom{\hat 0}} T^{C}_{\hphantom{C}\hat u}, \quad
\mathcal{T}^{a}=e^{a}_{\hphantom{a}B}\nabla_{C}^{\vphantom{a}}  T^{CB}=\nabla_{C}T^{Ca},
\end{equation}
and explicitly find:
\begin{eqnarray}
\label{newlongrel}
\mathcal{L}&=&\upupsilon (\varepsilon)+\theta \varepsilon+ \left(\hat\nabla_a+2\varphi_a\right)q^a
+\left(
\xi_{ab}+\frac{\theta}{d}\delta_{ab}
\right)\left(\tau^{ab}
+p\delta^{ab}
\right),
\\
\mathcal{T}^{a}&=&  \left(\hat\nabla_b+\varphi_b\right) 
\left(\tau^{ab}
+p\delta^{ab}
\right) +\varphi^a \varepsilon+ 2q_b\varpi^{ba}
+\frac{1}{k^2}\left(
\hat\nabla_{\upupsilon}q^a
+ \frac{d+1}{d} \theta q^a+\xi^{ab}q_b
\right)
.
\label{newtrrela}
\end{eqnarray}
In the conformal case, assuming thus $\varepsilon=dp$ and $\tau_a^{\hphantom{a} a}=0$ and canonical conformal weights $d+1$ for $\varepsilon$, $q^a$ and $\tau^{ab}$ (we are in Cartan' frame and the weights do not depend on the position of the indices), these equations are recast as:
\begin{eqnarray}
\label{newlongrelconf}
\mathcal{L}&=&\hat{\mathscr{D}}_{\upupsilon} \varepsilon+\hat{\mathscr{D}}_a q^a
+
\xi_{ab}\tau^{ab},
\\
\mathcal{T}^{a}&=& \frac{1}{d} \hat{\mathscr{D}}^a \varepsilon+ \hat{\mathscr{D}}_b
\tau^{ab}
 +  2q_b\varpi^{ba}
+\frac{1}{k^2}\left(
 \hat{\mathscr{D}}_{\upupsilon}q^a
+\xi^{ab}q_b
\right)
.
\label{newtrrelaconf}
\end{eqnarray}
As discussed extensively in Refs. \cite{BigFluid, CMPPS1}, the outcome of the Carrollian limit depends on the behaviour of $\varepsilon$, $q^a$ and $\tau^{ab}$ with respect to $k$. The equations at hand will be conceivably multiplied, leading to replicas. The same phenomenon occurs in the Galilean limit with the emergence of the continuity equation out of the relativistic longitudinal equation, besides the energy equation. 

We would like to close this chapter with some formulas that are useful when considering the zero-$k$ limit, leading in particular to the flux-balance equation \eqref{f2fluxeq}. In the following, we reduce the Riemannian Levi--Civita and Weyl covariant derivatives in terms of the Carrollian connections introduced earlier.
\begin{description}
\item[Levi--Civita] We will present the vector and the rank-two tensor:
\begin{description}
\item[\boldmath$V= V^A  \text{e}_A$\unboldmath ] ---  $V^a$ provide the components of a Carrollian vector and $V_{\hat u}=kV_{\hat 0}=-kV^{\hat 0}$  a Carrollian scalar
\begin{equation}
\begin{cases}
k^2 \nabla_{\hat 0}V^{\hat 0}=k\upupsilon\left(V^{\hat 0}\right)+k^2\varphi_a V^a
\\
k \nabla_{\hat 0}V^b=\hat{\nabla}_\upupsilon V^b+kV^{\hat 0} \varphi^b +k^2 V^a\varpi_{a}^{\hphantom{a}b}
\\
k \nabla_aV^{\hat 0}=k\hat{\text{e}}_a\left(V^{\hat 0}\right)+\left(
\xi_{ab}+\frac{\theta}{d}\delta_{ab} + k^2 \varpi_{ab}
\right)V^b
\\
\nabla_aV^b= \hat{\nabla}_aV^b+\frac{1}{k}\left(
\xi_{a}^{\hphantom{a}b}+\frac{\theta}{d}\delta_{a}^{b} + k^2 \varpi_{a}^{\hphantom{a}b}
\right)V^{\hat 0};
\end{cases}
\end{equation}
\item[\boldmath$T= T^{AB} \text{e}_A\otimes \text{e}_B$\unboldmath ]  --- $T^{ab}$ are farther interpreted as components of a Carrollian rank-two tensor, $T_{\hat u}^{\hphantom{\hat u}a}=kT_{\hat 0}^{\hphantom{\hat 0} a}=-kT^{\hat 0  a}$ and $T^a_{\hphantom{a} \hat u}=kT^a_{\hphantom{a}\hat 0}=-kT^{a \hat 0 }$ those of Carrollian vectors, while 
 $T_{\hat u \hat u}=k^2 T_{\hat 0\hat 0}=k^2 T^{\hat 0\hat 0}$  gives a Carrollian scalar
\begin{equation}
\begin{cases}
k^3\nabla_{\hat 0}T^{\hat 0\hat 0}=k^2\upupsilon\left(T^{\hat 0\hat 0}\right)+k^3\varphi_{a}\left(T^{a\hat 0}+T^{\hat 0 a}\right)
\\
k^2\nabla_{\hat 0}T^{b\hat 0}=k\hat{\nabla}_{\upupsilon}T^{b\hat 0}+k^2\varphi^{b}T^{\hat 0\hat 0}+k^2\varphi_{a}T^{ba}+k^3\varpi_{a}^{\hphantom{a}b}T^{a\hat 0}
\\
k\nabla_{\hat 0}T^{ab}=\hat{\nabla}_{\upupsilon}T^{ab}+k\left(\varphi^{a}T^{\hat 0b}+\varphi^{b}T^{a\hat 0}\right)+k^2\left(T^{ac}\varpi_{c}^{\hphantom{c}b}+T^{cb}\varpi_{c}^{\hphantom{c}a}\right)
\\
k\nabla_aT^{b\hat 0}=k\hat{\nabla}_aT^{b\hat 0}+\left(\xi_{ac}+\frac{\theta}{d}\delta_{ac}+k^2\varpi_{ac}\right)T^{bc}+\left(\xi_{a}^{\hphantom{a}b}+\frac{\theta}{d}\delta_{a}^{b}+k^2\varpi_{a}^{\hphantom{a}b}\right)T^{\hat 0\hat 0}
\\
k^2\nabla_{a}T^{\hat 0\hat 0}=k^2\hat{\text{e}}_a\left(T^{\hat 0\hat 0}\right)+k\left(\xi_{ac}+\frac{\theta}{d}\delta_{ac}+k^2\varpi_{ac}\right)T^{c\hat 0}+k\left(\xi_{ac}+\frac{\theta}{d}\delta_{ac}+k^2\varpi_{ac}\right)T^{\hat 0c}
\\
\nabla_{a}T^{bc}=\hat{\nabla}_aT^{bc}+\frac{1}{k}\left(\xi_{a}^{\hphantom{a}b}+\frac{\theta}{d}\delta_{a}^{b}+k^2\varpi_{a}^{\hphantom{a}b}\right)T^{\hat 0c}+\frac{1}{k}\left(\xi_{a}^{\hphantom{a}c}+\frac{\theta}{d}\delta_{a}^{c}+k^2\varpi_{a}^{\hphantom{a}c}\right)T^{b\hat 0};
\end{cases}
\end{equation}
\end{description}

\item[Weyl] similarly:
\begin{description}
\item[\boldmath$V= V^A \text{e}_A$\unboldmath ] 
\begin{equation}
\begin{cases}
k^2 \mathscr{D}_{\hat 0}V^{\hat 0}=k\hat{\mathscr{D}}_\upupsilon V^{\hat 0}
\\
k \mathscr{D}_{\hat 0}V^b=\hat{\mathscr{D}}_\upupsilon V^b+k^2 V^a\varpi_{a}^{\hphantom{a}b}
\\
k \mathscr{D}_aV^{\hat 0}=k\hat{\mathscr{D}}_a V^{\hat 0}+\left(
\xi_{ab}+ k^2 \varpi_{ab}
\right)V^b
\\
\mathscr{D}_aV^b= \hat{\mathscr{D}}_aV^b+\frac{1}{k}\left(
\xi_{a}^{\hphantom{a}b}+ k^2 \varpi_{a}^{\hphantom{a}b}
\right)V^{\hat 0}
;
\end{cases}
\end{equation}
\item[\boldmath$T= T^{AB} \text{e}_A\otimes \text{e}_B$\unboldmath ] 
\begin{equation}
\begin{cases}
k^3\mathscr{D}_{\hat 0}T^{\hat 0\hat 0}=\hat{\mathscr{D}}_{\upupsilon}\left(k^2T^{\hat 0\hat 0}\right)
\\
k^2\mathscr{D}_{\hat 0}T^{\hat 0b}=\hat{\mathscr{D}}_{\upupsilon}\left(kT^{\hat 0b}\right)+k^3T^{\hat 0a}\varpi_{a}^{\hphantom{a}b}
\\
k\mathscr{D}_{\hat 0}T^{ab}=\hat{\mathscr{D}}_{\upupsilon}T^{ab}+k^2\left(T^{cb}\varpi_{c}^{\hphantom{c}a}+T^{ac}\varpi_{c}^{\hphantom{c}b}\right)
\\
k^2\mathscr{D}_{a}T^{\hat 0\hat 0}=\hat{\mathscr{D}}_{a}\left(k^2T^{\hat 0\hat 0}\right)+\left(\xi_{ab}+k^2\varpi_{ab}\right)kT^{b\hat 0}+\left(\xi_{ab}+k^2\varpi_{ab}\right)kT^{\hat 0b}
\\
k\mathscr{D}_{a}T^{\hat 0b}=\hat{\mathscr{D}}_{a}\left(kT^{\hat 0b}\right)+\left(\xi_{ac}+k^2\varpi_{ac}\right)T^{cb}+\left(\xi^{\hphantom{a}b}_{a}+k^2\varpi^{\hphantom{a}b}_a\right)T^{\hat 0\hat 0}
\\
\mathscr{D}_{a}T^{bc}=\hat{\mathscr{D}}_{a}T^{bc}+\frac{1}{k}\left(\xi^{\hphantom{a}b}_{a}+k^2\varpi^{\hphantom{a}b}_a\right)T^{\hat 0c}+\frac{1}{k}\left(\xi^{\hphantom{a}c}_{a}+k^2\varpi^{\hphantom{a}c}_a\right)T^{b\hat 0}.
\end{cases}
\end{equation}
\end{description}

\end{description}

\section{The Carrollian Cotton tensors in three dimensions} \label{carcot3}

The Cotton tensor introduced in Sec. \ref{ESMNU} can be decomposed in terms of Carrollian descendants, which obey Carrollian identities resulting from \eqref{C-cons}. In Papapetrou--Randers' frame and for vanishing $\xi_{ab}$ a thorough exhibition is available in App. C of 
\cite{Mittal:2022ywl}. 
The Carrollian Cotton tensor will be investigated from a more general viewpoint in \cite{MOPRS}. 
Here we will summarize its  properties in Cartan' frame with $\xi_{ab}\neq0$.  Prior to this presentation we need to spend some time on $d=2$.

In three boundary spacetime dimensions, we pointed out that given a congruence $\text{u}$, a transverse Hodge duality can be designed mapping
transverse vectors to transverse vectors and symmetric, traceless and transverse two-tensors onto similar objects, Eqs. \eqref{eta2},  \eqref{eta2contr},  \eqref{eta2dualv},  \eqref{eta2dualw}. This procedure is readily extended to a Carroll structure $\mathscr{M}= \mathbb{R} \times \mathscr{S}$ and the duality coincides with the Hodge duality in the  $2$-dimensional basis $\mathscr{S}$: Carrollian vectors are mapped onto Carrollian vectors and Carrollian symmetric and traceless two-tensors onto the same class. In the relativistic Cartan frame we use here the antisymmetric pseudo-tensor is
$\epsilon_{ABC}$ or $\epsilon^{ABC}$ with $\epsilon_{\hat 0 \hat 1 \hat 2}=-\epsilon^{\hat 0\hat 1\hat 2}=1$, whereas using \eqref{veloform} in \eqref{eta2}
we find $\hat \eta_{AB}= - \epsilon_{\hat 0AB}$ so that only $\hat \eta_{ab}$ is non-zero with $\hat \eta_{\hat 1\hat 2}=-1$. We adopt  this convention for the Carrollian object, without introducing any further symbol. Now \eqref{eta2contr} translates into
\begin{equation}
\label{etaortho}
\hat\eta^{\vphantom{b}}_{ac}\hat\eta_{b}^{\hphantom{b}c}=\delta_{ab}, \quad \hat\eta^{ab}\hat\eta_{ab}=2,
\end{equation}
and  \eqref{eta2dualv},  \eqref{eta2dualw} give
\begin{equation}
\label{eta2dualvwcar}
\ast v^a=\hat\eta^b_{\hphantom{b}a}v_b, \quad  \ast w_{ab}=\hat\eta^c_{\hphantom{c}a}w_{c b},
\end{equation} 
for Carrollian vectors $v^a$ and Carrollian symmetric, traceless  tensors $w_{ab}$. We will often use the following identities , generalizable to any tensor:
\begin{equation}
\label{stup-id}
\ast\! \ast v^a=-v^a, \quad  \ast v^a w_{a}=-v^a \ast\! w_{a}.
\end{equation} 
The Carroll--Riemann, Carroll--Ricci and scalar \eqref{carricci-scalar}  curvatures read:
\begin{equation}
\label{carricci-expand}
\hat R_{abcd}= \hat K \left(\delta_{ac}\delta_{bd}-\delta_{ad}\delta_{bc}\right), \quad \hat R_{ab}=\hat K \delta_{ab}, \quad \hat R= 2 \hat K.
\end{equation}
The Carroll--Weyl--Riemann and Ricci tensors, the Carroll--Weyl--Ricci scalar (see \eqref{CWricci-scalar}) as well as the Carroll--Weyl tensor curvature \eqref{CWOme} are
\begin{equation}
\label{CWricci-dec}
\hat {\mathscr{S}}_{abcd}= \hat{\mathscr{K}} \left(\delta_{ac}\delta_{bd}-\delta_{ad}\delta_{bc}\right), \quad\hat{\mathscr{S}}_{ab}=\hat{\mathscr{K}} \delta_{ab}, \quad \Omega_{ab}= -\hat{\mathscr{A}} \hat\eta_{ab},
\end{equation}
expressed in terms of two weight-$2$ Weyl-covariant scalars: 
\begin{equation}
\label{CWscalar}
\hat{\mathscr{K}}=\hat{K}+ \hat{\nabla}_a \varphi^a
,\quad \hat{\mathscr{A}}=  \ast\varpi\theta- \ast \upvarphi
\end{equation}
with
\begin{equation}
\label{scalar} 
\ast\varpi=\frac{1}{2} \hat\eta^{ab} \varpi_{ab} \Leftrightarrow \varpi_{ab} =\hat\eta_{ab} \ast\varpi,
\end{equation}
and
\begin{equation}
\label{upvarphi} 
\ast\upvarphi=\frac{1}{2} \hat\eta^{ab} \upvarphi_{ab} \quad \text{where} \quad  \upvarphi_{ab} = \hat{\text{e}}_{a}(\varphi_{b}) - \hat{\text{e}}_{b}(\varphi_{a}).
\end{equation}
These obey Carroll--Bianchi identities:
\begin{eqnarray}
 2\hat{\mathscr{D}}_\upupsilon \ast\! \varpi +\hat{\mathscr{A}}&=&0
\label{Carroll-Bianchi1}
,\\
\hat{\mathscr{D}}_\upupsilon \hat{\mathscr{K}}
-\hat{\mathscr{D}}_a  \hat{\mathscr{R}}^a -  \hat{\mathscr{D}}_a \hat{\mathscr{D}}_b \xi^{ab} &=& 0 ,
\label{Carroll-Bianchi2}
\\
\hat{\mathscr{D}}_\upupsilon \hat{\mathscr{A}}+ \hat\eta^{ab}\hat{\mathscr{D}}_a  \hat{\mathscr{R}}_{b} &=&0
\label{Carroll-Bianchi3}
.
\end{eqnarray}

The Carroll reduction of the Cotton tensor is encrypted in the longitudinal, mixed and transverse components \eqref{cottdens} and \eqref{cotransp}, which encompass several weight-$3$ Carrollian scalars, vectors and symmetric, traceless two-tensors, dubbed ``Carrollian Cotton tensors.'' In Cartan' frame we obtain
\begin{eqnarray}
\label{cotsc}
c&=& c_{(-1)}k^2  + c_{(0)} +\frac{c_{(1)} }{k^2} + \frac{c_{(2)} }{k^4},
\\
\label{cotveca}
c^a &=& k^2 \psi^{a}  +\chi^{a}+ \frac{z^{a} }{k^2}, 
\\
\label{cottenab}
c^{ab}&=&k^2 \Psi^{ab} +X^{ab}+  \frac{Z^{ab} }{k^2}
\end{eqnarray}
with
\begin{itemize}
\item four Carroll scalars:
\begin{equation}
\label{cfr}
c_{(-1)}= 8\ast\!\varpi^3,
\quad
c_{(0)} =\left(\hat{\mathscr{D}}_a\hat{\mathscr{D}}^a+2\hat{\mathscr{K}}
\right)\ast\! \varpi,
\quad
c_{(1)}=\hat{\mathscr{D}}_a\hat{\mathscr{D}}_b \ast\! \xi^{ab},
\quad
c_{(2)} = \ast \xi_{ab}\hat{\mathscr{D}}_\upupsilon \xi^{ab};
\end{equation}

\item three Carroll vectors:
\begin{eqnarray}
\label{c7fr}
\psi^{a}&=& 3\hat\eta^{ba}\hat{\mathscr{D}}_b\ast\! \varpi^2
,
\\
\label{c6fr}
\chi^{a}&=& \frac{1}{2}\hat\eta^{ba}\hat{\mathscr{D}}_b\hat{\mathscr{K}}+ \frac{1}{2} \hat{\mathscr{D}}^a\hat{\mathscr{A}}-2\ast \!\varpi\left(\hat{\mathscr{R}}^a + 2 \hat{\mathscr{D}}_b  \xi^{ab}
\right)+3\hat{\mathscr{D}}_b \left(\ast \varpi \xi^{ab}\right),
\\
\label{c5fr}
z^{a} &=& \frac{1}{2}\hat\eta^{ab}\hat{\mathscr{D}}_b \xi^2 - \hat{\mathscr{D}}_b \hat{\mathscr{D}}_\upupsilon \ast\! \xi^{ab} - \ast \xi^{a}_{\hphantom{a}b}\hat{\mathscr{D}}_c \xi^{bc}
,
\end{eqnarray}
where we defined\footnote{Many identities of this sort are useful: 
$\xi^{ac}\ast\!\xi_{c}^{\hphantom{c}b}= \xi^2 \hat\eta^{ab}$, 
$\ast\xi^{ac}\ast\!\xi_{c}^{\hphantom{c}b}= \xi^2 \delta^{ab}$, 
$\varpi^{ac}\varpi_{c}^{\hphantom{c}b}=\ast \varpi^2 \delta^{ab}$.}
\begin{equation}
\label{xisquare}
 \xi^2 =\frac{1}{2}\xi^{ab}\xi_{ab}
\Leftrightarrow
\xi^{ac}\xi_{c}^{\hphantom{c}b}=
 \xi^2 \delta^{ab}
;
\end{equation}

\item three Carroll traceless and symmetric two-index tensors: 
\begin{eqnarray}
\label{c10fr}
\Psi^{ab} &=& -2 \ast \! \varpi^2  \ast \xi^{ab}
+\hat{\mathscr{D}}^a \hat{\mathscr{D}}^b\ast\! \varpi -\frac{1}{2}\delta^{ab} \hat{\mathscr{D}}_c \hat{\mathscr{D}}^c \ast\! \varpi -\hat\eta^{ab}   \hat{\mathscr{D}}_\upupsilon\ast\! \varpi^2,
\\
X^{ab}&=&\frac{1}{2}\hat\eta^{ca}\hat{\mathscr{D}}_c
\left(\hat{\mathscr{R}}^b+ \hat{\mathscr{D}}_d  \xi^{bd}\right)+
\frac{1}{2} \hat\eta^{cb}\hat{\mathscr{D}}^a
\left(\hat{\mathscr{R}}_c+ \hat{\mathscr{D}}^d  \xi_{cd}\right)\nonumber\\
&&-
\frac{3}{2}\hat{\mathscr{A}} \xi^{ab} -\hat{\mathscr{K}}\ast \! \xi^{ab} +3\ast\! \varpi \hat{\mathscr{D}}_\upupsilon \xi^{ab}
,
\label{c9fr}
\\
\label{c8fr}
Z^{ab} &=& 2 \ast\! \xi^{ab} \xi^2 - \hat{\mathscr{D}}_\upupsilon  \hat{\mathscr{D}}_\upupsilon \ast\!\xi^{ab}
.
\end{eqnarray}
\end{itemize}

As for the conservation equations \eqref{C-cons}, expressing them as in \eqref{newlongrelconf}, \eqref{newtrrelaconf}, they yield
\begin{eqnarray}
\label{Lcot}
\mathcal{L}_{\text{Cot}} &=& -k^3 \mathcal{D}_{\text{Cot}}
-k 
\mathcal{E}_{\text{Cot}} 
-\frac{\mathcal{F}_{\text{Cot}} }{k} 
-\frac{\mathcal{W}_{\text{Cot}} }{k^3} =0,
\\
\label{Tcota}
\mathcal{T}^{a}_{\text{Cot}}
&=&k^3 \mathcal{I}_{\text{Cot}}^a  + 
k\mathcal{G}_{\text{Cot}}^a
+\frac{\mathcal{H}_{\text{Cot}}^a }{k} 
+\frac{\mathcal{X}_{\text{Cot}}^a }{k^3}=0 
\end{eqnarray}
with
\begin{eqnarray}
\mathcal{D}_{\text{Cot}} &=&-\hat{\mathscr{D}}_\upupsilon c_{(-1)}-\hat{\mathscr{D}}_a \psi^{a}
,
 \label{carDcot} \\
 \mathcal{E}_{\text{Cot}} &=&-\hat{\mathscr{D}}_\upupsilon c_{(0)}-\hat{\mathscr{D}}_a \chi^{a}
+\Psi_{ab}\xi^{ab},
 \label{carEcot} \\
 \mathcal{F}_{\text{Cot}} &=&-\hat{\mathscr{D}}_\upupsilon c_{(1)}-\hat{\mathscr{D}}_a z^{a}
+X_{ab}\xi^{ab},
 \label{carFcot} \\
 \mathcal{W}_{\text{Cot}} &=&-\hat{\mathscr{D}}_\upupsilon c_{(2)}
+Z_{ab}\xi^{ab}
 \label{carWcot} 
\end{eqnarray}
and 
\begin{eqnarray}
\mathcal{I}_{\text{Cot}}^a &=& \frac{1}{2}\hat{\mathscr{D}}^a c_{(-1)}
+2 \ast\! \varpi   \ast\!\psi^{a}
,
  \label{carIcot}\\
  \mathcal{G}_{\text{Cot}}^a &=& \frac{1}{2}\hat{\mathscr{D}}^a c_{(0)}- \hat{\mathscr{D}}_b \Psi^{ab}
+2 \ast\! \varpi   \ast\!\chi^{a}
+ \hat{\mathscr{D}}_\upupsilon \psi^a +\xi^{ab} \psi_b
,
  \label{carGcot}\\
  \mathcal{H}_{\text{Cot}}^a &=& \frac{1}{2}\hat{\mathscr{D}}^a c_{(1)}- \hat{\mathscr{D}}_b X^{ab}
+2 \ast\! \varpi   \ast\! z^{a}
+ \hat{\mathscr{D}}_\upupsilon \chi^a +\xi^{ab} \chi_b
,
 \label{carHcot} \\
\mathcal{X}_{\text{Cot}}^a &=& \frac{1}{2}\hat{\mathscr{D}}^a c_{(2)}- \hat{\mathscr{D}}_b Z^{ab}
+ \hat{\mathscr{D}}_\upupsilon z^a +\xi^{ab} z_b
.
  \label{carXcot}
  \end{eqnarray}
The four couples of equations $\left\{\mathcal{D}_{\text{Cot}}=0, \mathcal{I}_{\text{Cot}}^a =0\right\}$, 
$\left\{\mathcal{E}_{\text{Cot}}=0, \mathcal{G}_{\text{Cot}}^a =0\right\}$, 
$\left\{\mathcal{F}_{\text{Cot}}=0, \mathcal{H}_{\text{Cot}}^a =0\right\}$
and $\left\{\mathcal{W}_{\text{Cot}}=0, \mathcal{X}_{\text{Cot}}^a =0\right\}$
originate from the different orders in $k$ in which the conservation of the Cotton tensor \eqref{C-cons} decomposes.
These are \emph{purely geometrical identities fulfilled on any three-dimensional Carroll structure $\mathscr{M}= \mathbb{R} \times \mathscr{S}$. } Moreover,
they are typical Carrollian conservation equations obtained as a consequence of general covariance applied to a Weyl-invariant action  
$S=-\frac{1}{2}\int_{\mathscr{M}} \hat\eta_{ab}\hat\uptheta^a \wedge \hat\uptheta^b\wedge \upmu \, \mathscr{L}$
defined on $\mathscr{M}= \mathbb{R} \times \mathscr{S}$:
\begin{eqnarray}
 \hat{\mathscr{D}}_\upupsilon\Pi
+\hat{\mathscr{D}}_a \Pi^{a}
+\Upsilon^{a}_{\hphantom{a}b}\xi_{a}^{\hphantom{a}b}&=& 0 ,
  \label{carEbiscon} 
  \\
\frac{1}{2}\hat{\mathscr{D}}_a \Pi+\hat{\mathscr{D}}_b \Upsilon^{b}_{\hphantom{b}a}+2\ast \! \varpi \ast \! \Pi^{a}
+ \hat{\mathscr{D}}_\upupsilon  P_a +\xi_{a}^{\hphantom{a} b}  P_b&=&0 .
  \label{carGcon} 
 \end{eqnarray}
The momenta   $\Pi$, $\Pi^{a}$, $ P_b$ and $ \Pi^{a}_{\hphantom{a}b} =  \Upsilon^{a}_{\hphantom{a}b}+\frac{1}{2}\Pi \delta^a_b$
are defined as variations of the action with respect to the triad $\left\{\upmu,\hat\uptheta^a\right\}$ (the explicit computation is accessible in Ref.\cite{BigFluid} for the Papapetrou--Randers frame,\footnote{Equations \eqref{carEbiscon} and \eqref{carGcon} were obtained for the first time in Ref. \cite{CMPPS1}. They have been recently rediscussed in \cite{Armas:2023dcz}.} where the organizing pattern is the subgroup of Carrollian diffeomorphisms instead of the subgroup of local orthogonal transformations in the tangent space). These are the energy density, the energy flux, the momentum and the stress.

For Carroll structures with vanishing Carrollian shear, $\xi_{ab}=0$, met e.g.  at null infinity of asymptotically flat spacetimes, six out of the ten Carroll Cotton tensors survive: $c_{(-1)}$, $c_{(0)}$, $\psi^a$  as in Eqs. \eqref{cfr}, \eqref{c7fr} and 
  $\chi^a$, $\Psi^{ab}$, $X^{ab}$. Using Eqs.  \eqref{c6fr},  \eqref{c10fr},  \eqref{c9fr} we find the simplified expressions of the latter: 
\begin{eqnarray}
\label{c6frnt}
\chi^{a}&=& \frac{1}{2} \ast \!\hat{\mathscr{D}}^a\hat{\mathscr{K}}+ \frac{1}{2} \hat{\mathscr{D}}^a\hat{\mathscr{A}}-2\ast \!\varpi\hat{\mathscr{R}}^a ,
\\
\label{c10frnt}
\Psi^{ab} &=&
\hat{\mathscr{D}}^a \hat{\mathscr{D}}^b\ast\! \varpi -\frac{1}{2}\delta^{ab} \hat{\mathscr{D}}_c \hat{\mathscr{D}}^c \ast\! \varpi -\hat\eta^{ab}   \hat{\mathscr{D}}_\upupsilon\ast\! \varpi^2,
\\
X^{ab}&=&\frac{1}{2}\ast\!\hat{\mathscr{D}}^a
\hat{\mathscr{R}}^b+
\frac{1}{2} \hat{\mathscr{D}}^a
\ast\!\hat{\mathscr{R}}^b
.
\label{c9frnt}
   \end{eqnarray}
These tensors now obey
\begin{eqnarray}
&&\mathcal{D}_{\text{Cot}} =-\hat{\mathscr{D}}_\upupsilon c_{(-1)}-\hat{\mathscr{D}}_a \psi^{a}=0
,
 \label{carDcotns} \\
 &&\mathcal{E}_{\text{Cot}} =-\hat{\mathscr{D}}_\upupsilon c_{(0)}-\hat{\mathscr{D}}_a \chi^{a}=0,
 \label{carEcotns}  
\end{eqnarray}
and
\begin{eqnarray}
&&\mathcal{I}_{\text{Cot}}^a = \frac{1}{2}\hat{\mathscr{D}}^a c_{(-1)}
+2 \ast\! \varpi   \ast\!\psi^{a}=0,
  \label{carIcotns}\\
 && \mathcal{G}_{\text{Cot}}^a = \frac{1}{2}\hat{\mathscr{D}}^a c_{(0)}- \hat{\mathscr{D}}_b \Psi^{ab}
+2 \ast\! \varpi   \ast\!\chi^{a}
+ \hat{\mathscr{D}}_\upupsilon \psi^a =0
,
  \label{carGcotns}\\
&&  \mathcal{H}_{\text{Cot}}^a =- \hat{\mathscr{D}}_b X^{ab}
+ \hat{\mathscr{D}}_\upupsilon \chi^a =0
.
 \label{carHcotns}
   \end{eqnarray}

On a Carroll manifold in Cartan frame, the degenerate metric is invariant under local Carroll-group transformations. Invariance of the action under its local orthogonal subgroup is in line with a symmetric $\Pi_{ab}$; invariance under local Carroll boosts demands  $\Pi^{a}=0$. This is not always met in Carrollian theories approached from relativistic theories at vanishing speed of light (see e.g. \cite{Rivera-Betancour:2022lkc})--- alternatively it can be imposed by hand as in \cite{Baiguera:2022lsw}. The Cotton tensor and the corresponding Chern--Simons dynamics 
\cite{MOPRS} 
admirably illustrate  this feature, which persists in the flux-balance equations of Ricci flat spacetimes, 
powered by gravitational radiation.

%\newpage

\end{document}